\documentclass[aps,PRB,reprint,showpacs,superscriptaddress,
footinbib,citeautoscript]{revtex4-1}

\usepackage{graphicx}

\usepackage[utf8]{inputenc}
\usepackage[export]{adjustbox}
\usepackage{csquotes}
\usepackage[american]{babel}
\usepackage[T1]{fontenc}
\usepackage{enumerate}
\usepackage{mdwlist}
\usepackage[activate=normal]{pdfcprot}
\usepackage{bbding}
\usepackage{color}
\usepackage{amsmath}
\usepackage{eqnarray,amsmath}
\usepackage{amssymb}
\usepackage{amsmath}
\usepackage{amsfonts}
\usepackage{mathrsfs}
\usepackage[titletoc,title]{appendix}
\usepackage{bm}
\usepackage{dcolumn}
\usepackage{color}
\usepackage[normalem]{ulem}
\usepackage[colorlinks=true,citecolor=blue]{hyperref}
\hypersetup{colorlinks=true,citecolor=blue,linkcolor=red
,urlcolor=blue}

\newcommand{\beq}{\begin{equation}}
\newcommand{\eeq}{\end{equation}}
\newcommand{\beqa}{\begin{eqnarray}}
\newcommand{\eeqa}{\end{eqnarray}}

\begin{document}

\title{Unidirectional valley-contrasting photo-current in strained transition metal dichalcogenide monolayers}
\author{Reza Asgari}
%\email{asgari@ipm.ir}
\affiliation{School  of  Physics,  University  of  New  South  Wales,  Kensington,  NSW  2052,  Australia}
\affiliation{ARC Centre of Excellence in Future Low-Energy Electronics Technologies,  UNSW Node,  Sydney 2052,  Australia}
\affiliation{School of Physics, Institute for Research in Fundamental Sciences, IPM, Tehran, 19395-5531, Iran}
\author{Dimitrie Culcer}
\affiliation{School  of  Physics,  University  of  New  South  Wales,  Kensington,  NSW  2052,  Australia}
\affiliation{ARC Centre of Excellence in Future Low-Energy Electronics Technologies,  UNSW Node,  Sydney 2052,  Australia}

\begin{abstract}
We examine the full static non-linear optical response of uniaxially strained transition metal dichalcogenide monolayers doped with a finite carrier density in the conduction band, in the presence of disorder. We find that the customary shift current is suppressed, yet we identify a strong, valley-dependent non-reciprocal response, which we term a \textit{unidirectional valley-contrasting photo-current} (UVCP). This DC current originates from the combined effect of strain and Kramers symmetry breaking by trigonal warping, while the contributions due to individual valleys can be separated by introducing an energy offset between them by means of a magnetization. This latter fact enables one to monitor inter-valley transitions. The UVCP is proportional to the mobility and is enhanced by the excitonic Coulomb interaction and inter-valley scattering, as well as by a top gate bias. We discuss detection strategies in state-of-the-art experiments.
\end{abstract}
\maketitle

\section{Introduction} 
The study of non-linear electromagnetic responses~\cite{papadopoulos2006non,Boyd2007} in solids has witnessed a strong resurgence recently thanks to the rise of topological materials~\cite{Hasan_Sci2009,Shen_Nat2010,Kane_RMP2010,Mele_PRL2007,Balents_PRB2007}. Non-linear responses typically require broken inversion symmetry, a condition most topological materials satisfy. In addition to spurring basic discoveries such as non-reciprocal currents~\cite{Nagaosa_NRR_NC2018, PhysRevLett.122.227402, tzuang2014non,kim2015non,shao2020non} and Hall effects~\cite{Zhang_PhyTod2010,Liu_AnnRev2016,Du_NC2019,Nandy_PRB2019,sinistyn_07} in time-reversal invariant systems, non-linear responses probe phenomena that are inaccessible in linear response, such as crystallographic and orientation information, and details of the band structure and grain boundaries \cite{yin2014edge, carvalho2019nonlinear}. Impressive experimental developments include generation of terahertz harmonics up to the seventh order in graphene at room temperature ~\cite{hafez2018extremely, PhysRevX.3.021014,PhysRevX.3.021014,soavi2018broadband,rostami2020many,PhysRevB.91.205407}.

The second-order response to an AC electric field contains a static, DC part responsible for rectification, shift and injection currents, and the resonant photovoltaic effect~\cite{
belinicher1980photogalvanic, belinicher1978space, Ivchenko_SovPhy1984, Khurgin_JOS1994, PhysRevB.23.5590,
fridkin2001bulk, PhysRevLett.119.067402, Nature493, PhysRevLett.7.118, Boyd2007,green2006third,PhysRevB.100.064301, Belinicher_SovPhy1978,Shengyuan_PRB2009,Sipe_PRB2010,PhysRevLett.109.116601,npj2016,PhysRevB.97.245143, morimoto2016topological, npj2016, autere2018nonlinear, mak2016photonics, green2020tracking, behura2019graphene, PhysRevB.91.125424, npj2016, cook2017design, Nagaosa_NRR_NC2018, PhysRevB.95.035134, RevModPhys.82.1959, ahn2021riemannian, wang2019ferroicity, ishizuka2017local, nakamura2017shift}. Such DC phenomena are intimately connected to topology and underlie photovoltaic devices~\cite{yang2010above, wang20182d, PhysRevB.79.081406}. For example, one mechanism is associated with a shift in the electron wave packet centre of mass during light-induced transitions, which is determined by the derivative of the momentum-space phase of the transition matrix element and by the Berry connection. On the other hand, the injection mechanism comes from the fact that electron and hole have different velocities and that the coherent ${\bf k}$ and ${-\bf k}$ excitations are unbalanced, resulting in a ${\bf k}$ and $-{\bf k}$ asymmetry in the steady-state population and a net current. Recently a large longitudinal photo-current peak originating from topological band crossings was identified by terahertz emission spectroscopy with tunable photon energy in the chiral topological semimetal CoSi~\cite{ni2021giant}. This is believed to indicate a strong injection current ~\cite{PhysRevB.93.115433, ma2019nonlinear} in materials without inversion symmetry. Moreover, a magnetic shift current ~\cite{PhysRevB.105.075123, wang2020nonlinear} can be finite in a magnetic parity-violating system, and diminishes
when the system is nonmagnetic~\cite{PhysRevX.11.011001}. Recent studies of such second-order magneto-optical effects ~\cite{PhysRevB.105.075123} have identified magnetic shift and injection currents as part of the rectification current.

Yet the simple picture of non-linear DC response above has considerable limitations. Firstly, it focuses on intrinsic contributions while ignoring the kinetic processes of relaxation of photoexcited electrons ~\cite{sturman2020ballistic}. Secondly, the study of non-linear DC phenomena has focused on intrinsic effects in single-valley topological materials, leaving vital questions unanswered: \textit{What is the dominant response of multi-valley materials, how is it affected by disorder, and what information can it yield on inter-valley processes?} Since valleys are time-reversed partners multi-valley systems support interactions absent for single valleys, where they would break Kramers degeneracy. Strong warping~\cite{PhysRevB.84.245435,
PhysRevLett.98.176806, PhysRevB.90.195413, saynatjoki2017ultra} also breaks particle-hole symmetry, such that terms dominant in e.g. topological insulators are suppressed, while novel contributions may arise. 

%%%%%%Fig1
\begin{figure}[tbp]
	\includegraphics[width=4cm]{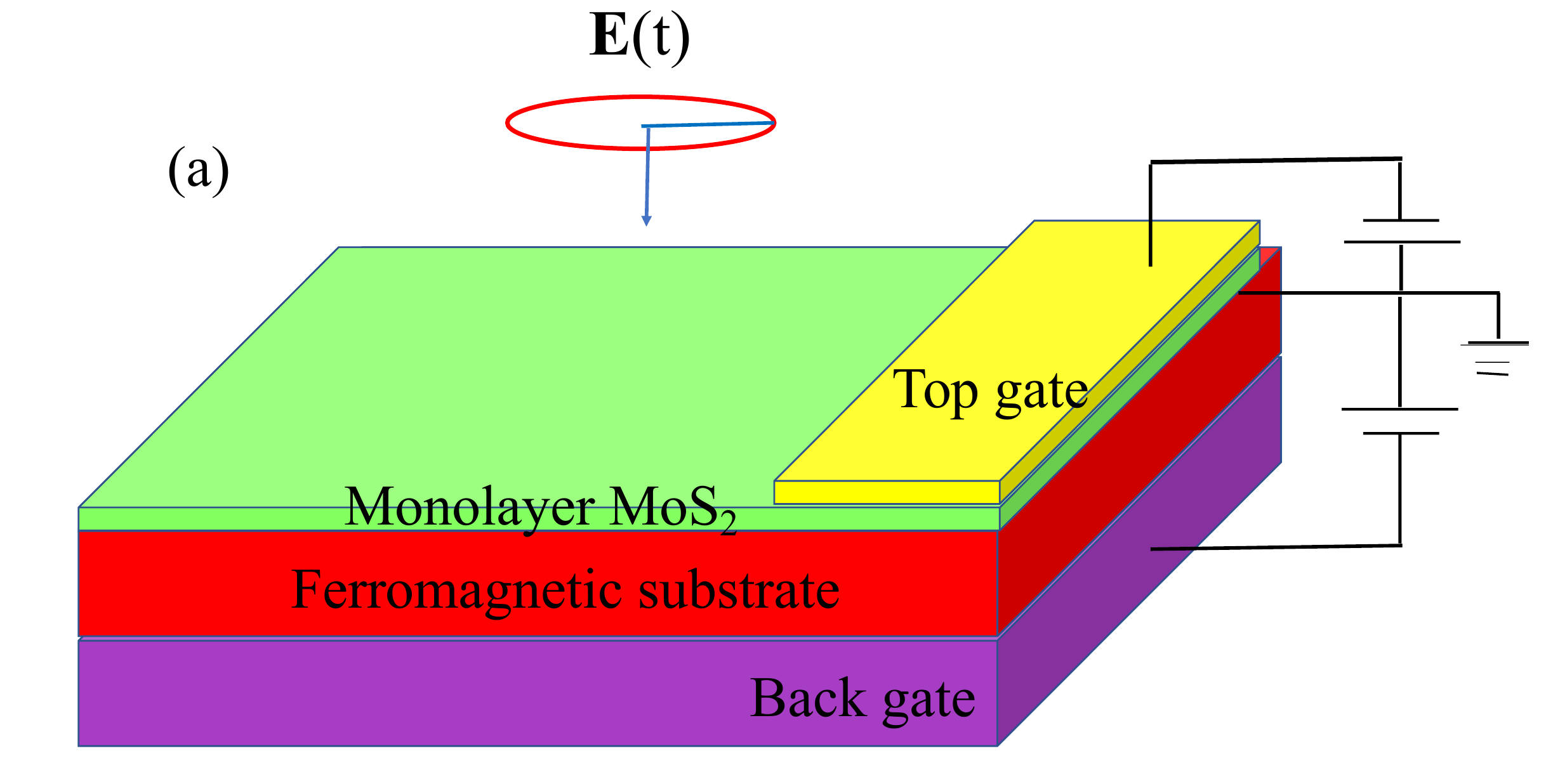}
	\includegraphics[width=4cm]{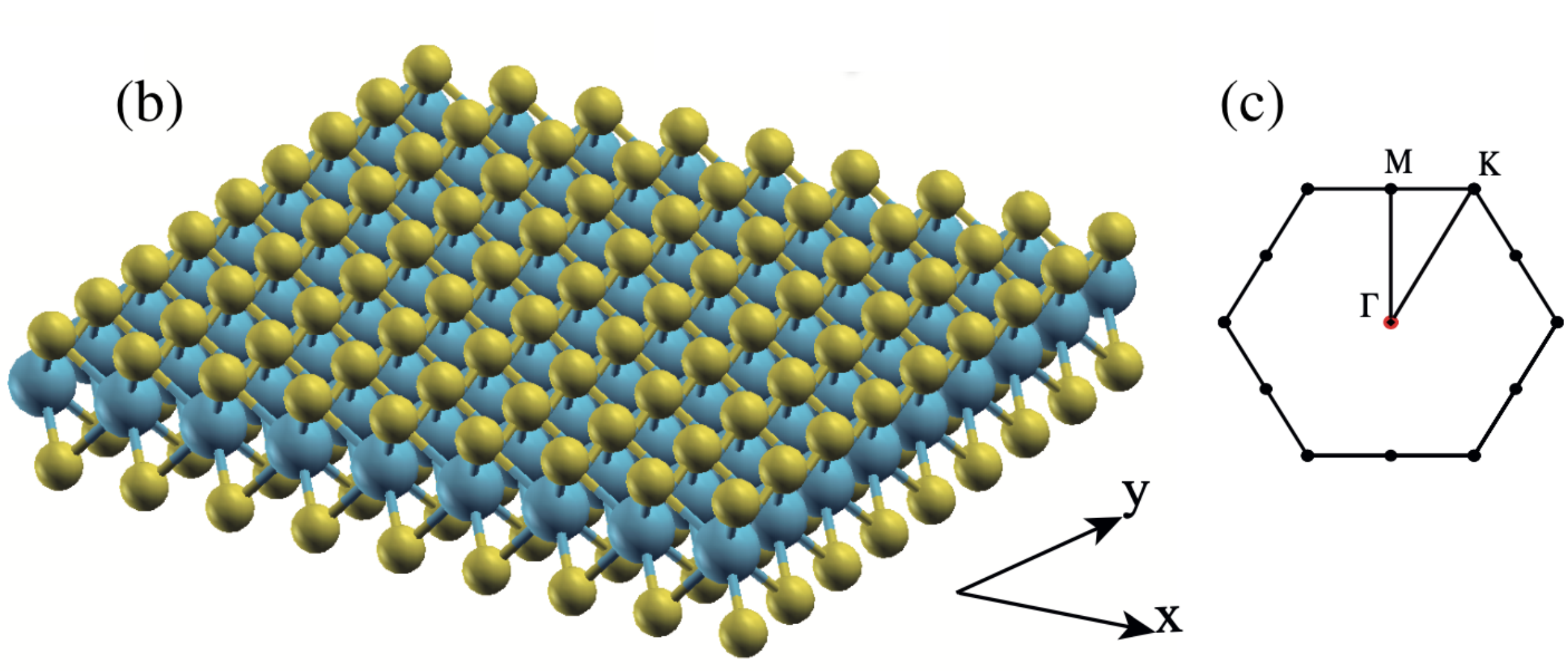}
	\caption{(Color online) (a) Illustration of a device consisting of gated monolayer MoS$_2$ on top of a magnetic substrate. (b) The lattice structure of monolayer MoS$_2$ as a representative of TMDCs monolayers. There are three atomic layers, and each atomic layer has a trigonal lattice. The top and bottom layers are Sulfur (chalcogen) atoms and the inner layer comprises Molybdenum (metal) atoms. (c)  1BZ structure of MoS$_2$ in momentum-space.  
}\label{fig0}
\end{figure}

To address these questions, in this work we provide a full theory of the static non-linear optical response of monolayer strained TMDs with broken inversion symmetry, whose development has leapfrogged in recent years ~\cite{saynatjoki2017ultra, hong2020structuring, PhysRevLett.114.097403, wang2015nonlinear, PhysRevB.94.245434, PhysRevB.99.085301, joshi2020localized, culcer2020transport}. We take as a prototype strained MoS$_2$ \cite{xiao_PRL2012, PhysRevB.88.085440, kormanyos2015k,PhysRevB.92.195402}, known for strong second-harmonic generation~\cite{liu2017high, PhysRevB.87.161403}, though the findings apply to all TMDs. Monolayer MoS$_2$ has two valley minima at ${\bm K}$ and ${\bm K}'$ and the symmetry point group at the ${\bm K}$-point is $C_{3h}$. In order to activate the DC non-linear optical response one needs to apply strain, breaking the three-fold symmetry \cite{PhysRevB.92.195402}. We demonstrate that the static non-linear response of the strained material to circularly polarized light is dominated by a non-reciprocal, unidirectional current due to the interplay of topological properties -- the Berry connection and valence band topology influenced by trigonal warping -- and scattering processes, and is enhanced by inter-valley scattering. We refer to this response as a \textit{unidirectional valley-contrasting photo-current} (UVCP). The theory incorporates the inter-band and inter-valley optical transitions in the presence of short-range impurity scattering, which also couples the valleys. The results show a large non-linear response with non-trivial features at optical light frequencies.

The UVCP is driven by scattering processes between the two valleys. Two additional factors enhance the effect and its utility, as shown in Fig.~\ref{fig0}. Firstly a bias voltage is employed to break mirror symmetry and increase the non-linear DC response. Secondly the doped MoS$_2$ is placed on a magnetic substrate which breaks the valley degeneracy, and this enables one to resolve the contributions of the two individual valleys. Consequently the effect could provide a method of monitoring intervalley transitions in transition metal dichalcogenides. 

In connection to the above, as part of this calculation we examine the intrinsic shift current for MoS$_2$ following the theory reported in~\cite{morimoto2016topological}. The extremely small optical current does not indicate any feature in terms of optical light frequency (Appendix: Fig~\ref{fig8}). This is not surprising: For almost non-polar materials such as transition metal dichalcogenides (TMDs), phosphorus, and gapped graphene, the shift is expected to be small. The delocalized states with more mobile charge carriers give a high shift current response whereas states with less mobile carriers yield a low shift current. Transition metal dichalcogenide monolayers are non-polar and the conduction and valence bands are made effectively by $d$-orbitals. We stress, however, that our formalism is completely general and we calculate the \textit{total} second order DC response to the applied electric field.  Terms leading to the shift current and any present asymmetries in the Fermi surface are built into our formalism from the start. Having completed the evaluation of the full non-linear DC response, we identify the UVCP term as the main physical process. 

This paper is organized as follows. First we discuss the low-energy model Hamiltonian describing strained MoS$_2$ monolayer system in Sec.~\ref{model}. The quantum kinetic theory and density matrix approach generalized for two valleys system and excitons physics for the MoS$_2$ with Keldysh potential and solving the Bethe-Salpeter equations for density matrix elements are discussed in Sec.~\ref{model} as well. Our main results are given in Sec.~\ref{nomeric} and finally we wrap up our discussions in Sec.~\ref{conc}.

\section{Model and theory}\label{model}

A low-energy ${\bf k}\cdot {\bf p}$ continuum model Hamiltonian around the K and K' points which describes a two-dimensional strained TMD system is provided by the terms ${\cal H}_0={\cal H}_1+{\mathcal H}_{w}$ \cite{PhysRevB.92.195402}, where 
\begin{eqnarray}\label{hami}
{\mathcal H}_1=&& \frac{\Delta_0+\lambda_0 s \tau_v}{2}+\frac{\Delta_{\tau_v}+\lambda s \tau_v}{2}\sigma_z+t_0 a_0 \left({\bf k}
+b_1{\bf A}\right)\cdot{\bm \sigma}_{\tau_v}\nonumber\\
&&+\frac{\hslash^2}{4m_0}\left(|{\bf k}
+b_2{\bf A}|^2\alpha
+|{\bf k}+b_3{\bf A}|^2\beta\sigma_z\right)+D+V_{\text{el}}
\end{eqnarray}
and trigonal warping ${\mathcal H}_{w}$ contribution is given by
\begin{equation}\label{warping}
{\mathcal H}_{w}=t_1 a_0^2 \left(({\bf k}
+b_1{\bf A})\cdot{\sigma}_{\tau_v}^*\right)\sigma_x\left(({\bf k}
+b_1{\bf A})\cdot{\sigma}_{\tau_v}^*\right)
\end{equation}
where the Pauli matrices ${\bf \sigma}_{\tau_v}=(\tau_v \sigma_x$, $\sigma_y)$ acts on the two component wave functions, $b_m=\frac{e}{\hslash}\tau \eta_m$ with $\eta_1=0.002$, $\eta_2=-5.655$ and $\eta_3=1.633$,  $\Delta_0=-0.11$ eV, $\lambda=-80$ meV, $t_0=2.34$ eV, $t_1=-0.14$ eV, $\alpha=-0.01$, $\beta=-1.54$, $D$ is a diagonal matrix with elements $a_1|\bf A|^2$ and $a_2|\bf A|^2$ where $a_1=15.95$, $a_2=-2.2$ eV.  The pseudovector field for uniaxial strain is given by ${\bf A}=(1-\nu, 0)\varepsilon$ and the elastic potential 
$V_{\text{el}}=(1+\nu)\varepsilon I$ with Poisson ratio $\nu=-0.125$ and strain $\varepsilon$. Here $\tau_v=\pm$ is the valley index, $s=\pm$ is the spin index and the modified band gap is $\Delta_{\tau_v}$. For the sake of simplicity, the spin-orbit effect in the conduction band is neglected. We consider a system on top of a ferromagnetic substrate with a finite perpendicular magnetization. Notice that in the presence of the magnetization, the interacting Hamiltonian is $s m \sigma_z$ where $s=\pm$ indicates the spin and $m$ is the magnetization. In the top valence band in valleys, spins are different, therefore, we have $\tau_v m \sigma_z$. Subsequently, a magnetization is induced by a proper substrate to break valley degeneracy and the band gap is modified as $\Delta_{\tau_v}=\Delta-\tau_v m$ with $\Delta=1.82$ eV. The detail of the strain and bias voltage on the low-energy Hamiltonian are provided in  
Appendix \ref{A:hamiltonian}.  

Trigonal warping causes the valence band dispersion to be strongly anisotropic, unlike that in the conduction band. The effective-mass approximation restricts our theory to a small energy range in the vicinity of the band edge, though the large effective masses ensure its applicability to high excitations. 

It is known that uniaxial and shear strain induces a shift of the band edges from the K points~\cite{PhysRevB.92.195402}, similar to the strained graphene. Both conduction and valence band edges shift in phase towards the $\Gamma$ point for compressive $\varepsilon\le 0$ strain, whereas they move in the opposite direction for tensile strain. The position of the conduction band minimum and the valence band maximum are given by $-{\bf A}\frac{\alpha \eta_2+\beta \eta_3+\gamma \eta_1}{\alpha+\beta+\gamma}$ and  $-{\bf A}\frac{\alpha \eta_2-\beta \eta_3-\gamma \eta_1}{\alpha-\beta-\gamma}$, respectively, where $\gamma=4m_0v^2/(\Delta_{\tau_v}-s\lambda)$.

\subsection{Quantum kinetic theory} 

The quantum kinetic theory based on the density matrix \cite{PhysRevB.96.035106, PhysRevB.96.235134, PhysRevLett.124.087402} successfully describes inter-band transitions in the presence of scattering terms.
For a system with ${\cal H}={\cal H}_0+{H}_E$, the single particle density matrix $\rho$ obeys the quantum Liouville equation:
\begin{equation}
\frac{\partial \langle \rho \rangle}{\partial t}+\frac{i}{\hslash}\langle [{\cal H}_0, \rho]\rangle+J(\langle \rho\rangle)=
-\frac{i}{\hslash}\langle [{\cal H}_E,\rho]\rangle
\end{equation}
where the scattering term $J$ is expressed in the Born approximation \cite{PhysRevB.96.235134} and we assume the correlation function $\langle U({\bf r})U({\bf r}')\rangle=n_i U_0^2\delta({\bf r}-{\bf r}')$ with $n_i$ is the impurity density disorder and $U$ is a disorder potential. The interaction with the time dependent external field is represented by ${\cal H}_E=\textrm {e}{\bf E}(t)\cdot {\bf r}$. The scattering term is given by
$
J(\langle \rho \rangle)=\int dt' [U ,[ U(t'), \langle \rho \rangle]]
$
where $U(t')=e^{-iT/\hslash} U e^{iT/\hslash}$ with time-evolution operator $T=\int_0^t ({\cal H}_0+{\cal H}_E(t')) dt'$ including the electric field term. Usually, ${\cal H}_E(t')$ does not contribute to the scattering term $J(\langle \rho \rangle)$ if ${\cal H}_0$ is a single-component Hamiltonian and $U$ represents scalar scattering, however, the electric field contribution is important for spin and pseudospin-dependent scattering.

To account for the two valleys we divide the Brillouin zone into 3 sectors: 1) ${\bf K}-{\bf \Lambda} \le {\bf k}_1\le {\bf K}+{\bf \Lambda}$, 2) ${\bf K}'-{\bf \Lambda} \le {\bf k}_2\le {\bf K}'+{\bf \Lambda}$ and 3) intervalley or intermediate region. The density matrix is accordingly divided into three sectors: $\langle \rho_1 \rangle$, $\langle \rho_2 \rangle$ around K and K' point and an intermediate, unoccupied region of large wave vectors which is outside our interest. Here $\langle \rho_{i} \rangle$, $i=1, 2$ are $2\times2$ matrices in the conduction and valence band representation, and we are selecting two sectors in ${\bf k}$-space. All the dynamics between these sectors, i. e. all the inter-valley dynamics, are contained in the scattering term. Inter-valley scattering is quantified by $\lambda U_0$, where we set $\lambda=0.7$ and $U_0^2=\tau^2/n_i$.

We can also divide the scattering term into two parts for $J_1(\langle \rho \rangle)$ and $J_2(\langle \rho \rangle)$ such that 
\begin{eqnarray}
&&\frac{\partial\langle \rho_1 \rangle}{\partial t}+\frac{i}{\hbar}\langle [{\mathcal H}_0,\rho_1]\rangle+J_1(\langle \rho \rangle)=-\frac{i}{\hbar}\langle[{\mathcal H}_E,\rho_1]\rangle\\
&&\frac{\partial\langle \rho_2 \rangle}{\partial t}+\frac{i}{\hbar}\langle [{\mathcal H}_0,\rho_2]\rangle+J_2(\langle \rho \rangle)=-\frac{i}{\hbar}\langle[{\mathcal H}_E,\rho_2]\rangle\nonumber
\end{eqnarray}

We can also write $J_1(\langle \rho \rangle)$ (similarly for $J_2(\langle \rho \rangle)$) \cite{PhysRevB.96.235134} as:
\begin{widetext}
\begin{eqnarray}
&&J_1(\langle \rho_{{\bf k}_1} \rangle)=\\
&&\frac{1}{\hbar^2}\int dt' \{\sum_{{{\bf k}'_1}\in {{\text sector} 1}}\{ U_{{\bf k}_1{\bf k}'_1}U_{{\bf k}'_1{\bf k}_1}(t') \langle \rho_{{\bf k}_1} \rangle
+\langle \rho_{{\bf k}_1} \rangle U_{{\bf k}_1{\bf k}'_1}(t')U_{{\bf k}'_1{\bf k}_1}-U_{{\bf k}'_1{\bf k}_1}\langle \rho_{{\bf k}'_1} \rangle U_{{\bf k}'_1{\bf k}_1}(t')-U_{{\bf k}_1{\bf k}'_1}(t')\langle \rho_{{\bf k}'_1} \rangle U_{{\bf k}'_1{\bf k}_1}\}\nonumber\\
&&+\sum_{{{\bf k}'_2}\in {{\text sector} 2}}\{ U_{{\bf k}_1{\bf k}'_2}U_{{\bf k}'_2{\bf k}_1}(t') \langle \rho_{{\bf k}_1} \rangle
+\langle \rho_{{\bf k}_1} \rangle U(t')_{{\bf k}_1{\bf k}'_2}U_{{\bf k}'_2{\bf k}_1}-U_{{\bf k}'_2{\bf k}_1}\langle \rho_{{\bf k}'_2} \rangle U(t')_{{\bf k}'_2{\bf k}_1}-U_{{\bf k}_1{\bf k}'_2}(t')\langle \rho_{{\bf k}'_2} \rangle U_{{\bf k}'_2{\bf k}_1}\}\}\nonumber
\end{eqnarray}
\end{widetext}
There is $\sum_{{{\bf k}'_1}\in {{\text sector} 1}}$ which is the usual both scattering term for one valley, and $\sum_{{{\bf k}'_2}\in {{\text sector} 2}}$ which represents intervalley scattering. There is an equivalent term for $J_2(\langle \rho_{{\bf k}} \rangle)$ with proper momentum dependence. 
Now both $\rho_{{\bf k}_1}$ and $\rho_{{\bf k}_2}$ are $2\times 2$ matrices and $J_1$ and $J_2$ allow all relevant interband transitions.

Here, two scattering terms will have contributions from ${\cal H}_E$, so:
\begin{equation}
J=J_0+J_E,
\end{equation}
where $J_0$ is the bare scattering term with $e^{-i{{\cal H}_0} t/\hbar} U e^{i{{\cal H}_0} t/\hbar}$ in the time integral. The matrix element of $J_0$ is then given by
\begin{widetext}
\begin{eqnarray}
&&\hbar^2J^{mn}_{0,{\bf k}_1}(\langle \rho_{} \rangle)=\\
&&\int dt' \{\sum_{{{\bf k}'_1}\in I}\{ U^{mm'}_{{\bf k}_1{\bf k}'_1}U^{m'm''}_{{\bf k}'_1{\bf k}_1} (t')\langle \rho_{{\bf k}_1} \rangle^{m''n}
+\langle \rho_{{\bf k}_1} \rangle^{mm'} U^{m'm''}_{{\bf k}_1{\bf k}'_1}(t')U^{m''n}_{{\bf k}'_1{\bf k}_1}-U^{mm'}_{{\bf k}'_1{\bf k}_1}\langle \rho_{{\bf k}'_1} \rangle^{m'm''} U^{m''n}_{{\bf k}'_1{\bf k}_1}(t')-U^{mm'}_{{\bf k}_1{\bf k}'_1}(t')\langle \rho_{{\bf k}'_1} \rangle^{m'm''} U^{m''n}_{{\bf k}'_1{\bf k}_1}\}\nonumber\\
&&+\sum_{{{\bf k}'_2}\in II}\{ U^{mm'}_{{\bf k}_1{\bf k}'_2}U^{m'm''}_{{\bf k}'_2{\bf k}_1}(t') \langle \rho_{{\bf k}_1} \rangle^{m''n}
+\langle \rho_{{\bf k}_1} \rangle^{mm'} U^{m'm''}_{{\bf k}_1{\bf k}'_2}(t')U^{m''n}_{{\bf k}'_2{\bf k}_1}-U^{mm'}_{{\bf k}'_2{\bf k}_1}\langle \rho_{{\bf k}'_2} \rangle^{m'm''} U(t')^{m''n}_{{\bf k}'_2{\bf k}_1}-U^{mm'}_{{\bf k}_1{\bf k}'_2}(t')\langle \rho_{{\bf k}'_2} \rangle^{m'm''} U^{m''n}_{{\bf k}'_2{\bf k}_1}\}\}\nonumber
\end{eqnarray}
\end{widetext}
where sum over inner index $m'$ and $m''$ are taken. In Eq. (7), $I$ and $II$ refer to the sectors 1 and 2, respectively.

To work out $J_E$, we define $g=\rho-\langle \rho \rangle$ and express the quantum Liouville equation as:
\begin{equation}
\frac{\partial g}{\partial t}+\frac{i}{\hbar}[{\mathcal H}_0+{\mathcal H}_E,g]+\frac{i}{\hbar}[U,g]-\frac{i}{\hbar}\langle[U,g]\rangle=-\frac{i}{\hbar}[U, \langle \rho \rangle]
\end{equation}  
In the Born approximation, we can ignore the last two terms on the left hand side of above equation and what remains is a leading order in time inhomogeneous linear differential equation for $g$ which can be integrated yield
\begin{equation}\label{g0}
g_0=-\frac{i}{\hbar}\int_0^{\infty} dt' e^{-i{{\cal H}_0} t'/\hbar} [U, \langle \rho \rangle] e^{i{{\cal H}_0} t'/\hbar}
\end{equation}
which gives the scattering term $J_0(\langle \rho \rangle)$ and next term is given by~\cite{PhysRevB.96.035106}
\begin{equation}
\frac{\partial g_E}{\partial t}+\frac{i}{\hbar}[{\mathcal H}_0,g_E]=-\frac{i}{\hbar}[H_E, g_0]
\end{equation}
Therefore, the solution of $g_E$ is given by
\begin{equation}\label{ge}
g_E=-\frac{i}{\hbar}\int_0^{\infty} dt'' e^{-i{{\cal H}_0} t''/\hbar} [{\cal H}_E(t''), g_0] e^{i{{\cal H}_0} t''/\hbar}
\end{equation}
and the scattering term due to $g_E$ is given by
\begin{eqnarray}
\langle m,{\bf k} |J_E(\langle \rho \rangle) |n, {\bf k} \rangle=\frac{i}{\hbar}&&\sum_{m',{\bf k}'} \{U^{mm'}_{{\bf k}{\bf k}'} \langle m',{\bf k}'|g_E |n, {\bf k} \rangle\nonumber\\
&& - \langle m,{\bf k}|g_E | m', {\bf k}'\rangle U^{m'n}_{{\bf k}'{\bf k}} \}
\end{eqnarray}
Notice that optical inter-valley transition in the theory goes beyond the Fermi golden rule transition since the Fermi golden rule does not possess the matrix element for two different states~\cite{merzbacher1998quantum} and its matrix is diagonal, however, in our theory, off-diagonal elements of the scattering terms defined in $J_E$, are nonzero.

The density matrix can be expanded in the powers of the electric field~\cite{PhysRevLett.124.087402}, and thus the quantum kinetic equation at K point for diagonal $f^n_d=\langle \rho^n_{1d} \rangle$ and off-diagonal $f^n_{od}=\langle \rho^n_{1od} \rangle$ terms can be simplified as
\begin{eqnarray}\label{f}
&&\frac{\partial f^n_d}{\partial t}+\frac{f^n_d}{\tau_1}+J^d_{E}(f^{(n-1)})=-\frac{i}{\hslash}\langle[{\mathcal H}_E,f^{(n-1)}_d]\rangle-J^{'}_0(\langle \rho \rangle)\nonumber\\
&&\frac{\partial f^n_{od}}{\partial t}+\frac{i}{\hslash}\langle |[{\mathcal H}_0,f^n_{od}]|\rangle+\frac{f^n_{od}}{\tau_2}+J^{od}_{E}(f^{(n-1)})=\nonumber\\
&&-\frac{i}{\hslash}\langle |[{\mathcal H}_E,f^{(n-1)}_{od}]| \rangle
-J^{''}_{0}(\langle \rho \rangle) 
\end{eqnarray}
where relaxation times are given by Eqs. (\ref{tau}-\ref{tau2}) and $J^{'}_0=J_0(\langle \rho \rangle)-f^n_d/\tau_1$ (see Eq. (\ref{Jtau1}), and equally $J^{''}_0=J_0(\langle \rho \rangle)-f^n_{od}/\tau_2$, Eq. (\ref{Jtau2}). The time $\tau_1$ describes the relaxation of an arbitrary initial distribution function to a non-equilibrium Fermi function, and $1/\tau_2$ is the damping of the oscillation of the transition amplitude and hence of the macroscopic polarization function. In general $\tau_{\textrm i=1,2}$ could be a function of the frequency and external fields. The frequency dependence might affect the non-linear optics in low-frequency or intra-band process~\cite{rostami2020many}. Since we are interested in the optical transition, for the sake of simplicity, we assume a constant $\tau=\tau_2$=$\tau_1$ and hence the theory is valid when $E_{\rm F}-\Delta_{\tau_v}/2\gg \hslash/\tau$ where $E_{\rm F}$ is the electron-doped Fermi energy.  

In the system studied here the recombination rate is smaller than the excitation rate. The reasons lie in the fact that first in non-centrosymmetric crystals the principle of detailed balancing is broken for non-equilibrium photo-excited carriers~\cite{fridkin2001bulk}. Second, the electron mobility is greater than the hole mobility implying different band masses. Finally, a separation between the center of the electron wave-packet and hole packet in real space occurs, which becomes larger when trigonal warping is included. 

Solving Eq.~(\ref{f}), the DC part of the optical current (summing over valleys) is 
$
{\bf j}=-\textrm{e}\int \frac{d{\bf k}}{4\pi^2} {\text Tr}[{\bf v} {\bf f}_{{\bf k}}]
$
where the velocity tensor $\hbar{\bf v}=\frac{D {\cal H}_0}{D{\bf k}} \equiv \frac{d {\cal H}_0}{d{\bf k}} - i[{\bm R}, {\cal H}_0]$, and $D/D{\bf k}$ is the covariant derivative. The Berry connection ${\cal R}^{mm'}_{kk'}=i\langle\psi^{m}_k|\nabla_{\bf k'}|\psi^{m'}_{k'}\rangle$ for each valley and ${\cal R}^{m'm}_{\bf k}={{\cal R}^*}^{mm'}_{\bf k}$ in which $m(m')=\pm1$ refers to the conduction or valence band. Explicitly,
${\bf v}=(\nabla_{\bf k} \varepsilon_{\bf k}-i[{\cal R},{\cal H}_0])/\hslash$ where $\varepsilon_{\bf k}$ is the energy, and
\begin{equation}
{\bf j}^m=-\frac{\textrm {e}}{\hslash}\int \frac{d{\bf k}}{4\pi^2} \{ \nabla_{\bf k} \varepsilon^m_{\bf k} f^{mm'}_{{\bf k}}\delta_{mm'}-i {\cal R}^{m'm}_{\bf k} [\varepsilon^m_{\bf k}-\varepsilon^{m'}_{\bf k}]f^{mm'}_{{\bf k}}\}.
\end{equation}
and thus the interplay between the Berry connection and band topology plays essential role in the UVCP. 

It is important to note that a naive application of the reduced two-valley model will result in an unphysical non-zero DC current to second order in the electric field despite the system possessing C$_{3h}$ symmetry. This is because the model is restricted to two valleys, whereas the system has six valleys, which all contribute to the second-order current. When considering the contribution of all six valleys it is evident that the second-order DC current will vanish, as is expected for C$_{3h}$ symmetry. However, the addition of uniaxial strain causes the contribution of the two valleys parallel to the strain direction to be different from that of the remaining valleys. Hence a nonlinear DC signal survives. In order to obtain the correct optical current, we calculate $\sigma^{(2)}=[J(\varepsilon)-J(\varepsilon=0)]/I_0$ where $I_0$ represents the intensity of the incident light and $\varepsilon$ is the strain. It should be noted that due to the deformation of the crystal lattice, there is a possibility that a small difference remains between separate pairs of valleys when strain is applied. However, this difference between pairs should be tiny since the strain and Poisson's ratio are small and so we ignore these contributions.

% No idea what the above means.

\subsection{Excitons and the Bethe-Salpeter equation} 
Since an absorbed photon results in the creation of an electron-hole pair a coupled electron-hole state emerges owing to the Coulomb interaction, which can be viewed as a non-charged exciton. This new state leads to additional absorption peaks shifted from the fundamental absorption edge by the coupling energies~\cite{scharf2019dynamical}. MoS$_2$ possesses relatively large effective band masses and its charge carriers are confined to a single atomic layer. Accordingly the electron-hole interactions are much stronger than in conventional semiconductors~\cite{PhysRevLett.111.216805, quintela2020colloquium}. Moreover, finite momentum excitons are optically inactive but can play an important role in valley dynamics. Exciton states can be obtained by solving a two-body problem with attractive interactions between one conduction band electron and one valence band hole. We use an interaction potential of the Keldysh form~\cite{PhysRevB.88.045318} 
\begin{equation}
V_{R}=\frac{\pi e^2}{2\epsilon r_0}[H_0(R/r_0)-Y_0(R/r_0)]
\end{equation}
to account for the finite width of the system and the spatial inhomogeneity of the dielectric screening environment. The Bessel function of the second kind is defined by
\begin{equation}
Y_n(x)=\frac{J_n(x)\cos(n x)-J_{-n}(x)}{\sin(nx)}
\end{equation}
where $J_n(x)$ is the Bessel function of the first kind. The Struve function, $H_n(x)$ solves the inhomogeneous Bessel equation. 
Here $r_0=33.87 {\AA}/\epsilon$, the averaged environment dielectric constant is $\epsilon$ and the Fourier transform of the bare potential is given by $V_q=\frac{2 \pi e^2}{\epsilon q} F(q)$ where $F(q)=\frac{1}{1+r_0 q}$ . We are interested in the response of the system near the absorption edge when $|\hslash \omega-{\Delta}_{\tau_v}| \ll {\Delta}_{\tau_v}$ with the effective optical band gap ${\Delta}_{\tau_v}$ for an electron doped system. In the Hartree-Fock approximation, the first order interactive approximation yields (see detailed discussions in Appendix \ref{A:excitons})
\begin{eqnarray}
&&f^{(1)}_{od}({\bf k})=\frac{\textrm {e}}{\omega}\frac{[1-\theta(\varepsilon_{\rm F}-\varepsilon_{\bf k}] (\varepsilon^c_{\bf k}-\varepsilon^{\nu}_{\bf k}){\bf E}\cdot {{\bf \cal R}^{c \nu }}({\bf k})}{(\varepsilon^c_{\bf k}-\varepsilon^{\nu}_{\bf k}-\hslash \omega-i \hslash \tau)}\nonumber\\
&&+\frac{\textrm {e}}{\omega((\varepsilon^c_{\bf k}-\varepsilon^{\nu}_{\bf k}-\hslash \omega-i \hslash \tau))} {\bf E}\cdot {\bf K}({\bf k})
\end{eqnarray}
where ${\bf K}({\bf k})$ is given by 
\begin{widetext}
\begin{eqnarray}
{\bf K}({\bf k})=\int k'dk' &&[(1+\cos \theta_k)(1+\cos \theta_{k'})g(0)+2\sin \theta_k \sin \theta_{k'} g(s)
+(1-\cos \theta_k)(1-\cos \theta_{k'})g(2s)]\nonumber\\
&&\times\frac{[1-\theta(\varepsilon_{\rm F}-\varepsilon^c({k}')] (\varepsilon^c({k}')-\varepsilon^{\nu}({k}')){{\bf \cal R}^{c \nu }}({k}')}{(\varepsilon^c({k}')-\varepsilon^{\nu}({k}')-\hbar \omega-i \hbar \tau)}
\end{eqnarray}
\end{widetext}
and 
\begin{equation}
g(m)= \frac{1}{2(2\pi)^2}\int_0^{2\pi} d\phi
V(|{\bf k}-{\bf k}'|)\cos (m\phi)
\end{equation}
here $\cos \theta_k=\Delta/\varepsilon_k$, and $\cos \phi={\bf k}\cdot{\bf k}'/({|{\bf k}|}{|{\bf k}'|})$.
Having calculated first order density matrix, the second order off-diagonal density matrix element can be obtained.

The spectral function and the optical linear susceptibility are given by the real-space Green's function for which we define the Sommerfeld factor~\cite{Vasko2005}. In addition, $f^{(1)}_{od}({\bf k})$ can be obtained from the ${\bm k}-$space Green's function. The Sommerfeld factor implies a peak around the optical transition in the density of states, hence we expect a jump in the current near the optical transition, which indicates the large density of states i.e. the existence of the Sommerfeld factor. The latter is given by $ |\psi^{n}_{{\bf k}}({\bf r}=0)|^2\sim \frac{8}{\pi a^{*2}_B}$ where the effective Bohr radius $a^*_B=\hslash^2 \epsilon/m^* \textrm {e}^2$ with $m^*\sim 0.25 m_e$. This expression increases as $m^*/\epsilon $ and the energy is $ \hslash^2 k^2/2 m^*$. The peak in optical absorption near the band edge originates from the Sommerfeld factor~\cite{schafer2013semiconductor} and its amplitude depends on material parameters.

%%%%%%Fig2%%%%%%Fig2%%%%%%Fig2%%%%%%Fig2%%%
\begin{figure}[tbp]
	\includegraphics[width=7cm]{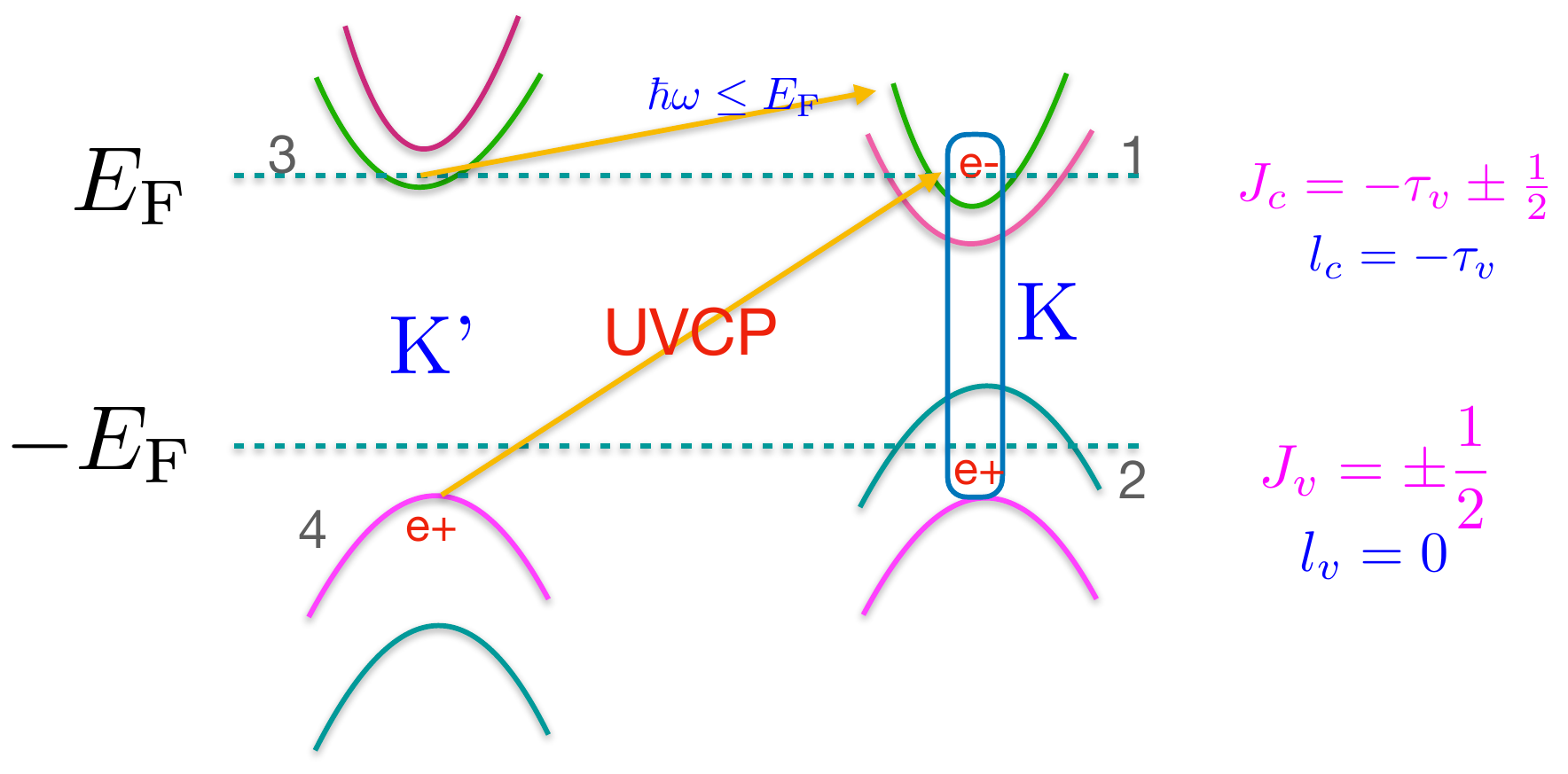}
\caption{(Color online) The interband transition in the same valley and intervalley transition between various valleys are demonstrated. Because inversion symmetry is broken in MoS$_2$, the coupling is allowed between real spin and valley pseudospin and gives rise to valley-dependent optical selection rules. $J_{c(v)}$ refers to the total angular momentum of the bands $c(v)$. }\label{fig2}
\end{figure}

%%Fig1
\begin{figure}[htp]
	\includegraphics[width=4cm]{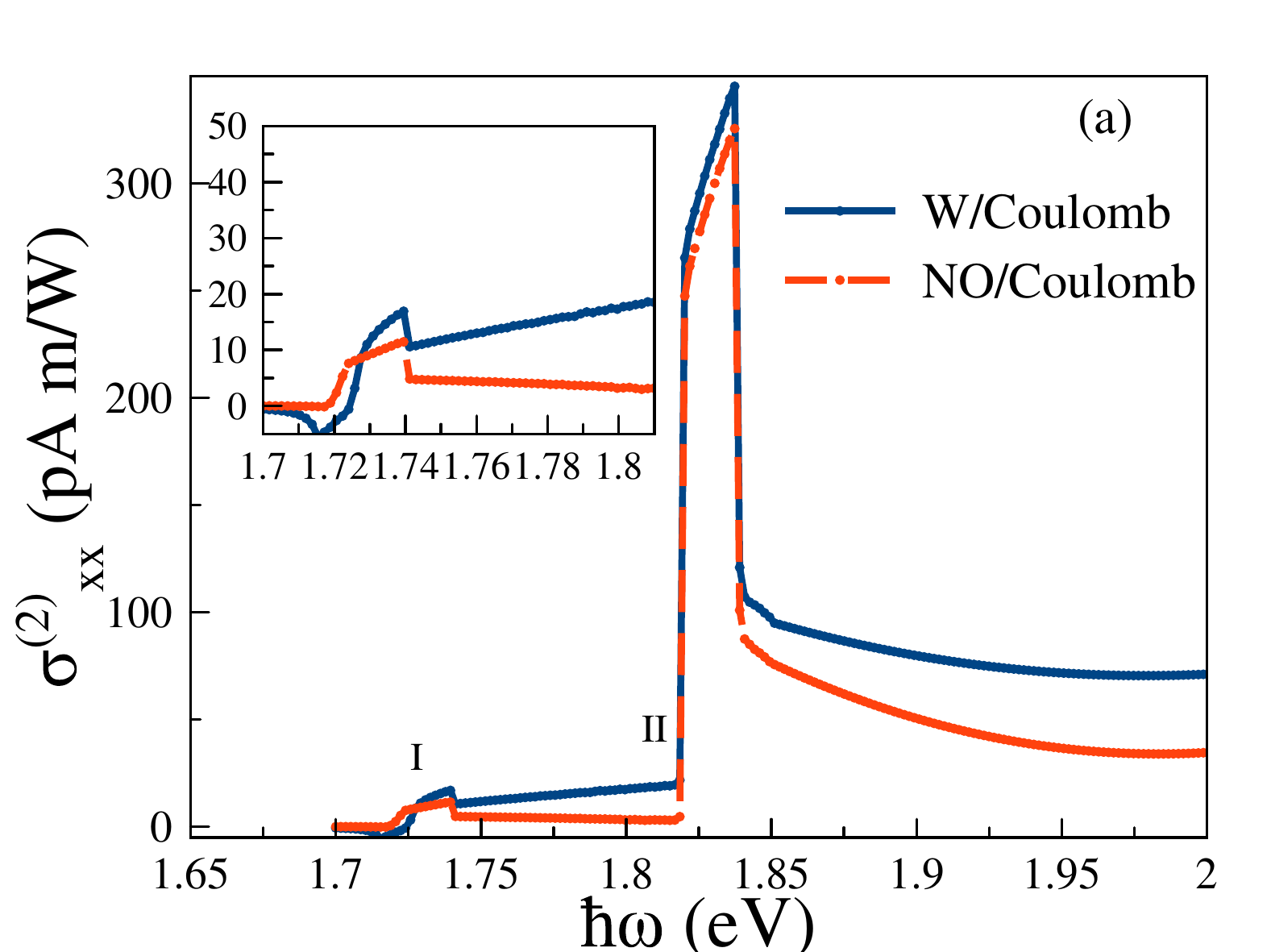}
	\includegraphics[width=4cm]{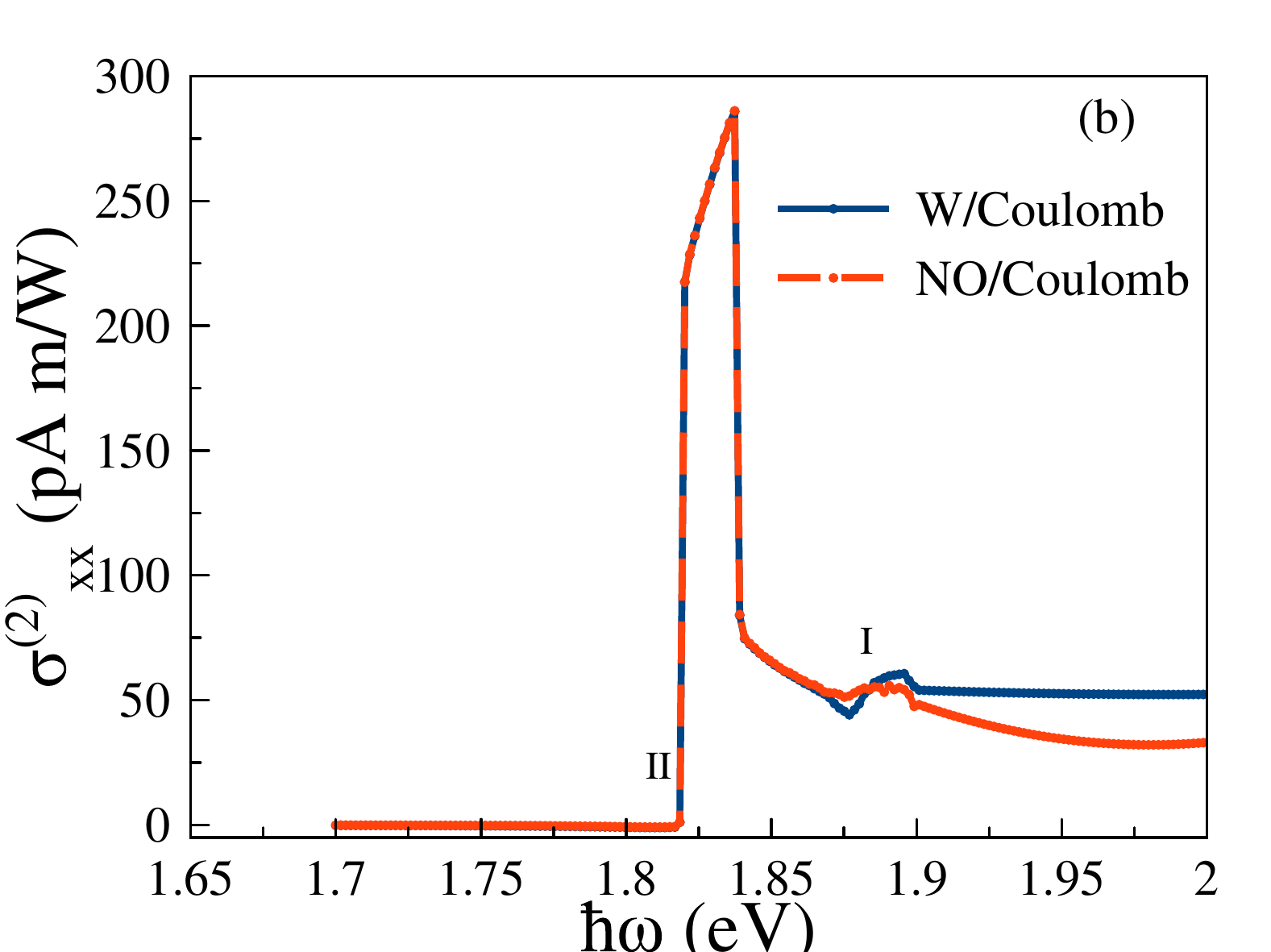}
\caption{Second-order DC photo-current at the ${\bm K}$-point as a function of energy $\hslash \omega$ for circularly polarized light, with a relaxation time $\tau=2$ ps, $U_{ele}=-0.02$ eV, $\varepsilon=0.005$ and magnetization $m=0.02$ eV, showing the (a) spin-up ($\uparrow)$ and (b) spin-down ($\downarrow$) components with and without Coulomb potential for excitons. The inter-band transition in the same valley (I) and inter-valley transition (II) are shown. The solid (dashed-dotted) lines represent UVCP with~\cite{Vasko2005} (without) the Coulomb interaction. Notice that the band gap is controlled by $U_{ele}$, $m$ and strain.
}\label{fig1}
\end{figure}

\section{Numerical results and discussion}\label{nomeric}
    
\begin{figure}[htp]
\centering
	\includegraphics[width=4cm]{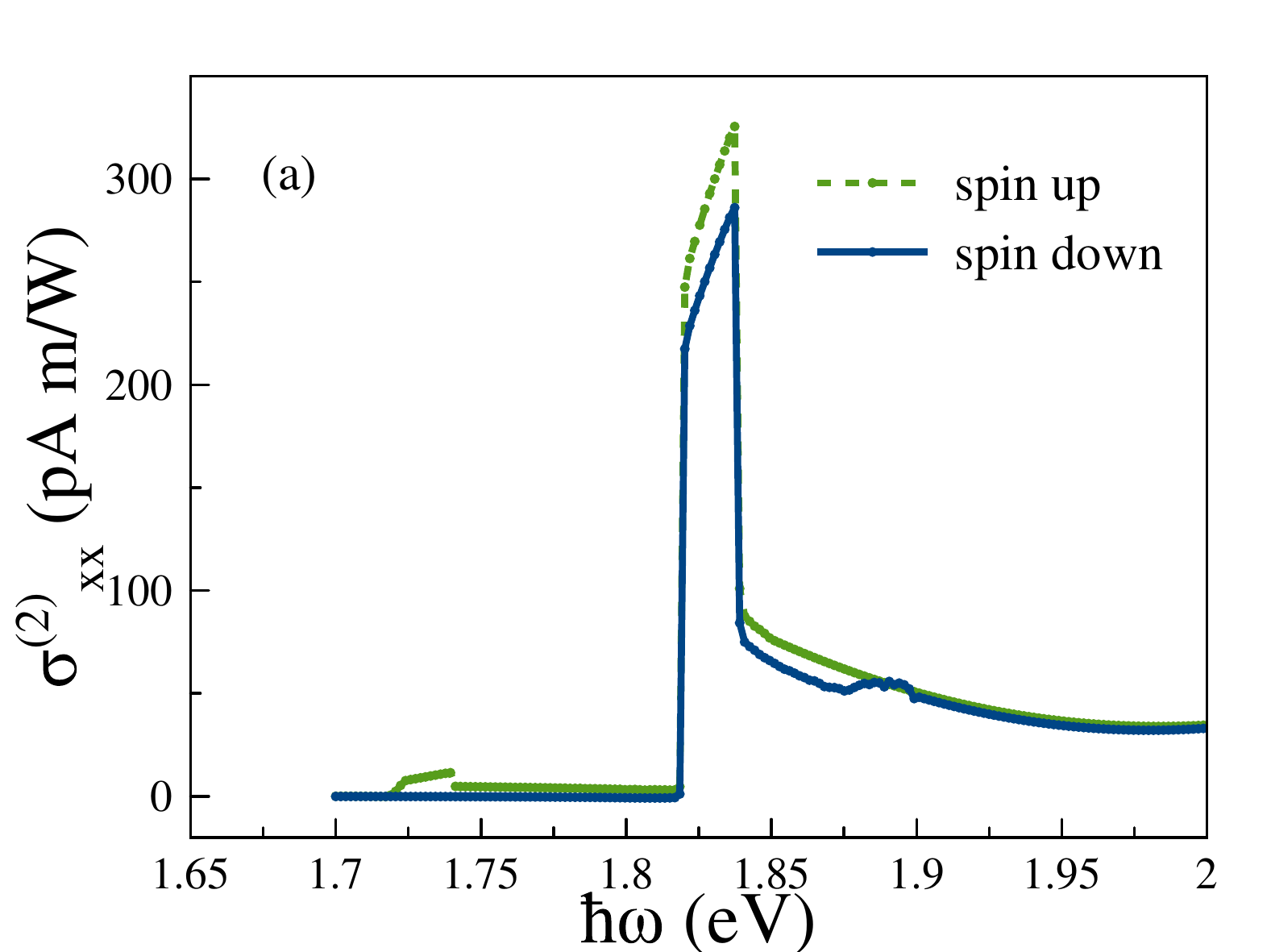}
\includegraphics[width=4.5cm]{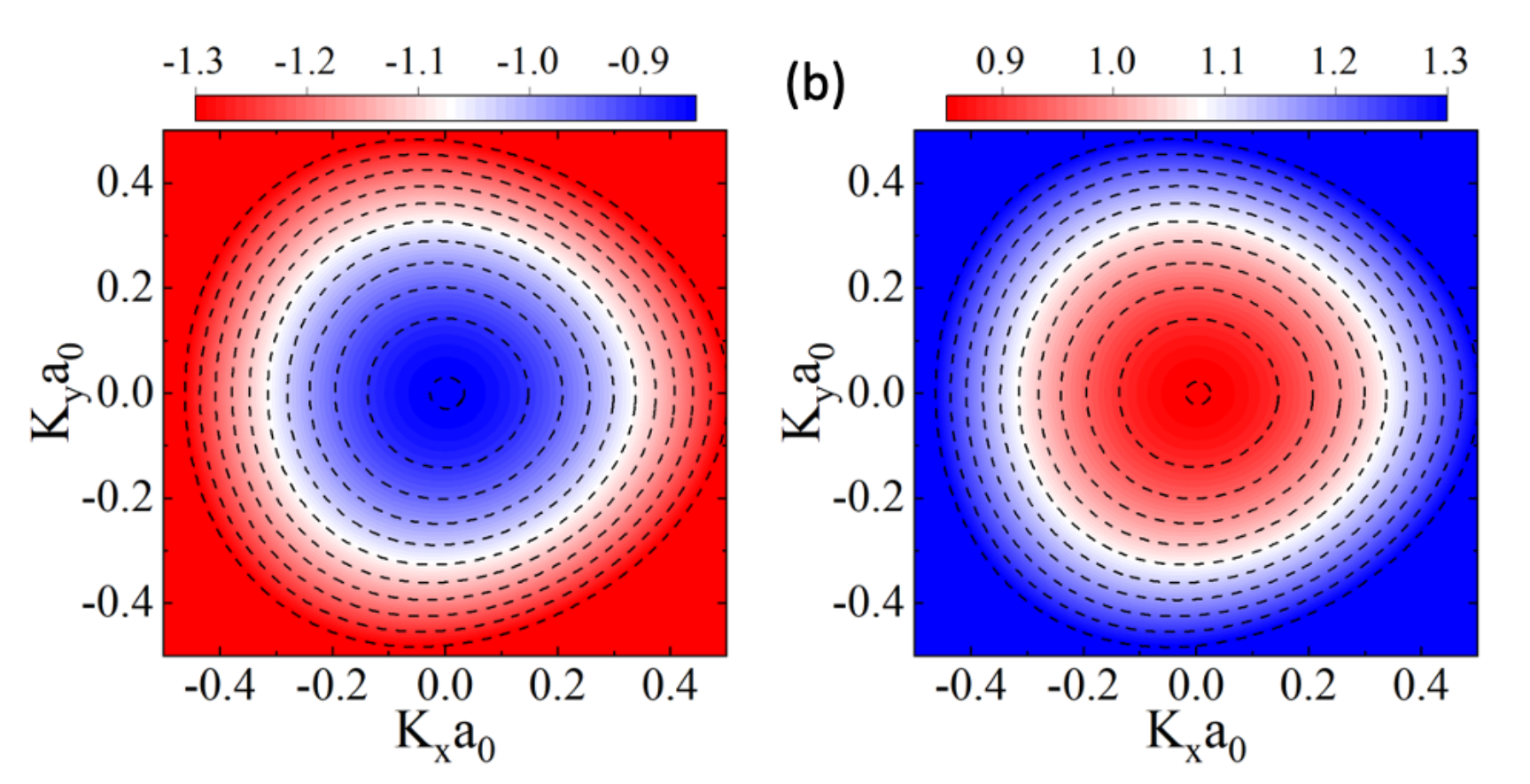}
	\includegraphics[width=4cm]{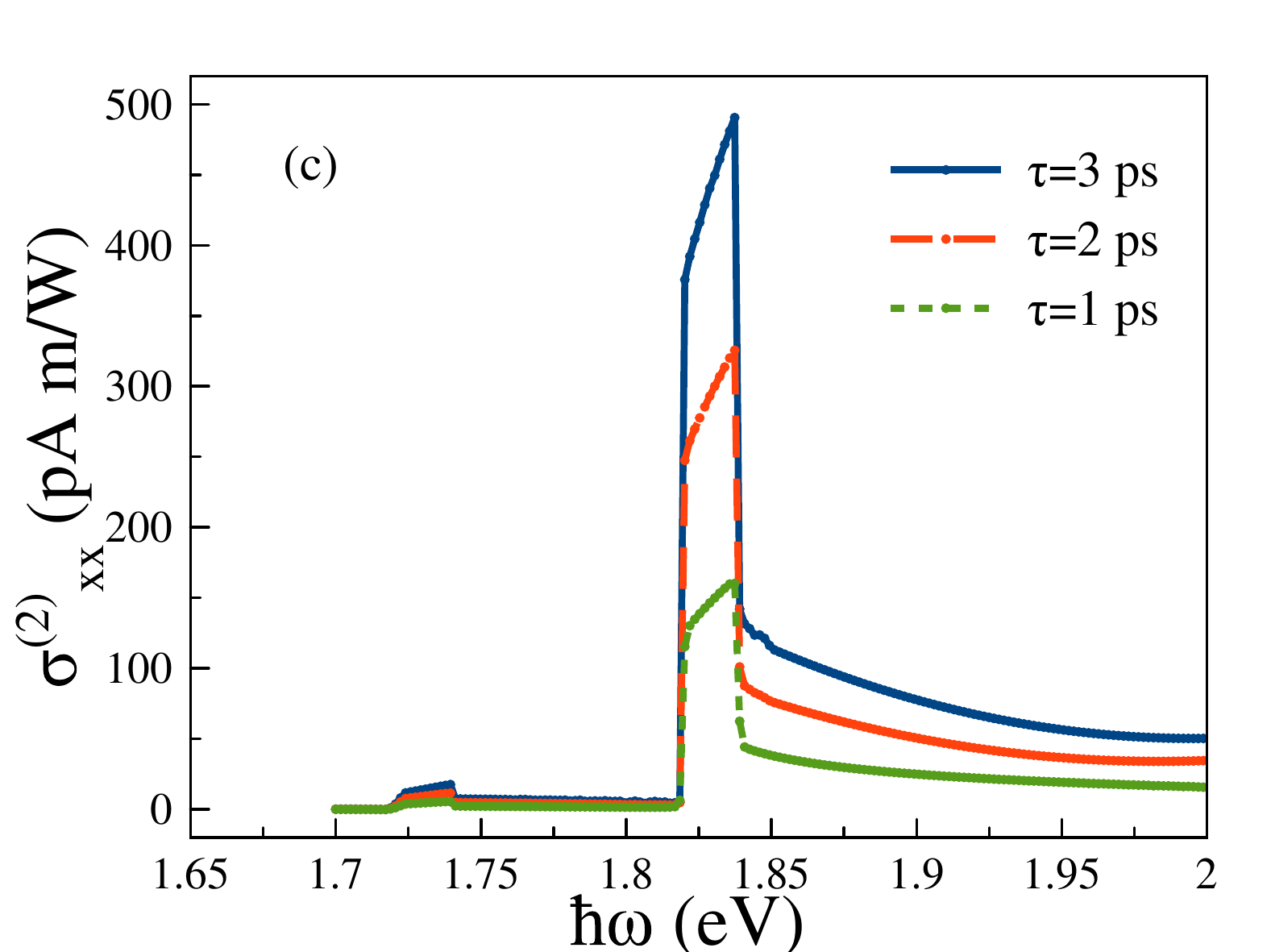}
	\includegraphics[width=4cm]{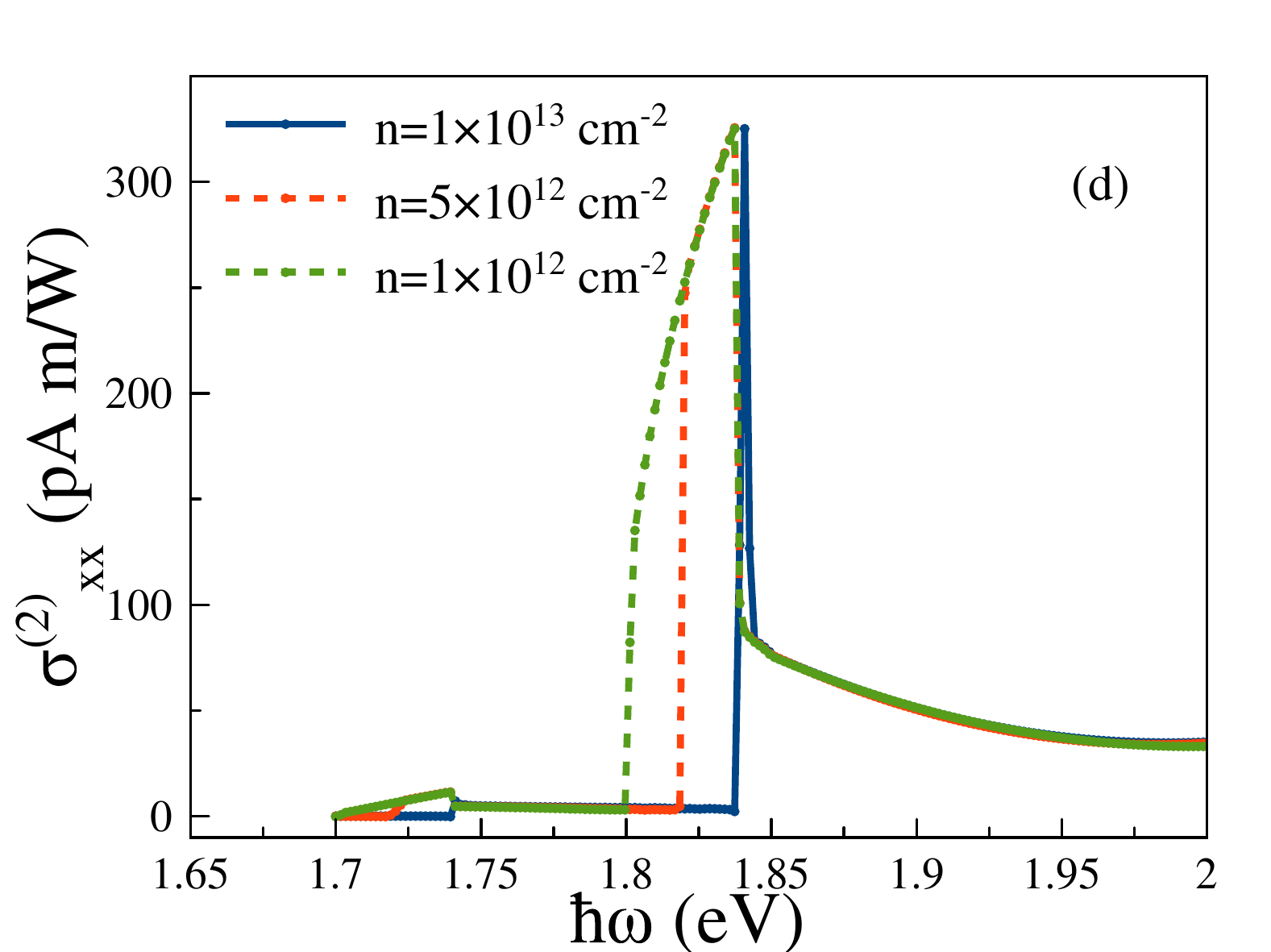}
	\caption{(Color online) (Top panel) (a): Second-order optical response (in units of pA m/W) as a function of $\hslash \omega$ (in units of eV) for circularly polarized light, $\theta_{\textrm p}=\pi/4$. The relaxation time $\tau=2$ ps, $U_{ele}=-0.02$ eV, $\varepsilon=0.005$ and $m=0.02$ eV for spin-up and down and without Coulomb interaction. The first UVCP bump appears at 1.7234 eV due to the off-diagonal density matrix contribution. A jump at around 1.823 eV stems from the $J_E$ contribution.  (b): Contour plot of the energy dispersion of the valence and conduction bands in the presence of strain $\varepsilon=0.005$ around K point for spin-up component. Different effective masses cause a discrepancy between the valence and conduction band dispersions. The energy step of the isoenergy is 0.05 eV. (Bottom panel) Same as a (a) for varying $\tau$ in ps (c) and electron density, $n$ (d). We consider spin-up in this example. The magnitude of the bump is insensitive to the electron density, although the self energy is renormalised by electron-electron interactions, while the band-gap experiences a large, nonlinear renormalization upon adding free carriers to the conduction band. Therefore, the position of the peak jump is renormalized by quantum many-body effects.}\label{fig3}
\end{figure}

MoS$_2$ exhibits circular dichroism \cite{zeng2012valley, mak2012control, cao2012valley, PhysRevB.91.035402}. Carriers in different valleys are associated with different angular momenta suggesting the possibility of controlling and pumping different valleys by controlling the circular polarization of the incident light. This means that the orbital angular momentum in the conduction band is $l_c=-\tau_v$. The spin and valley (K and K') degrees of freedom are locked. As shown in Fig. \ref{fig2}, the band gap is $\Delta_{\tau_v} = \Delta - \tau_v m$, where $\Delta$ is the MoS$_2$ optical band gap, $m$ is the magnetization and $\tau_v=\pm$ refers to valleys. Bands are labeled by 1, 2, 3, and 4. For spin-up, the direct band gap between 1 and 2 is 1.72 eV however the indirect band gap between 4 and 1 is $(1.92+1.72)/2 = 1.82$ eV at given $m=0.02$ eV and $U_{ele}=-0.02$ eV as illustrated in Fig. \ref{fig2} for small strain values. Notice that those values change for spin down since the band gap changes.  This will be seen in the second UVCP peak in Fig. \ref{fig3}. We focus on all possible transitions to the conduction band at K. Notice that optical transitions between the conduction bands in the two valleys can occur at situation when the photon energy is smaller than the Fermi energy, $\hslash \omega \le \varepsilon_{\textrm {F}}$, thus there is a finite current associated with this scattering term at low energies. To satisfy momentum conservation, short-range scattering potential due to defects or impurity is needed. Since we are interested in energies exceeding the band gap, we do not consider this term.

The central result of our work is presented in Fig.~\ref{fig1}. The non-linear DC current bumps emerge at the inter-band absorption threshold in the same valley and at a transition point between the valence and conduction bands of different valleys in the electron doped system. The conventional intrinsic shift current is negligible, and the resonant photovoltaic effect~\cite{PhysRevLett.124.087402} is also negligible owing to a strong particle-hole asymmetry. To capture the non-reciprocal current the electric field needs to be incorporated into the time evolution operator leading to the scattering term. In addition to intrinsic and extrinsic contributions, our theory accounts for exciton effects such as the electron-hole interaction, band gap renormalization and the Sommerfeld factor~\cite{Vasko2005, schafer2013semiconductor}. To emphasize this Fig.~\ref{fig1} shows the UVCP found using two different methods, namely the quantum kinetic theory and a related method incorporating the Bethe-Salpeter equation for excitons. 
Once again, the first order of density matrix can be evaluated by using the Bethe-Salpeter equation incorporating the electron-hole Coulomb interaction in the level of Hartree-Fock approximation and then the second-order of density matrix components incorporates the scattering terms can be evaluated by using the kinetic theory. Although it is known that the optical
edge will be shifted by electron-hole interactions~\cite{PhysRevLett.105.136805} as obtained by Eq.~\ref{A:e-h}, for the sake of comparison, we ignore the position of the shift to show the strength of the optical peak due to the many-body interaction.
Our numerical results, reported in Fig. \ref{fig1}, illustrate the current is greater in this approach with respect to those results obtained without the attractive Coulomb interaction. The location of the optical band edges vary by imposing external electric and mechanical fields. Accordingly, the optical band edges are shifted and there is a noticeable bumps in    
$J(\varepsilon)-J(\varepsilon=0)$ at the band edges.
Two bumps are associated with optical transitions from the inter-band and inter-valley transitions, respectively, while the current is visibly enhanced by many-body effects~\cite{PhysRevB.94.035117}. Such interplay between non-linear optics and inter-valley scattering enables monitoring of non-equilibrium inter-valley dynamics and probing their strength, opening new directions in non-linear light-matter engineering~\cite{vitale2018valleytronics}.

%%%%%%%%%Fig3%%Fig3%%Fig3%%Fig3%

The second order steady-state current as a function of photon energy is shown in Fig. \ref{fig3}(a).  The UVCP shows $\omega \theta(\hbar \omega-\varepsilon^c+\varepsilon^v)$ in the vicinity of the band edge as similar to the optical absorption coefficient. 
For spin-up, the first peak is associated with the transition between the valence and the conduction band at the K point, while the second peak originates from the inter-valley transition between the valence band at K' to the conduction band at K. However, for spin-down, the inter-valley transition between the valence band at K' and the conduction band at K takes place first and then the inter-band transition occurs.  By increasing the light frequency, electrons deeper in valence band, with stronger warping, are excited to the conduction band, and finite value of the UVCP is observed. Moreover, intervalley scattering processes are proportional to $\lambda^2$.

The UVCP is the result of two factors of the electric field {\bf E}. The first comes from the non-equilibrium distribution function $f_{od}^{(1)}$, the off-diagonal density matrix element of  $\langle \rho_1 \rangle$. Essentially this is the distribution of excited electrons in the conduction band. The second factor of {\bf E} comes when we consider $\textrm {e}\nabla_{\bf k} {\bf E}$ in second order. This represents the acceleration of excited carriers $\dot{\bm k} = -\textrm {e}{\bm E}(t)/\hslash$ under the action of the time-dependent electric field $\bf E$. The non-equilibrium distribution to second order in $\bf E$ has some time dependence, and part of this time dependence is in phase with $\bf E$, so that their product has a nonzero time average, resulting in a DC term. Even though the acceleration oscillates in time, the distribution of excited carriers also oscillates in time so that the time average of the total acceleration is nonzero. This nonzero time average represents a net, constant acceleration. Thanks to the anisotropy of the excitation, which comes from warping, the angular average of this acceleration is nonzero as well. So there is a net acceleration of excited carriers, and this needs to be halted by scattering process, represented here generically by the relaxation time, $\tau$. The presence of $\tau$ can also be viewed as a reflection of Kramers symmetry breaking by the warping term, which causes the excited carrier distribution to be asymmetric on the two sides of the conduction band. This argument applies for inter-valley excitation as well, since the warping is the same in the two valleys. 
 Thus the UVCP is a result of: (i) topological effects through the Berry connections (ii) band mass discrepancy between the electron and hole, which leads to a change in the electron wave packet with respect to the hole wave packet (iii) trigonal warping in the valence band which makes the wave packet wider than in the conduction band and (iv) the band off-set due to strain at the optical band edges. Based on our formalism, the intraband transition primarily originates from an expression given by Eq. (\ref{eq:f22d}), although the terms in Eq. (\ref{eq:f21od}) dominates for the intra-valley transition and the term ${\cal R}^{44} (f^1-f^4) /(\hslash \omega -\varepsilon^v_{K'}+\varepsilon^c_{K}-i\hslash \eta)$, including the Pauli blocking factor, is principally responsible for the inter-valley transition, where $f^i$ is the Fermi-Dirac distribution function and ${\cal R}^{44}$ represents the Berry connection.
 
 As we discussed earlier, the contribution of the shift current in our formalism is negligible. Our results show that injection current~\cite{PhysRevX.11.011001}, which is proportional to $({\hat i}\cdot{\bf {\cal R}}^{1m}_{\bf k} {\hat j}\cdot{\bf {\cal R}}^{m1}_{\bf k}-{\hat j}\cdot{\bf {\cal R}}^{1m}_{\bf k} {\hat i}\cdot{\bf {\cal R}}^{m1}_{\bf k} )(\nabla \varepsilon_{\bf k}^m-\nabla \varepsilon_{\bf k}^1)$ with $m=1$ and 4, contributes to the nonlinear DC optical response in the system, but it is not a dominant contributor as we discussed previously. 

We explore the effect of varying the relaxation time, $\tau$ and the electron density $n$. The current changes significantly by changing the relaxation time and when $\tau$ is large the UVCP can be large, which is advantageous for photovoltaic solar cell applications. Terms of the form $(f^i-f^j)$ are always present in $f^{(2)}_{od}$, where $f^{(2)}_{od}$ is the off-diagonal element of the density matrix. However, the denominator contains expressions of the form $(-\hslash\omega+ (\varepsilon^c-\varepsilon^v)+i\hslash/\tau)^2$, which tend to a smaller value as $\tau$ increases, and thus the peak becomes stronger although its width does not change due to the $(f^i-f^j)$ term. Therefore, integrating over $k$, a larger current emerges owing to the stronger peak. Interestingly, the current does not change with the Fermi energy, apart from the optical transition point owing to $\Delta_{\tau_v}$.

% Mathematically $\tau$ enters through the term $\nabla_{\bf k} \delta(-\hslash\omega+ (\varepsilon^c-\varepsilon^v)+i\hslash/\tau)$.
\section{Conclusion}\label{conc}

To summarize we have studied the static non-linear optical response of monolayer strained MoS$_2$ with broken inversion symmetry. The theory incorporates the inter-band and inter-valley optical transitions in the presence of short-range impurity scattering, which also couples the valleys. In addition to intrinsic and extrinsic contributions, our theory includes exciton effects such as the electron-hole interaction and the Sommerfeld factor. The results show a large non-linear response with meaningful features at optical light frequencies and identified a new, unidirectional response termed \textit{non-reciprocal valley photo-current}, with no equivalent in single-valley systems. Its direction is set by trigonal warping and strain, and it increases with the mobility and trigonal warping coefficient. Two bumps were associated with optical transitions from the inter-band and inter-valley transitions, respectively, while the current is visibly enhanced by many-body effects. We have shown that the optical current changes significantly by changing the relaxation time and the UVCP can be large when the relaxation time is large, which is advantageous for photovoltaic solar cell applications.

With the intrinsic shift current suppressed, our approach predicts a large UVCP, which is accessible in experiment and can monitor inter-valley transitions.

\section{Acknowledgment:} 
This work was supported by the Australian Research Council Centre of Excellence in Future Low-Energy Electronics Technologies (project number CE170100039).

\bibliography{ref1}

\newpage
\begin{widetext}
\appendix
\section{Model Hamiltonian and Theory}\label{A:hamiltonian}

To begin with, we start of considering a generic low-energy ${\bf k}\cdot{\bf p}$  continuum model Hamiltonian around K and K' points which describes pristine two-dimensional TMD system~\cite{PhysRevB.88.085440}, such as Mo$_2$ or WS$_2$ 
\begin{equation}\label{hami}
{\mathcal H}_1= \gamma \frac{\Delta_0+\lambda_0 S \tau_v}{2}+\frac{\Delta_{\tau_v}+\lambda S \tau_v}{2}\sigma_z+t_0 a_0 {\bf k}\cdot{\sigma_{\tau_v}} +\frac{\hbar^2 k^2}{4 m_0}(\alpha+\beta \sigma_z)
\end{equation}
and trigonal warping ${\mathcal H}_{w}$ contribution is given by
\begin{equation}
{\mathcal H}_{w}=t_1 a_0^2 ({\bf k}\cdot{\sigma}_{\tau_v}^*)\sigma_x({\bf k}\cdot{\sigma}_{\tau_v}^*)+t_2a_0^3\tau_v(k_x^3-3k_xk_y^2)(\alpha'+\beta' \sigma_z)
\end{equation}
where the Pauli matrices ${\bf \sigma}_{\tau_v}=(\tau_v \sigma_x$, $\sigma_y)$ acts on the two component wave functions. The spin-orbit couplings in the valence and conduction bands are considered and the sample degrees of freedom is accounted by the degeneracy factor $\gamma=\pm$. Here $\tau_v=\pm$ is a valley index, $S=\pm$ refers to a spin index, Notice that ${\bf k}=k(\cos\theta,\sin\theta)$. All terms in the Hamiltonian are related to broken spatial inversion symmetry in monolayer TMD. In general, we could consider a system which is on top of a Ferromagnetic substrate with a finite perpendicular magnetization. A magnetization is induced by a proper substrate in order to break valley degeneracy and thus $\Delta_{\tau_v}=\Delta-\tau_v m$ where $m$ is the magnetization.
The contribution to the band dispersion owing to trigonal warping has the character form $Z_{\pm}\cos 3\phi$ where $z_{\pm}=t_2(\alpha'+\beta')+4t_1t_2/[2\Delta_{\tau_v}-(\lambda_0-\lambda) s\tau]$ and $\pm$ refers to the conduction and valence bands, respectively. Note that the conduction band is nearly isotropic while the valence band is strongly warped due to the trigonal warping term.

In the case of monolayer MoS$_2$, all parameters are: $a_0=a/\sqrt{3}$, $a=3.16$ \AA, $\Delta_0=-0.11$ eV, $\Delta=1.82$ eV, $\lambda_0=69$ meV, $\lambda=-80$ meV, $t_0=2.34$ eV, $t_1=-0.14$ eV, $t_2=1$ eV, $\alpha=-0.01$, $\beta=-1.54$, $\alpha'=0.44$ and $\beta=-0.53$.

The Hamiltonian can be written as
\begin{equation}
{\mathcal H}=
\begin{pmatrix}
    h_{11}+h_z & h_{12}   \\
    h_{12} & h_{11}-h_z
\end{pmatrix}
\end{equation}
where
\begin{eqnarray}
&&h_{11}=\gamma\frac{\Delta_0+\lambda_0 S\tau}{2}+\frac{\hbar^2 k^2}{4m_0}\alpha+t_2a_0^3\tau_v (k_x^3-3k_xk_y^2)\alpha' ,\\
&&h_z=\frac{\Delta+\lambda S\tau_v}{2}+\frac{\hbar^2 k^2}{4m_0}\beta+t_2a_0^3\tau_v (k_x^3-3k_xk_y^2)\beta',\nonumber\\
&&h_{12}=t_0 a_0(\tau_v k_x-i k_y)+t_1a_0^2(\tau_v k_x+i k_y)^2 ,\nonumber
\end{eqnarray}
where $k_x^3-3k_xk_y^2=k^3 \cos 3\theta$. The dispersion relations are given by $| {\mathcal H}-\varepsilon |=0$ and thus
\begin{eqnarray}
&&\varepsilon^{s}({\bf k})=h_{11}+s \sqrt{h_z^2+|h_{12}|^2},\nonumber
\end{eqnarray}
where $s=\pm$ denotes the conduction and valence bands, respectively.  The eigenvector of the system, ${\mathcal H}|\psi^s\rangle=\varepsilon^{s}({\bf k})|\psi^s\rangle$ can be easily obtained as

\begin{equation}
|\psi^{s}\rangle=
\begin{pmatrix}
    \psi_1   \\
    \psi_2 
\end{pmatrix}
=\frac{1}{\sqrt{D{^{s}}^2+|h_{12}|^2}}
\begin{pmatrix}
  -h_{12}\\
  D^{s}
\end{pmatrix}
\end{equation}
where $D^{s}=h_{z}-s \sqrt{h_z^2+|h_{12}|^2}$.
In order to calculate the Berry connection, we do need to calculate the $\nabla_{\bf k} |\psi^s \rangle$. To do so, we make use of cylindrical coordinate and 

\begin{eqnarray}
&&\partial_k h_{11}=\frac{\hbar^2 k}{2m_0}\alpha+3 t_2a_0^3\tau_v \alpha'  k^2 \cos 3\theta\nonumber\\
&&\partial_k h_z=\frac{\hbar^2 k}{2m_0}\beta+3 t_2a_0^3\tau_v \beta' k^2 \cos 3\theta,\nonumber\\
&&\partial_k h_{12}=t_0 a_0(\tau_v \cos \theta-i \sin \theta)+2 t_1a_0^2 k(\tau_v \cos \theta+i \sin \theta)^2 ,\nonumber
\end{eqnarray}
and 
\begin{eqnarray}
&&\partial_{\theta} h_{11}=-3t_2a_0^3 \tau_v \alpha' k^3 \sin 3\theta\nonumber\\
&&\partial_{\theta} h_z=-3 t_2a_0^3\tau_v \beta' k^3 \sin 3\theta,\nonumber\\
&&\partial_{\theta} h_{12}=t_0 a_0 k(-\tau_v \sin \theta-i \cos \theta)+2 t_1a_0^2 k^2(i\tau_v \cos 2\theta- \sin 2\theta) ,\nonumber
\end{eqnarray}
finally, it leads to

\begin{eqnarray}\label{nabla}
&&\nabla_{\bf k} |\psi^{s}\rangle=\partial_k |\psi^{s}\rangle {\hat k}+\frac{1}{k}\partial_{\theta}|\psi^{s}\rangle{\hat \theta}\\
&=&
-\frac{{\bf f}^s({\bf k})}{(D{^{s}}^2+|h_{12}|^2)^{3/2}}
\begin{pmatrix}
  -h_{12}\\
  D^{s}
\end{pmatrix}
+\frac{1}{\sqrt{D{^{s}}^2+|h_{12}|^2}}\left[\partial_k
\begin{pmatrix}
  -h_{12}\\
  D^{s}
\end{pmatrix}
{\hat k}+\frac{1}{k}\partial_{\theta}
\begin{pmatrix}
  -h_{12}\\
  D^{s}
\end{pmatrix}
{\hat \theta}\right]
\end{eqnarray}
where we define ${\bf f}^s({\bf k})=\nabla_{\bf k} (D{^{s}}^2+|h_{12}|^2)/2$. Notice that ${\hat k}=\cos \theta {\hat i}+\sin \theta {\hat j}$ and ${\hat \theta}=-\sin \theta {\hat i}+\cos \theta {\hat j}$. 

We can generalize the formalism by considering inter valley process. To do so, we should consider $|\psi \rangle=|\psi^{s,\tau=+} \rangle \otimes |\psi^{s,\tau=-} \rangle$. Therefore, by making use of  all derivatives, the Berry connection part for different band indices combination is

\begin{eqnarray}
&& {\cal R}^{ss',\tau \tau'}_{kk'}=i\langle\psi^{s,\tau}_k|\nabla_{\bf k'}|\psi^{s'\tau'}_{k'}\rangle,\\
&&=-i \frac{{\bf f}^{s'\tau'}({\bf k}')}{(D{^{s}}^2+|h_{12}|^2)^{1/2}_{\tau}(D{^{s'}}^2+|h_{12}|^2)^{3/2}_{\tau'}}
\left[h_{12}^{*\tau} h_{12}^{\tau'}+D^{s\tau} D^{s'\tau'} \right]
\nonumber\\
&&+i\frac{1}{(D{^{s}}^2+|h_{12}|^2)^{1/2}_{\tau}(D{^{s'}}^2+|h_{12}|^2)^{1/2}_{\tau'}} \left[  
(h_{12}^{*\tau}\partial_k h_{12}^{\tau'}+D^{s\tau}\partial_k D^{s'\tau'}){\hat k}+
\frac{1}{k}(h_{12}^{*\tau}\partial_{\theta} h_{12}^{\tau'}+D^{s\tau}\partial_{\theta} D^{s'\tau'}){\hat \theta}
\right]
\end{eqnarray}
it tells that each quantity might be evaluated in its own valley.
We also consider the disorder as $U(r) =U_0\sum_i\delta({\bf r}-{\bf r}_i)$ and define matrix elements of $U^{ss'}_{kk'}$ as
\begin{eqnarray}
U^{ss'}_{kk'}&&=\langle\psi^{s\tau}_k|U(r)|\psi^{s'\tau'}_{k'}\rangle\\
&&=U_0\frac{h_{12}^{*\tau}(k)h_{12}^{\tau'}(k')+D_k{^{s\tau}}D_{k'}{^{s'\tau'}}}{(D_k{^{s}}^2+|h_{12}(k)|^2)^{1/2}_{\tau}(D_{k'}{^{s'}}^2+|h_{12}(k')|^2)^{1/2}_{\tau'}}
\nonumber
\end{eqnarray}

\subsection{External gate potential}

It is also important to investigate the effect of a perpendicular external electric field on the optical response. The vertical bias breaks the mirror symmetry, $\sigma_h$ and thus modifies the on-site energies of atoms in three sublayers of TMDs. We assume a single-gate device in which the induced potentials take the values $U^b=0$ and $U^t=2U_{ele}$ for layers. Using simple electronic arguments, the induced potentials for an applied vertical bias $V$ can be estimated as $U_{ele}=e \frac{d}{L}\frac{\epsilon'}{\epsilon} V$ where $\epsilon$, $d$, $\epsilon'$ and $L$ denote the dielectric constants and thickness of ML-MDS and the substrate, respectively. Based on Ref. [\onlinecite{PhysRevB.88.085440}] we do have 
$$
\delta \Delta=-0.1+0.2 (U_{ele}+0.5),~~~~\delta t_0=0.055-0.1(U_{ele}+0.55),~~~~\delta \alpha=-0.15 U_{ele}/{\text eV}, ~~~~~\delta \beta=-1.95 U_{ele}/{\text eV}
$$
The effect of the applied vertical voltage has been discussed in Ref. [\onlinecite{PhysRevB.88.085440}].

\subsection{The effect of strained MoS$_2$}
More often, there is a crystal lattice mismatch between substrate and the system and hence strain is inevitably exist and it leads to change hopping terms and also breaks three-fold symmetry. Having approximated the strained trigonal warping, ignoring the triangle trigonal warping term $t_2$ and using $\eta_1$ for modified $t_1$ term, the strain-dependent Hamiltonian around K point~\cite{PhysRevB.92.195402}, up to second order in strain and momentum 
can be written as
\begin{eqnarray}
{\mathcal H}_0=
\begin{pmatrix}
    h_{11}+h_z & h_{12}   \\
    h_{12} & h_{11}-h_z
\end{pmatrix}
+\begin{pmatrix}
    a_1 |{\bf A}|^2 & 0   \\
    0 & a_1 |{\bf A}|^2
\end{pmatrix}
+\begin{pmatrix}
    a_2 V_{\text {elastic}} & 0   \\
    0 & -a_2 V_{\text {elastic}}
    \end{pmatrix}\nonumber
\end{eqnarray}
where  
\begin{eqnarray}
&&h_{11}=\gamma\frac{\Delta_0+\lambda_0 S\tau}{2}+\frac{\hbar^2 ({\bf k}+\eta_2 \tau_v{\bf A})^2}{4m_0}\alpha ,\nonumber\\
&&h_z=\frac{\Delta+\lambda S\tau_v}{2}+\frac{\hbar^2 ({\bf k}+\eta_3 \tau_v{\bf A})^2}{4m_0}\beta,\nonumber\\
&&h_{12}=t_0 a_0(\tau_v ({k_x}+\eta_1 \tau_v{ A_x})-i ({k_y}+\eta_1 \tau_v{ A_y}))+t_1a_0^2(\tau_v ({k_x}+\eta_1 \tau_v{ A_x})+i ({k_y}+\eta_1 \tau_v{ A_y}))^2 ,\nonumber
\end{eqnarray}
where $a_1=15.95$, $a_2=-2.2$ eV and $\eta_1=0.002$, $\eta_2=-5.655$ and $\eta_3=1.633$.
The pseudovector filed for uniaxial strain is given by ${\bf A}=(1-\nu, 0)\varepsilon$ and the elastic potential 
$V_{\text{elastic}}=(1+\nu)\varepsilon$ with Poisson ration $\nu=-0.125$ and strain $\varepsilon$. 
The effect of the voltage and strain appear in the eigenvalue and eigenvector of strained system and thus the Berry connection and matrix elements of impurity might be calculated accordingly. Figure 5 shows the dispersion relation of the
conduction and valence bands for various vertical voltage and strain.

\begin{figure}
	\includegraphics[width=8.2cm]{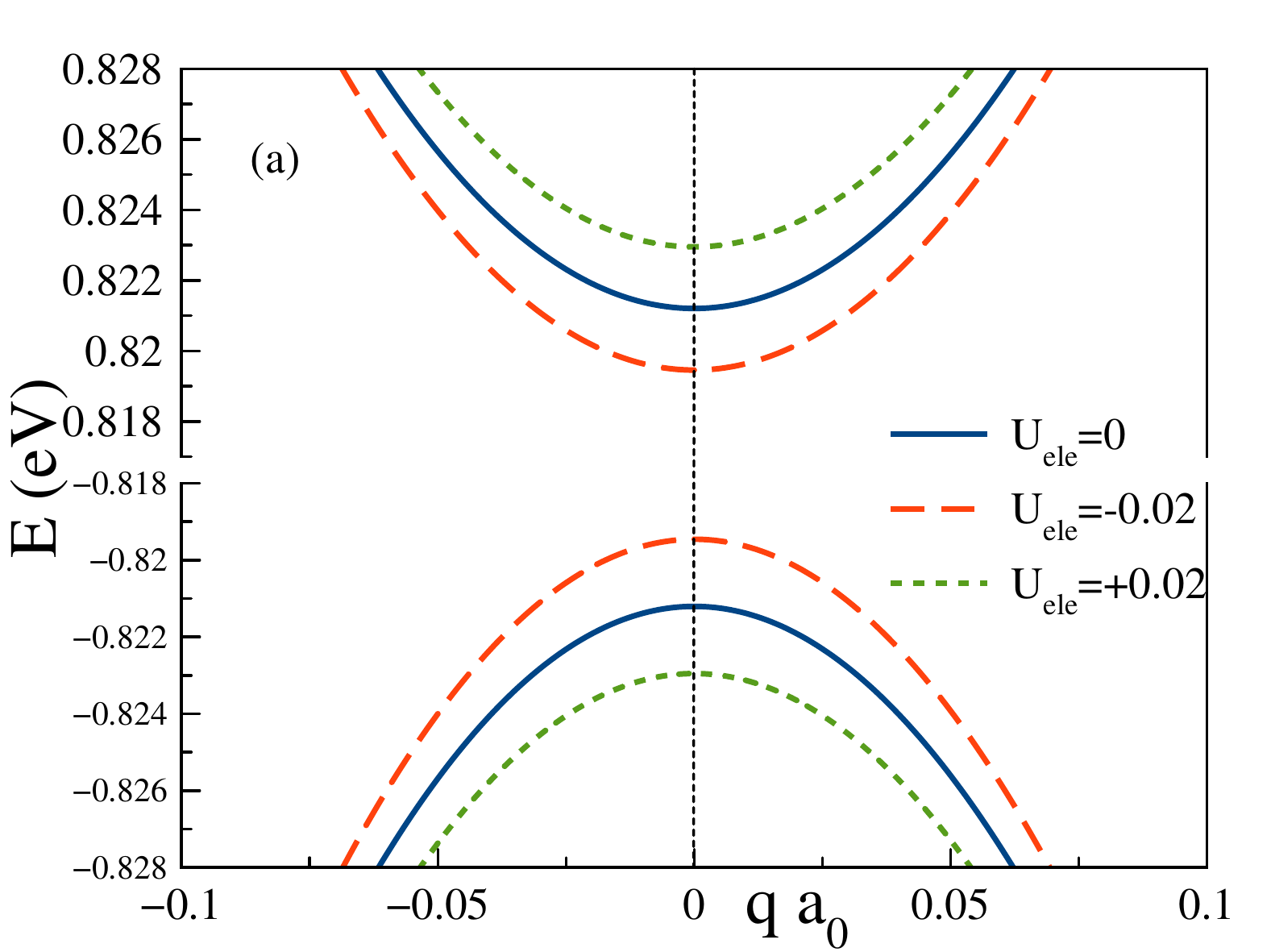}
	\includegraphics[width=8.2cm]{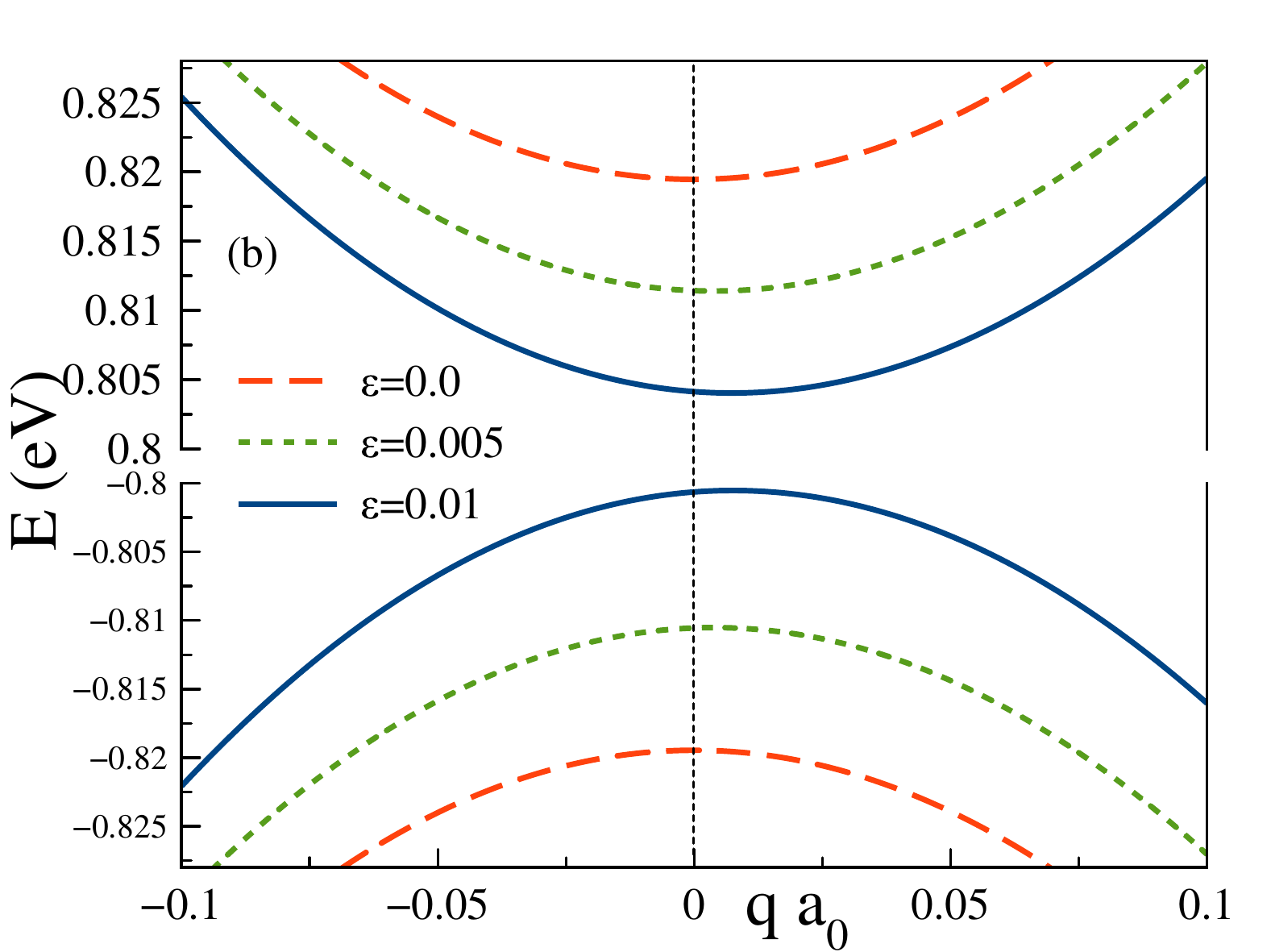}
	\caption{(Color online) The energy dispersion of the system for (a) $\varepsilon=0$ and different vertical voltage in units of eV and (b) $U_{ele}=-0.02$ eV and varies strain for given $m=0.02$ eV.}\label{fig8}
\end{figure}

Based on the DM equation, the density matrix is given by $\rho=| \psi \rangle \langle \psi |$. The dynamic of the density matrix obeys quantum Liouville equation:
\begin{equation}
\frac{\partial\rho}{\partial t}+\frac{i}{\hbar}[{\mathcal H}_0,\rho]+[U, \rho]=-\frac{i}{\hbar}[{\mathcal H}_E,\rho]
\end{equation} 
where $J[\rho]$ is the scattering term which takes the form with in the Born approximation and we assume the correlation function $\langle U({\bf r})U({\bf r}')\rangle=n_i U_0^2\delta({\bf r}-{\bf r}')$ with $n_i$ the impurity density. The scattering term, in general form, is given by
\begin{equation}
J(\langle \rho \rangle)=\frac{1}{\hbar^2}\int dt' [U ,[ U(t'), \langle \rho \rangle]]
\end{equation}
where $U(t')=e^{-iA/\hbar} U e^{iA/\hbar}$ with time-evolution operator $A=\int_0^t ({\cal H}_0+{\cal H}_E(t')) dt'$ including the electric field term. Usually ${\cal H}_E(t')$ does not contribute to the scattering term $J(\langle \rho \rangle)$ if ${\cal H}_0$ is the single component Hamiltonian and $U$ represents scaler scattering, however, this is
important for spin, pseudospin-dependent scattering.

Two scattering terms will have contributions from ${\cal H}_E$, so:
\begin{equation}
J=J_0+J_E,
\end{equation}
where $J_0$ is the bare scattering term is given by
\begin{eqnarray}
-i\hbar J^{mn}_{0,{\bf k}_1}(\langle \rho \rangle)=&&\sum_{m'm''{{\bf k}'_1}\in I}[ U^{mm'}_{{\bf k}_1{\bf k}'_1}U^{m'm''}_{{\bf k}'_1{\bf k}_1} \frac{\langle \rho_{{\bf k}_1} \rangle^{m''n}}{-\varepsilon^{m'}_{{\bf k}'_1}+\varepsilon^{m''}_{{\bf k}_1}+i\eta}+\frac{\langle \rho_{{\bf k}_1} \rangle^{mm'}}{-\varepsilon^{m'}_{{\bf k}_1}+\varepsilon^{m''}_{{\bf k}'_1}+i\eta} U^{m'm''}_{{\bf k}_1{\bf k}'_1}U^{m''n}_{{\bf k}'_1{\bf k}_1}\\
&&-U^{mm'}_{{\bf k}'_1{\bf k}_1}\frac{\langle \rho_{{\bf k}'_1} \rangle^{m'm''}}{-\varepsilon^{m''}_{{\bf k}'_1}+\varepsilon^{n}_{{\bf k}_1}+i\eta} U^{m''n}_{{\bf k}'_1{\bf k}_1}-U^{mm'}_{{\bf k}_1{\bf k}'_1}\frac{\langle \rho_{{\bf k}'_1} \rangle^{m'm''}}{-\varepsilon^{m}_{{\bf k}_1}+\varepsilon^{m'}_{{\bf k}'_1}+i\eta}U^{m'n}_{{\bf k}'_1{\bf k}_1}]\nonumber\\
&&+\sum_{m'm''{{\bf k}'_2}\in II}[ U^{mm'}_{{\bf k}_1{\bf k}'_2}U^{m'm''}_{{\bf k}'_2{\bf k}_1} \frac{\langle \rho_{{\bf k}_1} \rangle^{m''n}}{-\varepsilon^{m'}_{{\bf k}'_1}+\varepsilon^{m''}_{{\bf k}'_2}+i\eta}
+\frac{\langle \rho_{{\bf k}_1} \rangle^{mm'}}{-\varepsilon^{m'}_{{\bf k}_1}+\varepsilon^{m''}_{{\bf k}'_2}+i\eta} U^{m'm''}_{{\bf k}_1{\bf k}'_2}U^{m''n}_{{\bf k}'_2{\bf k}_1}\nonumber\\
&&-U^{mm'}_{{\bf k}'_2{\bf k}_1}\frac{\langle \rho_{{\bf k}'_2} \rangle^{m'm''}}{-\varepsilon^{m''}_{{\bf k}'_2}+\varepsilon^{n}_{{\bf k}_1}+i\eta} U^{m''n}_{{\bf k}'_2{\bf k}_1}-U^{mm'}_{{\bf k}_1{\bf k}'_2}\frac{\langle \rho_{{\bf k}'_2} \rangle^{m'm''}}{-\varepsilon^{m}_{{\bf k}_1}+\varepsilon^{m'}_{{\bf k}'_2}+i\eta} U^{m''n}_{{\bf k}'_2{\bf k}_1}]\nonumber
\end{eqnarray}

so that 
\begin{eqnarray}
-i\hbar J^{11}_{0,{\bf k}_1}(\langle \rho \rangle)=&&\sum_{m'm''{{\bf k}'_1}\in I}[ U^{1m'}_{{\bf k}_1{\bf k}'_1}U^{m'm''}_{{\bf k}'_1{\bf k}_1} \frac{\langle \rho_{{\bf k}_1} \rangle^{m''1}}{-\varepsilon^{m'}_{{\bf k}'_1}+\varepsilon^{m''}_{{\bf k}_1}+i\eta}+\frac{\langle \rho_{{\bf k}_1} \rangle^{1m'}}{-\varepsilon^{m'}_{{\bf k}_1}+\varepsilon^{m''}_{{\bf k}'_1}+i\eta} U^{m'm''}_{{\bf k}_1{\bf k}'_1}U^{m''1}_{{\bf k}'_1{\bf k}_1}\\
&&-U^{1m'}_{{\bf k}'_1{\bf k}_1}\frac{\langle \rho_{{\bf k}'_1} \rangle^{m'm''}}{-\varepsilon^{m''}_{{\bf k}'_1}+\varepsilon^{1}_{{\bf k}_1}+i\eta} U^{m''1}_{{\bf k}'_1{\bf k}_1}-U^{1m'}_{{\bf k}_1{\bf k}'_1}\frac{\langle \rho_{{\bf k}'_1} \rangle^{m'm''}}{-\varepsilon^{1}_{{\bf k}_1}+\varepsilon^{m'}_{{\bf k}'_1}+i\eta}U^{m'1}_{{\bf k}'_1{\bf k}_1}]\nonumber\\
&&+\sum_{m'm''{{\bf k}'_2}\in II}[ U^{1m'}_{{\bf k}_1{\bf k}'_2}U^{m'm''}_{{\bf k}'_2{\bf k}_1} \frac{\langle \rho_{{\bf k}_1} \rangle^{m''1}}{-\varepsilon^{m'}_{{\bf k}'_1}+\varepsilon^{m''}_{{\bf k}'_2}+i\eta}
+\frac{\langle \rho_{{\bf k}_1} \rangle^{1m'}}{-\varepsilon^{m'}_{{\bf k}_1}+\varepsilon^{m''}_{{\bf k}'_2}+i\eta} U^{m'm''}_{{\bf k}_1{\bf k}'_2}U^{m''1}_{{\bf k}'_2{\bf k}_1}\nonumber\\
&&-U^{1m'}_{{\bf k}'_2{\bf k}_1}\frac{\langle \rho_{{\bf k}'_2} \rangle^{m'm''}}{-\varepsilon^{m''}_{{\bf k}'_2}+\varepsilon^{1}_{{\bf k}_1}+i\eta} U^{m''1}_{{\bf k}'_2{\bf k}_1}-U^{1m'}_{{\bf k}_1{\bf k}'_2}\frac{\langle \rho_{{\bf k}'_2} \rangle^{m'm''}}{-\varepsilon^{1}_{{\bf k}_1}+\varepsilon^{m'}_{{\bf k}'_2}+i\eta} U^{m''1}_{{\bf k}'_2{\bf k}_1}]\nonumber
\end{eqnarray}
We are at this stage to define relaxation-time of the system where we have
\begin{eqnarray}\label{Jtau1}
-i\hbar J^{11}_{0,{\bf k}_1} (\langle \rho\rangle)=&&i\hbar\frac{\langle \rho_{{\bf k}_1} \rangle^{11}}{\tau_1}+\\
&&+\sum_{m'{{\bf k}'_1}\in I}[ U^{1m'}_{{\bf k}_1{\bf k}'_1}U^{m'2}_{{\bf k}'_1{\bf k}_1} \frac{\langle \rho_{{\bf k}_1} \rangle^{21}}{-\varepsilon^{m'}_{{\bf k}'_1}+\varepsilon^{2}_{{\bf k}_1}+i\eta}+\frac{\langle \rho_{{\bf k}_1} \rangle^{12}}{-\varepsilon^{2}_{{\bf k}_1}+\varepsilon^{m'}_{{\bf k}'_1}+i\eta} U^{2m'}_{{\bf k}_1{\bf k}'_1}U^{m'1}_{{\bf k}'_1{\bf k}_1}]\\
&&+\sum_{{{\bf k}'_1}\in I}[-U^{11}_{{\bf k}'_1{\bf k}_1}\frac{\langle \rho_{{\bf k}'_1} \rangle^{11}}{-\varepsilon^{1}_{{\bf k}'_1}+\varepsilon^{1}_{{\bf k}_1}+i\eta} U^{11}_{{\bf k}'_1{\bf k}_1}-U^{11}_{{\bf k}_1{\bf k}'_1}\frac{\langle \rho_{{\bf k}'_1} \rangle^{11}}{-\varepsilon^{1}_{{\bf k}_1}+\varepsilon^{1}_{{\bf k}'_1}+i\eta}U^{11}_{{\bf k}'_1{\bf k}_1}]\nonumber\\
%%m'=1 m''=2
&&+\sum_{{{\bf k}'_1}\in I}[-U^{11}_{{\bf k}'_1{\bf k}_1}\frac{\langle \rho_{{\bf k}'_1} \rangle^{12}}{-\varepsilon^{2}_{{\bf k}'_1}+\varepsilon^{1}_{{\bf k}_1}+i\eta} U^{21}_{{\bf k}'_1{\bf k}_1}-U^{11}_{{\bf k}_1{\bf k}'_1}\frac{\langle \rho_{{\bf k}'_1} \rangle^{12}}{-\varepsilon^{1}_{{\bf k}_1}+\varepsilon^{1}_{{\bf k}'_1}+i\eta}U^{11}_{{\bf k}'_1{\bf k}_1}]\nonumber\\
%%m'=2 m''=1
&&+\sum_{{{\bf k}'_1}\in I}[-U^{12}_{{\bf k}'_1{\bf k}_1}\frac{\langle \rho_{{\bf k}'_1} \rangle^{21}}{-\varepsilon^{1}_{{\bf k}'_1}+\varepsilon^{1}_{{\bf k}_1}+i\eta} U^{11}_{{\bf k}'_1{\bf k}_1}-U^{12}_{{\bf k}_1{\bf k}'_1}\frac{\langle \rho_{{\bf k}'_1} \rangle^{21}}{-\varepsilon^{1}_{{\bf k}_1}+\varepsilon^{2}_{{\bf k}'_1}+i\eta}U^{21}_{{\bf k}'_1{\bf k}_1}]\nonumber\\
&&+\sum_{m'=3,4,{{\bf k}'_2}\in II}[ \frac{\langle \rho_{{\bf k}_1} \rangle^{11}}{-\varepsilon^{1}_{{\bf k}_1}+\varepsilon^{m'}_{{\bf k}'_2}+i\eta} U^{1m'}_{{\bf k}_1{\bf k}'_2}U^{m'1}_{{\bf k}'_2{\bf k}_1}+\frac{\langle \rho_{{\bf k}_1} \rangle^{12}}{-\varepsilon^{2}_{{\bf k}_1}+\varepsilon^{m'}_{{\bf k}'_2}+i\eta} U^{2m'}_{{\bf k}_1{\bf k}'_2}U^{m'1}_{{\bf k}'_2{\bf k}_1}]\nonumber\\
&&+\sum_{m'm''{{\bf k}'_2}\in II}[-U^{1m'}_{{\bf k}'_2{\bf k}_1}\frac{\langle \rho_{{\bf k}'_2} \rangle^{m'm''}}{-\varepsilon^{m''}_{{\bf k}'_2}+\varepsilon^{1}_{{\bf k}_1}+i\eta} U^{m''1}_{{\bf k}'_2{\bf k}_1}-U^{1m'}_{{\bf k}_1{\bf k}'_2}\frac{\langle \rho_{{\bf k}'_2} \rangle^{m'm''}}{-\varepsilon^{1}_{{\bf k}_1}+\varepsilon^{m'}_{{\bf k}'_2}+i\eta} U^{m''1}_{{\bf k}'_2{\bf k}_1}]
\nonumber\end{eqnarray}

Furthermore, the off-diagonal component of $J_0$ is given by
\begin{eqnarray}
-i\hbar J^{12}_{0,{\bf k}_1}(\langle \rho \rangle)=&&\sum_{m'm''{{\bf k}'_1}\in I}[ U^{1m'}_{{\bf k}_1{\bf k}'_1}U^{m'm''}_{{\bf k}'_1{\bf k}_1} \frac{\langle \rho_{{\bf k}_1} \rangle^{m''2}}{-\varepsilon^{m'}_{{\bf k}'_1}+\varepsilon^{m''}_{{\bf k}_1}+i\eta}+\frac{\langle \rho_{{\bf k}_1} \rangle^{1m'}}{-\varepsilon^{m'}_{{\bf k}_1}+\varepsilon^{m''}_{{\bf k}'_1}+i\eta} U^{m'm''}_{{\bf k}_1{\bf k}'_1}U^{m''2}_{{\bf k}'_1{\bf k}_1}\\
&&-U^{1m'}_{{\bf k}'_1{\bf k}_1}\frac{\langle \rho_{{\bf k}'_1} \rangle^{m'm''}}{-\varepsilon^{m''}_{{\bf k}'_1}+\varepsilon^{2}_{{\bf k}_1}+i\eta} U^{m''2}_{{\bf k}'_1{\bf k}_1}-U^{1m'}_{{\bf k}_1{\bf k}'_1}\frac{\langle \rho_{{\bf k}'_1} \rangle^{m'm''}}{-\varepsilon^{1}_{{\bf k}_1}+\varepsilon^{m'}_{{\bf k}'_1}+i\eta}U^{m'2}_{{\bf k}'_1{\bf k}_1}]\nonumber\\
&&+\sum_{m'm''{{\bf k}'_2}\in II}[ U^{1m'}_{{\bf k}_1{\bf k}'_2}U^{m'm''}_{{\bf k}'_2{\bf k}_1} \frac{\langle \rho_{{\bf k}_1} \rangle^{m''2}}{-\varepsilon^{m'}_{{\bf k}'_1}+\varepsilon^{m''}_{{\bf k}'_2}+i\eta}
+\frac{\langle \rho_{{\bf k}_1} \rangle^{1m'}}{-\varepsilon^{m'}_{{\bf k}_1}+\varepsilon^{m''}_{{\bf k}'_2}+i\eta} U^{m'm''}_{{\bf k}_1{\bf k}'_2}U^{m''2}_{{\bf k}'_2{\bf k}_1}\nonumber\\
&&-U^{1m'}_{{\bf k}'_2{\bf k}_1}\frac{\langle \rho_{{\bf k}'_2} \rangle^{m'm''}}{-\varepsilon^{m''}_{{\bf k}'_2}+\varepsilon^{2}_{{\bf k}_1}+i\eta} U^{m''2}_{{\bf k}'_2{\bf k}_1}-U^{1m'}_{{\bf k}_1{\bf k}'_2}\frac{\langle \rho_{{\bf k}'_2} \rangle^{m'm''}}{-\varepsilon^{1}_{{\bf k}_1}+\varepsilon^{m'}_{{\bf k}'_2}+i\eta} U^{m''2}_{{\bf k}'_2{\bf k}_1}]\nonumber
\end{eqnarray}
or equivalently
\begin{eqnarray}\label{Jtau2}
-i\hbar J^{12}_{0,{\bf k}_1}(\langle \rho \rangle)=&&-i\hbar \frac{\langle \rho_{{\bf k}_1} \rangle^{12}}{\tau_2}+\\
&&+\sum_{m'{{\bf k}'_1}\in I}[ U^{1m'}_{{\bf k}_1{\bf k}'_1}U^{m'2}_{{\bf k}'_1{\bf k}_1} \frac{\langle \rho_{{\bf k}_1} \rangle^{22}}{-\varepsilon^{m'}_{{\bf k}'_1}+\varepsilon^{2}_{{\bf k}_1}+i\eta}+\frac{\langle \rho_{{\bf k}_1} \rangle^{11}}{-\varepsilon^{1}_{{\bf k}_1}+\varepsilon^{m'}_{{\bf k}'_1}+i\eta} U^{1m'}_{{\bf k}_1{\bf k}'_1}U^{m'2}_{{\bf k}'_1{\bf k}_1}\\
%%m'=1 m''=2
&&+\sum_{{{\bf k}'_1}\in I}[-U^{11}_{{\bf k}'_1{\bf k}_1}\frac{\langle \rho_{{\bf k}_1} \rangle^{12}}{-\varepsilon^{2}_{{\bf k}'_1}+\varepsilon^{2}_{{\bf k}_1}+i\eta} U^{22}_{{\bf k}'_1{\bf k}_1}-U^{11}_{{\bf k}_1{\bf k}'_1}\frac{\langle \rho_{{\bf k}_1} \rangle^{12}}{-\varepsilon^{1}_{{\bf k}_1}+\varepsilon^{1}_{{\bf k}'_1}+i\eta}U^{12}_{{\bf k}'_1{\bf k}_1}]\nonumber\\
&&+\sum_{{{\bf k}'_1}\in I}[-U^{11}_{{\bf k}'_1{\bf k}_1}\frac{\langle \rho_{{\bf k}'_1} \rangle^{11}}{-\varepsilon^{1}_{{\bf k}'_1}+\varepsilon^{2}_{{\bf k}_1}+i\eta} U^{12}_{{\bf k}'_1{\bf k}_1}-U^{11}_{{\bf k}_1{\bf k}'_1}\frac{\langle \rho_{{\bf k}'_1} \rangle^{11}}{-\varepsilon^{1}_{{\bf k}_1}+\varepsilon^{1}_{{\bf k}'_1}+i\eta}U^{12}_{{\bf k}'_1{\bf k}_1}]\nonumber\\
%%%
&&+\sum_{{{\bf k}'_1}\in I}[-U^{12}_{{\bf k}'_1{\bf k}_1}\frac{\langle \rho_{{\bf k}'_1} \rangle^{22}}{-\varepsilon^{2}_{{\bf k}'_1}+\varepsilon^{2}_{{\bf k}_1}+i\eta} U^{22}_{{\bf k}'_1{\bf k}_1}-U^{12}_{{\bf k}_1{\bf k}'_1}\frac{\langle \rho_{{\bf k}'_1} \rangle^{22}}{-\varepsilon^{1}_{{\bf k}_1}+\varepsilon^{2}_{{\bf k}'_1}+i\eta}U^{22}_{{\bf k}'_1{\bf k}_1}]\nonumber\\
&&+\sum_{m'=3,4,{{\bf k}'_2}\in II}[ \frac{\langle \rho_{{\bf k}_1} \rangle^{11}}{-\varepsilon^{1}_{{\bf k}_1}+\varepsilon^{m'}_{{\bf k}'_2}+i\eta} U^{1m'}_{{\bf k}_1{\bf k}'_2}U^{m'2}_{{\bf k}'_2{\bf k}_1}+\frac{\langle \rho_{{\bf k}_1} \rangle^{12}}{-\varepsilon^{2}_{{\bf k}_1}+\varepsilon^{m'}_{{\bf k}'_2}+i\eta} U^{2m'}_{{\bf k}_1{\bf k}'_2}U^{m'2}_{{\bf k}'_2{\bf k}_1}]\nonumber\\
&&+\sum_{m'm''{{\bf k}'_2}\in II}[-U^{1m'}_{{\bf k}'_2{\bf k}_1}\frac{\langle \rho_{{\bf k}'_2} \rangle^{m'm''}}{-\varepsilon^{m''}_{{\bf k}'_2}+\varepsilon^{2}_{{\bf k}_1}+i\eta} U^{m''2}_{{\bf k}'_2{\bf k}_1}-U^{1m'}_{{\bf k}_1{\bf k}'_2}\frac{\langle \rho_{{\bf k}'_2} \rangle^{m'm''}}{-\varepsilon^{1}_{{\bf k}_1}+\varepsilon^{m'}_{{\bf k}'_2}+i\eta} U^{m''2}_{{\bf k}'_2{\bf k}_1}]\nonumber
\end{eqnarray}

where relaxation-times are defined as
\begin{eqnarray}\label{tau}
\frac{-i\hbar}{\tau_1}=&&\sum_{m'{{\bf k}'_1}\in I}[ U^{1m'}_{{\bf k}_1{\bf k}'_1}U^{m'1}_{{\bf k}'_1{\bf k}_1} \frac{1}{-\varepsilon^{m'}_{{\bf k}'_1}+\varepsilon^{1}_{{\bf k}_1}+i\eta}+\frac{1}{-\varepsilon^{1}_{{\bf k}_1}+\varepsilon^{m'}_{{\bf k}'_1}+i\eta} U^{1m'}_{{\bf k}_1{\bf k}'_1}U^{m'1}_{{\bf k}'_1{\bf k}_1}]\\
&&+\sum_{m'{{\bf k}'_2}\in II}[ U^{1m'}_{{\bf k}_1{\bf k}'_2}U^{m'1}_{{\bf k}'_2{\bf k}_1} \frac{1}{-\varepsilon^{m'}_{{\bf k}'_1}+\varepsilon^{1}_{{\bf k}'_2}+i\eta}
+\frac{1}{-\varepsilon^{1}_{{\bf k}_1}+\varepsilon^{m'}_{{\bf k}'_2}+i\eta} U^{1m'}_{{\bf k}_1{\bf k}'_2}U^{m'1}_{{\bf k}'_2{\bf k}_1}]\nonumber\\\label{tau2}
%%%
\frac{-i\hbar}{\tau_2}=&&\sum_{m'{{\bf k}'_1}\in I}[ U^{1m'}_{{\bf k}_1{\bf k}'_1}U^{m'1}_{{\bf k}'_1{\bf k}_1} \frac{1}{-\varepsilon^{m'}_{{\bf k}'_1}+\varepsilon^{1}_{{\bf k}_1}+i\eta}+\frac{1}{-\varepsilon^{2}_{{\bf k}_1}+\varepsilon^{m'}_{{\bf k}'_1}+i\eta} U^{2m'}_{{\bf k}_1{\bf k}'_1}U^{m'2}_{{\bf k}'_1{\bf k}_1}]\\
&&+\sum_{m'{{\bf k}'_2}\in II}[ U^{1m'}_{{\bf k}_1{\bf k}'_2}U^{m'1}_{{\bf k}'_2{\bf k}_1} \frac{1}{-\varepsilon^{m'}_{{\bf k}'_1}+\varepsilon^{1}_{{\bf k}'_2}+i\eta}
+\frac{1}{-\varepsilon^{2}_{{\bf k}_1}+\varepsilon^{m'}_{{\bf k}'_2}+i\eta} U^{2m'}_{{\bf k}_1{\bf k}'_2}U^{m'2}_{{\bf k}'_2{\bf k}_1}]\nonumber
\end{eqnarray}
Let us consider the generic formula for the external field as
\begin{equation}
{\bf E}(t)={E}({\hat i}\cos \theta_p {\frac{e^{i\omega t}+e^{-i\omega t}}{2}} +{\hat j}\sin \theta_p {\frac{e^{i\omega t}-e^{-i\omega t}}{2i}} )
\end{equation}
and thus
\begin{eqnarray}\label{ge1}
g_E=&-&\frac{eE\cos \theta_p }{2\hbar^2}\int_0^{\infty} dt'' \int_0^{\infty} dt'(e^{i\omega t''}+e^{-i\omega t''}) e^{-i{{\cal H}_0} t''/\hbar} [{\hat i}\cdot {\bf r},e^{-i{{\cal H}_0} t'/\hbar} [U, \langle \rho \rangle] e^{i{{\cal H}_0} t'/\hbar}e^{-\eta t'}] e^{i{{\cal H}_0} t''/\hbar}e^{-\eta t''}\\
&-&\frac{eE\sin \theta_p }{2i\hbar^2}\int_0^{\infty} dt'' \int_0^{\infty} dt'(e^{i\omega t''}-e^{-i\omega t''}) e^{-i{{\cal H}_0} t''/\hbar} [{\hat j}\cdot {\bf r},e^{-i{{\cal H}_0} t'/\hbar} [U, \langle \rho \rangle] e^{i{{\cal H}_0} t'/\hbar}e^{-\eta t'}] e^{i{{\cal H}_0} t''/\hbar}e^{-\eta t''}
\end{eqnarray}
let us calculate the matrix element of $g_E$ which gives us
\begin{eqnarray}
\langle m,{\bf k} |g_E |n, {\bf k}' \rangle=&+&\frac{eE\cos \theta_p }{2\hbar^2}\sum_{m' {\bf k}''} \frac{U^{m'n}_{{\bf k}''{\bf k}'}{\hat i}\cdot {\bf r}^{mm'}_{{\bf k}{\bf k}''} (f^{n}_{{\bf k}'}-f^{m'}_{{\bf k}''})}{i(\varepsilon^n_{{\bf k}'}-\varepsilon^{m'}_{{\bf k}''})/\hbar-\eta}\{\frac{e^{i\omega t}}{i\omega-i(\varepsilon^m_{{\bf k}}-\varepsilon^n_{{\bf k}'})/\hbar-\eta}+\frac{e^{-i\omega t}}{-i\omega-i(\varepsilon^m_{{\bf k}}-\varepsilon^n_{{\bf k}'})/\hbar-\eta}\}\\
&-&\frac{eE\cos \theta_p }{2\hbar^2}\sum_{m' {\bf k}''} \frac{U^{mm'}_{{\bf k}{\bf k}''}{\hat i}\cdot {\bf r}^{m'n}_{{\bf k}''{\bf k}'} (f^{m'}_{{\bf k}''}-f^{m}_{{\bf k}})}{i(\varepsilon^n_{{\bf k}'}-\varepsilon^{m}_{{\bf k}})/\hbar-\eta}\{\frac{e^{i\omega t}}{i\omega-i(\varepsilon^m_{{\bf k}}-\varepsilon^n_{{\bf k}'})/\hbar-\eta}+\frac{e^{-i\omega t}}{-i\omega-i(\varepsilon^m_{{\bf k}}-\varepsilon^n_{{\bf k}'})/\hbar-\eta}\}\nonumber\\
&+&\frac{eE\sin \theta_p }{2i\hbar^2}\sum_{m' {\bf k}''} \frac{U^{m'n}_{{\bf k}''{\bf k}'}{\hat j}\cdot {\bf r}^{mm'}_{{\bf k}{\bf k}''} (f^{n}_{{\bf k}'}-f^{m'}_{{\bf k}''})}{i(\varepsilon^n_{{\bf k}'}-\varepsilon^{m'}_{{\bf k}''})/\hbar-\eta}\{\frac{e^{i\omega t}}{i\omega-i(\varepsilon^m_{{\bf k}}-\varepsilon^n_{{\bf k}'})/\hbar-\eta}-\frac{e^{-i\omega t}}{-i\omega-i(\varepsilon^m_{{\bf k}}-\varepsilon^n_{{\bf k}'})/\hbar-\eta}\}\\
&-&\frac{eE\sin \theta_p }{2i\hbar^2}\sum_{m' {\bf k}''} \frac{U^{mm'}_{{\bf k}{\bf k}''}{\hat j}\cdot {\bf r}^{m'n}_{{\bf k}''{\bf k}'} (f^{m'}_{{\bf k}''}-f^{m}_{{\bf k}})}{i(\varepsilon^n_{{\bf k}'}-\varepsilon^{m}_{{\bf k}})/\hbar-\eta}\{\frac{e^{i\omega t}}{i\omega-i(\varepsilon^m_{{\bf k}}-\varepsilon^n_{{\bf k}'})/\hbar-\eta}-\frac{e^{-i\omega t}}{-i\omega-i(\varepsilon^m_{{\bf k}}-\varepsilon^n_{{\bf k}'})/\hbar-\eta}\}\nonumber
\end{eqnarray} 
We now calculate the matrix elements of the position operator
\begin{equation}
\langle m {\bf k} | {\bf r} | n {\bf k}' \rangle={\bf r}^{mn}_{{\bf k}{\bf k}'}=[i\nabla_{\bf k} \delta_{mn}+{\cal R}^{mn}_{\bf k}]\delta({\bf k}'-{\bf k})
\end{equation}
therefore, 
\begin{eqnarray}
&&\langle m,{\bf k} |g_E |n, {\bf k}' \rangle=\\
&+&\frac{eE\cos \theta_p }{2\hbar^2}\sum_{m' {\bf k}''} \frac{U^{m'n}_{{\bf k}''{\bf k}'}{\hat i}\cdot [i\nabla_{\bf k} \delta_{mm'}+{\cal R}^{mm'}_{\bf k}]\delta({\bf k}''-{\bf k}) (f^{n}_{{\bf k}'}-f^{m'}_{{\bf k}''})}{i(\varepsilon^n_{{\bf k}'}-\varepsilon^{m'}_{{\bf k}''})/\hbar-\eta}\{\frac{e^{i\omega t}}{i\omega-i(\varepsilon^m_{{\bf k}}-\varepsilon^n_{{\bf k}'})/\hbar-\eta}+\frac{e^{-i\omega t}}{-i\omega-i(\varepsilon^m_{{\bf k}}-\varepsilon^n_{{\bf k}'})/\hbar-\eta}\}\nonumber\\
&-&\frac{eE\cos \theta_p }{2\hbar^2}\sum_{m' {\bf k}''} \frac{U^{mm'}_{{\bf k}{\bf k}''}{\hat i}\cdot [i\nabla_{{\bf k}'} \delta_{m'n}+{\cal R}^{m'n}_{\bf k}]\delta({\bf k}''-{\bf k}') (f^{m'}_{{\bf k}''}-f^{m}_{{\bf k}})}{i(\varepsilon^n_{{\bf k}'}-\varepsilon^{m}_{{\bf k}})/\hbar-\eta}\{\frac{e^{i\omega t}}{i\omega-i(\varepsilon^m_{{\bf k}}-\varepsilon^n_{{\bf k}'})/\hbar-\eta}+\frac{e^{-i\omega t}}{-i\omega-i(\varepsilon^m_{{\bf k}}-\varepsilon^n_{{\bf k}'})/\hbar-\eta}\}\nonumber\\
&+&\frac{eE\sin \theta_p }{2i\hbar^2}\sum_{m' {\bf k}''} \frac{U^{m'n}_{{\bf k}''{\bf k}'}{\hat j}\cdot [i\nabla_{\bf k} \delta_{mm'}+{\cal R}^{mm'}_{\bf k}]\delta({\bf k}-{\bf k}'') (f^{n}_{{\bf k}'}-f^{m'}_{{\bf k}''})}{i(\varepsilon^n_{{\bf k}'}-\varepsilon^{m'}_{{\bf k}''})/\hbar-\eta}\{\frac{e^{i\omega t}}{i\omega-i(\varepsilon^m_{{\bf k}}-\varepsilon^n_{{\bf k}'})/\hbar-\eta}-\frac{e^{-i\omega t}}{-i\omega-i(\varepsilon^m_{{\bf k}}-\varepsilon^n_{{\bf k}'})/\hbar-\eta}\}\nonumber\\
&-&\frac{eE\sin \theta_p }{2i\hbar^2}\sum_{m' {\bf k}''} \frac{U^{mm'}_{{\bf k}{\bf k}''}{\hat j}\cdot [i\nabla_{{\bf k}'} \delta_{m'n}+{\cal R}^{m'n}_{\bf k}]\delta({\bf k}''-{\bf k}') (f^{m'}_{{\bf k}''}-f^{m}_{{\bf k}})}{i(\varepsilon^n_{{\bf k}'}-\varepsilon^{m}_{{\bf k}})/\hbar-\eta}\{\frac{e^{i\omega t}}{i\omega-i(\varepsilon^m_{{\bf k}}-\varepsilon^n_{{\bf k}'})/\hbar-\eta}-\frac{e^{-i\omega t}}{-i\omega-i(\varepsilon^m_{{\bf k}}-\varepsilon^n_{{\bf k}'})/\hbar-\eta}\}\nonumber
\end{eqnarray} 
with more simplification, we have
\begin{eqnarray}
&&\langle m,{\bf k} |g_E |n, {\bf k}' \rangle=\\
&-&\frac{eE\cos \theta_p }{2}\sum_{m' } \frac{U^{m'n}_{{\bf k}{\bf k}'}{\hat i}\cdot {\cal R}^{mm'}_{\bf k} (f^{n}_{{\bf k}'}-f^{m'}_{{\bf k}})}{(\varepsilon^n_{{\bf k}'}-\varepsilon^{m'}_{{\bf k}})+i\hbar\eta}\{\frac{e^{i\omega t}}{\hbar\omega-(\varepsilon^m_{{\bf k}}-\varepsilon^n_{{\bf k}'})+i\hbar\eta}+\frac{e^{-i\omega t}}{-\hbar\omega-(\varepsilon^m_{{\bf k}}-\varepsilon^n_{{\bf k}'})+i\hbar\eta}\}\nonumber\\
&+&\frac{eE\cos \theta_p }{2}\sum_{m' } \frac{U^{mm'}_{{\bf k}{\bf k}'}{\hat i}\cdot {\cal R}^{m'n}_{\bf k} (f^{m'}_{{\bf k}'}-f^{m}_{{\bf k}})}{(\varepsilon^n_{{\bf k}'}-\varepsilon^{m}_{{\bf k}})+i\hbar\eta}\{\frac{e^{i\omega t}}{\hbar\omega-(\varepsilon^m_{{\bf k}}-\varepsilon^n_{{\bf k}'})+i\hbar\eta}+\frac{e^{-i\omega t}}{-\hbar\omega-(\varepsilon^m_{{\bf k}}-\varepsilon^n_{{\bf k}'})+i\hbar\eta}\}\nonumber\\
&-&\frac{eE\sin \theta_p }{2i}\sum_{m'} \frac{U^{m'n}_{{\bf k}{\bf k}'}{\hat j}\cdot {\cal R}^{mm'}_{\bf k}(f^{n}_{{\bf k}'}-f^{m'}_{{\bf k}})}{(\varepsilon^n_{{\bf k}'}-\varepsilon^{m'}_{{\bf k}})+i\hbar\eta}\{\frac{e^{i\omega t}}{\hbar\omega-(\varepsilon^m_{{\bf k}}-\varepsilon^n_{{\bf k}'})+i\hbar\eta}-\frac{e^{-i\omega t}}{-\hbar\omega-(\varepsilon^m_{{\bf k}}-\varepsilon^n_{{\bf k}'})+i\hbar\eta}\}\nonumber\\
&+&\frac{eE\sin \theta_p }{2i}\sum_{m'} \frac{U^{mm'}_{{\bf k}{\bf k}'}{\hat j}\cdot {\cal R}^{m'n}_{\bf k} (f^{m'}_{{\bf k}'}-f^{m}_{{\bf k}})}{(\varepsilon^n_{{\bf k}'}-\varepsilon^{m}_{{\bf k}})+i\hbar\eta}\{\frac{e^{i\omega t}}{\hbar\omega-(\varepsilon^m_{{\bf k}}-\varepsilon^n_{{\bf k}'})+i\hbar\eta}-\frac{e^{-i\omega t}}{-\hbar\omega-(\varepsilon^m_{{\bf k}}-\varepsilon^n_{{\bf k}'})+i\hbar\eta}\}\nonumber\\
&+&\frac{eE\cos \theta_p }{2}\sum_{ {\bf k}''} \frac{U^{mn}_{{\bf k}''{\bf k}'}(f^{n}_{{\bf k}'}-f^{m}_{{\bf k}''}){\hat i}\cdot \nabla_{\bf k} \delta({\bf k}''-{\bf k}) }{(\varepsilon^n_{{\bf k}'}-\varepsilon^{m}_{{\bf k}''})+i\hbar\eta}\{\frac{e^{i\omega t}}{i\hbar\omega-i(\varepsilon^m_{{\bf k}}-\varepsilon^n_{{\bf k}'})-\eta}+\frac{e^{-i\omega t}}{-i\hbar\omega-i(\varepsilon^m_{{\bf k}}-\varepsilon^n_{{\bf k}'})-\eta}\}\nonumber\\
&-&\frac{eE\cos \theta_p }{2}\sum_{ {\bf k}''} \frac{U^{mn}_{{\bf k}{\bf k}''}(f^{n}_{{\bf k}''}-f^{m}_{{\bf k}}){\hat i}\cdot \nabla_{{\bf k}'} \delta({\bf k}''-{\bf k}') }{(\varepsilon^n_{{\bf k}'}-\varepsilon^{m}_{{\bf k}})+i\hbar\eta}\{\frac{e^{i\omega t}}{i\hbar\omega-i(\varepsilon^m_{{\bf k}}-\varepsilon^n_{{\bf k}'})-\eta}+\frac{e^{-i\omega t}}{-i\hbar\omega-i(\varepsilon^m_{{\bf k}}-\varepsilon^n_{{\bf k}'})-\eta}\}\nonumber\\
&+&\frac{eE\sin \theta_p }{2i}\sum_{ {\bf k}''} \frac{U^{mn}_{{\bf k}''{\bf k}'}(f^{n}_{{\bf k}'}-f^{m}_{{\bf k}''}){\hat j}\cdot \nabla_{\bf k} \delta({\bf k}-{\bf k}'') }{(\varepsilon^n_{{\bf k}'}-\varepsilon^{m}_{{\bf k}''})+i\hbar\eta}\{\frac{e^{i\omega t}}{i\hbar\omega-i(\varepsilon^m_{{\bf k}}-\varepsilon^n_{{\bf k}'})-\eta}-\frac{e^{-i\omega t}}{-i\hbar\omega-i(\varepsilon^m_{{\bf k}}-\varepsilon^n_{{\bf k}'})-\eta}\}\nonumber\\
&-&\frac{eE\sin \theta_p }{2i}\sum_{ {\bf k}''} \frac{U^{mn}_{{\bf k}{\bf k}''}(f^{n}_{{\bf k}''}-f^{m}_{{\bf k}}){\hat j}\cdot \nabla_{{\bf k}'} \delta({\bf k}''-{\bf k}') }{(\varepsilon^n_{{\bf k}'}-\varepsilon^{m}_{{\bf k}})+i\hbar\eta}\{\frac{e^{i\omega t}}{i\hbar\omega-i(\varepsilon^m_{{\bf k}}-\varepsilon^n_{{\bf k}'})-\eta}-\frac{e^{-i\omega t}}{-i\hbar\omega-i(\varepsilon^m_{{\bf k}}-\varepsilon^n_{{\bf k}'})-\eta}\}\nonumber
\end{eqnarray} 
Since $\eta$ is an infinitesimal number, I expect terms contain $\frac{f^{n}_{{\bf k}'}-f^{m'}_{{\bf k}}}{(\varepsilon^n_{{\bf k}'}-\varepsilon^{m'}_{{\bf k}})+i\hbar\eta}=-\pi^{-1}(f^{n}_{{\bf k}'}-f^{m'}_{{\bf k}})\delta(\varepsilon^n_{{\bf k}'}-\varepsilon^{m'}_{{\bf k}})$ can be ignored. 
\begin{eqnarray}
&&\langle m,{\bf k} |g_E |n, {\bf k}' \rangle=\\
&-&\frac{eE\cos \theta_p }{2}\sum_{m' } \frac{U^{m'n}_{{\bf k}{\bf k}'}{\hat i}\cdot {\cal R}^{mm'}_{\bf k} (f^{n}_{{\bf k}'}-f^{m'}_{{\bf k}})}{(\varepsilon^n_{{\bf k}'}-\varepsilon^{m'}_{{\bf k}})+i\hbar\eta}\{\frac{e^{i\omega t}}{\hbar\omega-(\varepsilon^m_{{\bf k}}-\varepsilon^n_{{\bf k}'})+i\hbar\eta}+\frac{e^{-i\omega t}}{-\hbar\omega-(\varepsilon^m_{{\bf k}}-\varepsilon^n_{{\bf k}'})+i\hbar\eta}\}\nonumber\\
&+&\frac{eE\cos \theta_p }{2}\sum_{m' } \frac{U^{mm'}_{{\bf k}{\bf k}'}{\hat i}\cdot {\cal R}^{m'n}_{\bf k} (f^{m'}_{{\bf k}'}-f^{m}_{{\bf k}})}{(\varepsilon^n_{{\bf k}'}-\varepsilon^{m}_{{\bf k}})+i\hbar\eta}\{\frac{e^{i\omega t}}{\hbar\omega-(\varepsilon^m_{{\bf k}}-\varepsilon^n_{{\bf k}'})+i\hbar\eta}+\frac{e^{-i\omega t}}{-\hbar\omega-(\varepsilon^m_{{\bf k}}-\varepsilon^n_{{\bf k}'})+i\hbar\eta}\}\nonumber\\
&+&\frac{eE\cos \theta_p }{2}\sum_{ {\bf k}''} \frac{U^{mn}_{{\bf k}''{\bf k}'}(f^{n}_{{\bf k}'}-f^{m}_{{\bf k}''}){\hat i}\cdot \nabla_{\bf k} \delta({\bf k}''-{\bf k}) }{(\varepsilon^n_{{\bf k}'}-\varepsilon^{m}_{{\bf k}''})+i\hbar\eta}\{\frac{e^{i\omega t}}{i\hbar\omega-i(\varepsilon^m_{{\bf k}}-\varepsilon^n_{{\bf k}'})-\eta}+\frac{e^{-i\omega t}}{-i\hbar\omega-i(\varepsilon^m_{{\bf k}}-\varepsilon^n_{{\bf k}'})-\eta}\}\nonumber\\
&-&\frac{eE\cos \theta_p }{2}\sum_{ {\bf k}''} \frac{U^{mn}_{{\bf k}{\bf k}''}(f^{n}_{{\bf k}''}-f^{m}_{{\bf k}}){\hat i}\cdot \nabla_{{\bf k}'} \delta({\bf k}''-{\bf k}') }{(\varepsilon^n_{{\bf k}'}-\varepsilon^{m}_{{\bf k}})+i\hbar\eta}\{\frac{e^{i\omega t}}{i\hbar\omega-i(\varepsilon^m_{{\bf k}}-\varepsilon^n_{{\bf k}'})-\eta}+\frac{e^{-i\omega t}}{-i\hbar\omega-i(\varepsilon^m_{{\bf k}}-\varepsilon^n_{{\bf k}'})-\eta}\}\nonumber\\
&&+(\frac{eE\cos \theta_p }{2} \rightarrow \frac{E\sin \theta_p }{2i} ~{\text {and}}~ ({\hat i} \rightarrow {\hat j}) ~{\text {and}}~(...+... )\rightarrow (...-...))\nonumber
\end{eqnarray} 

Furthermore the matrix element of $g_0$ is

\begin{equation}\label{g02}
\langle m,{\bf k} |g_0 |n, {\bf k}' \rangle=-U^{mn}_{{\bf k}{\bf k}'}\frac{(f^n_{{\bf k}'}-f^m_{{\bf k}})}{\varepsilon^m_{{\bf k}}-\varepsilon^{n}_{{\bf k}'}-i\hbar \eta}
\end{equation}

With that, we can solve the equation of $\langle\rho \rangle$ as
\begin{equation}
\frac{\partial \langle \rho \rangle}{\partial t}+\frac{i}{\hbar}[{\mathcal H}_0+{\mathcal H}_E, \langle \rho \rangle ]=-\frac{i}{\hbar}\langle[U, g_0+g_E]\rangle
\end{equation}
where
\begin{equation}
\langle m,{\bf k}| [U, g_0]n,{\bf k}\rangle=-U^{mm'}_{{\bf k}{\bf k}'}U^{m'n}_{{\bf k}'{\bf k}}\frac{f^n_{\bf k}-f^{m'}_{\bf k'}}{\varepsilon^{m'}_{{\bf k}'}-\varepsilon^{n}_{{\bf k}}-i\hbar \eta}
+U^{mm'}_{{\bf k}{\bf k}'}U^{m'n}_{{\bf k}'{\bf k}}\frac{f^{m'}_{\bf k'}-f^{m}_{\bf k}}{\varepsilon^{m}_{{\bf k}}-\varepsilon^{m'}_{{\bf k}'}-i\hbar \eta}
\end{equation}
and finally the $J_E$ is
\begin{equation}
\langle m,{\bf k} |J_E(\langle \rho \rangle) |n, {\bf k} \rangle=\frac{i}{\hbar}\sum_{m',{\bf k}'} \{U^{mm'}_{{\bf k}{\bf k}'} \langle m',{\bf k}'|g_E |n, {\bf k} \rangle - \langle m,{\bf k}|g_E | m', {\bf k}'\rangle U^{m'n}_{{\bf k}'{\bf k}} \}
\end{equation}
We look at the matrix elements of $J_E$ in explicit form:
\begin{eqnarray}
\langle 1,{\bf k} |J_E(\langle \rho \rangle) |1, {\bf k} \rangle=\\
&&\frac{i}{\hbar}\sum_{{\bf k}'\in I} \{U^{11}_{{\bf k}{\bf k}'} \langle 1,{\bf k}'|g_E |1, {\bf k} \rangle - \langle 1,{\bf k}|g_E | 1, {\bf k}' \rangle U^{11}_{{\bf k}'{\bf k}} \}+\{U^{12}_{{\bf k}{\bf k}'} \langle 2,{\bf k}'|g_E |1, {\bf k} \rangle - \langle 1,{\bf k}|g_E |2, {\bf k}' \rangle U^{21}_{{\bf k}'{\bf k}} \}\nonumber\\
&&+\frac{i}{\hbar}\sum_{{\bf k}'\in II} \{U^{13}_{{\bf k}{\bf k}'} \langle 3,{\bf k}'|g_E |1, {\bf k} \rangle - \langle 1,{\bf k}|g_E | 3, {\bf k}' \rangle U^{31}_{{\bf k}'{\bf k}} \}+\{U^{14}_{{\bf k}{\bf k}'} \langle 4,{\bf k}'|g_E |1, {\bf k} \rangle - \langle 1,{\bf k}|g_E | 4, {\bf k}' \rangle U^{41}_{{\bf k}'{\bf k}} \}\nonumber\\
%......
\langle 1,{\bf k} |J_E(\langle \rho \rangle) |2, {\bf k} \rangle=\nonumber\\
&&\frac{i}{\hbar}\sum_{{\bf k}'} \{U^{11}_{{\bf k}{\bf k}'} \langle 1,{\bf k}'|g_E |2, {\bf k} \rangle - \langle 1,{\bf k}|g_E | 1, {\bf k}'\rangle U^{12}_{{\bf k}'{\bf k}} \}+\{U^{12}_{{\bf k}{\bf k}'} \langle 2,{\bf k}'|g_E |2, {\bf k} \rangle - \langle 1,{\bf k}|g_E | 2, {\bf k}'\rangle U^{22}_{{\bf k}'{\bf k}} \}\nonumber\\
&&+\{U^{13}_{{\bf k}{\bf k}'} \langle 3,{\bf k}'|g_E |2, {\bf k} \rangle - \langle 1,{\bf k}|g_E | 3, {\bf k}'\rangle U^{32}_{{\bf k}'{\bf k}} \}+\{U^{14}_{{\bf k}{\bf k}'} \langle 4,{\bf k}'|g_E |2, {\bf k} \rangle - \langle 1,{\bf k}|g_E | 4, {\bf k}'\rangle U^{42}_{{\bf k}'{\bf k}} \}\nonumber\\
%.....
\langle 1,{\bf k}' |J_E(\langle \rho \rangle) |3, {\bf k}' \rangle=\nonumber\\
&&\frac{i}{\hbar}\sum_{m',{\bf k}} \{U^{11}_{{\bf k}'{\bf k}} \langle 1,{\bf k}|g_E |3, {\bf k}' \rangle - \langle 1,{\bf k}'|g_E | 1, {\bf k}\rangle U^{13}_{{\bf k}{\bf k}'} \}+ \{U^{12}_{{\bf k}'{\bf k}} \langle 2,{\bf k}|g_E |3, {\bf k}' \rangle - \langle 1,{\bf k}'|g_E | 2, {\bf k}\rangle U^{23}_{{\bf k}{\bf k}'} \}\nonumber\\
&& \{U^{13}_{{\bf k}'{\bf k}} \langle 3,{\bf k}|g_E |3, {\bf k}' \rangle - \langle 1,{\bf k}'|g_E | 3, {\bf k}\rangle U^{33}_{{\bf k}{\bf k}'} \}+ \{U^{14}_{{\bf k}'{\bf k}} \langle 4,{\bf k}|g_E |3, {\bf k}' \rangle - \langle 1,{\bf k}'|g_E | 4, {\bf k}\rangle U^{43}_{{\bf k}{\bf k}'} \}\nonumber\\
%.....
\langle 3,{\bf k}' |J_E(\langle \rho \rangle) |1, {\bf k}' \rangle=\nonumber\\
&&\frac{i}{\hbar}\sum_{m',{\bf k}} \{U^{31}_{{\bf k}'{\bf k}} \langle 1,{\bf k}|g_E |1, {\bf k}' \rangle - \langle 3,{\bf k}'|g_E | 1, {\bf k}\rangle U^{13}_{{\bf k}{\bf k}'} \}+ \{U^{32}_{{\bf k}'{\bf k}} \langle 2,{\bf k}|g_E |1, {\bf k}' \rangle - \langle 3,{\bf k}'|g_E | 2, {\bf k}\rangle U^{21}_{{\bf k}{\bf k}'} \}\nonumber\\
&& \{U^{33}_{{\bf k}'{\bf k}} \langle 3,{\bf k}|g_E |1, {\bf k}' \rangle - \langle 3,{\bf k}'|g_E | 3, {\bf k}\rangle U^{31}_{{\bf k}{\bf k}'} \}+ \{U^{34}_{{\bf k}'{\bf k}} \langle 4,{\bf k}|g_E |1, {\bf k}' \rangle - \langle 3,{\bf k}'|g_E | 4, {\bf k}\rangle U^{41}_{{\bf k}{\bf k}'} \}\nonumber
\end{eqnarray}
If we look at closely to $\langle 1,{\bf k}' |J_E(\langle \rho \rangle) |1, {\bf k}' \rangle$ and $\langle 1,{\bf k}' |J_E(\langle \rho \rangle) |3, {\bf k}' \rangle$, we should perceive which intervalley effects add, and which ones cancel out between bands 1 and 3. 

Thus, to calculate $J_E$ we do need to evaluate $g_E$ first. All terms of $g_E$ which contains regular expression in denominator can be ignored. We also need to evaluate $U^{11}_{{\bf k}{\bf k}''} \langle 1,{\bf k}''|g_E |1, {\bf k} \rangle, U^{11}_{{\bf k}'{\bf k}} \langle 1,{\bf k}|g_E |2, {\bf k}' \rangle$, together with $\langle 1,{\bf k}|g_E |2, {\bf k}'' \rangle U^{21}_{{\bf k}''{\bf k}}, \langle 1,{\bf k}|g_E | 3, {\bf k}'' \rangle U^{31}_{{\bf k}''{\bf k}}, \langle 1,{\bf k}|g_E | 4, {\bf k}'' \rangle U^{41}_{{\bf k}''{\bf k}}$. Therefore, 
\begin{eqnarray}
&&\sum_{ {\bf k}'}U^{mm''}_{{\bf k}{\bf k}'}\langle m'',{\bf k}' |g_E |n, {\bf k} \rangle=\\
&-&\frac{eE\cos \theta_p }{2}\sum_{m' {\bf k}'} \frac{U^{mm''}_{{\bf k}{\bf k}'}U^{m'n}_{{\bf k}'{\bf k}}{\hat i}\cdot {\cal R}^{m''m'}_{\bf k'} (f^{n}_{{\bf k}}-f^{m'}_{{\bf k}'})}{(\varepsilon^n_{{\bf k}}-\varepsilon^{m'}_{{\bf k}'})+i\hbar\eta}\{\frac{e^{i\omega t}}{\hbar\omega-(\varepsilon^{m''}_{{\bf k}'}-\varepsilon^n_{{\bf k}})+i\hbar\eta}+\frac{e^{-i\omega t}}{-\hbar\omega-(\varepsilon^{m''}_{{\bf k}'}-\varepsilon^n_{{\bf k}})+i\hbar\eta}\}\nonumber\\
&+&\frac{eE\cos \theta_p }{2}\sum_{m' {\bf k}'} \frac{U^{mm''}_{{\bf k}{\bf k}'}U^{m''m'}_{{\bf k}'{\bf k}}{\hat i}\cdot {\cal R}^{m'n}_{{\bf k}' }(f^{m'}_{{\bf k}}-f^{m''}_{{\bf k}'})}{(\varepsilon^n_{{\bf k}}-\varepsilon^{m''}_{{\bf k}'})+i\hbar\eta}\{\frac{e^{i\omega t}}{\hbar\omega-(\varepsilon^{m''}_{{\bf k}'}-\varepsilon^n_{{\bf k}})+i\hbar\eta}+\frac{e^{-i\omega t}}{-\hbar\omega-(\varepsilon^{m''}_{{\bf k}'}-\varepsilon^n_{{\bf k}})+i\hbar\eta}\}\nonumber\\
&+&\frac{eE\cos \theta_p }{2}\sum_{{\bf k}' {\bf k}''} \frac{U^{mm''}_{{\bf k}{\bf k}'}U^{m''n}_{{\bf k}''{\bf k}}(f^{n}_{{\bf k}}-f^{m''}_{{\bf k}''}){\hat i}\cdot \nabla_{\bf k}' \delta({\bf k}''-{\bf k}') }{(\varepsilon^n_{{\bf k}}-\varepsilon^{m''}_{{\bf k}''})+i\hbar\eta}\{\frac{e^{i\omega t}}{i\hbar\omega-i(\varepsilon^{m''}_{{\bf k}'}-\varepsilon^n_{{\bf k}})-\eta}+\frac{e^{-i\omega t}}{-i\hbar\omega-i(\varepsilon^{m''}_{{\bf k}'}-\varepsilon^n_{{\bf k}})-\eta}\}\nonumber\\
&-&\frac{eE\cos \theta_p }{2}\sum_{{\bf k}' {\bf k}''} \frac{U^{mm''}_{{\bf k}{\bf k}'}U^{m''n}_{{\bf k}'{\bf k}''}(f^{n}_{{\bf k}''}-f^{m''}_{{\bf k}'}){\hat i}\cdot \nabla_{{\bf k}} \delta({\bf k}''-{\bf k}) }{(\varepsilon^n_{{\bf k}}-\varepsilon^{m''}_{{\bf k}'})+i\hbar\eta}\{\frac{e^{i\omega t}}{i\hbar\omega-i(\varepsilon^{m''}_{{\bf k}'}-\varepsilon^n_{{\bf k}})-\eta}+\frac{e^{-i\omega t}}{-i\hbar\omega-i(\varepsilon^{m''}_{{\bf k}'}-\varepsilon^n_{{\bf k}})-\eta}\}\nonumber\\
&&+(\frac{eE\cos \theta_p }{2} \rightarrow \frac{E\sin \theta_p }{2i} ~{\text {and}}~ ({\hat i} \rightarrow {\hat j}) ~{\text {and}}~(...+... )\rightarrow (...-...))\nonumber
\end{eqnarray} 
 and
 \begin{eqnarray}
&-&\sum_{{\bf k}'}\langle m,{\bf k} |g_E |n, {\bf k}' \rangle U^{nm''}_{{\bf k}'{\bf k}}=\\
&+&\frac{eE\cos \theta_p }{2}\sum_{m' {\bf k}'} \frac{U^{m'n}_{{\bf k}{\bf k}'}U^{nm''}_{{\bf k}'{\bf k}}{\hat i}\cdot {\cal R}^{mm'}_{\bf k} (f^{n}_{{\bf k}'}-f^{m'}_{{\bf k}})}{(\varepsilon^n_{{\bf k}'}-\varepsilon^{m'}_{{\bf k}})+i\hbar\eta}\{\frac{e^{i\omega t}}{\hbar\omega-(\varepsilon^m_{{\bf k}}-\varepsilon^n_{{\bf k}'})+i\hbar\eta}+\frac{e^{-i\omega t}}{-\hbar\omega-(\varepsilon^m_{{\bf k}}-\varepsilon^n_{{\bf k}'})+i\hbar\eta}\}\nonumber\\
&-&\frac{eE\cos \theta_p }{2}\sum_{m' {\bf k}'} \frac{U^{mm'}_{{\bf k}{\bf k}'}U^{nm''}_{{\bf k}'{\bf k}}{\hat i}\cdot {\cal R}^{m'n}_{\bf k} (f^{m'}_{{\bf k}'}-f^{m}_{{\bf k}})}{(\varepsilon^n_{{\bf k}'}-\varepsilon^{m}_{{\bf k}})+i\hbar\eta}\{\frac{e^{i\omega t}}{\hbar\omega-(\varepsilon^m_{{\bf k}}-\varepsilon^n_{{\bf k}'})+i\hbar\eta}+\frac{e^{-i\omega t}}{-\hbar\omega-(\varepsilon^m_{{\bf k}}-\varepsilon^n_{{\bf k}'})+i\hbar\eta}\}\nonumber\\
&-&\frac{eE\cos \theta_p }{2}\sum_{ {\bf k}'{\bf k}''} \frac{U^{mn}_{{\bf k}''{\bf k}'}U^{nm''}_{{\bf k}'{\bf k}}(f^{n}_{{\bf k}'}-f^{m}_{{\bf k}''}){\hat i}\cdot \nabla_{\bf k} \delta({\bf k}''-{\bf k}) }{(\varepsilon^n_{{\bf k}'}-\varepsilon^{m}_{{\bf k}''})+i\hbar\eta}\{\frac{e^{i\omega t}}{i\hbar\omega-i(\varepsilon^m_{{\bf k}}-\varepsilon^n_{{\bf k}'})-\eta}+\frac{e^{-i\omega t}}{-i\hbar\omega-i(\varepsilon^m_{{\bf k}}-\varepsilon^n_{{\bf k}'})-\eta}\}\nonumber\\
&+&\frac{eE\cos \theta_p }{2}\sum_{{\bf k}' {\bf k}''} \frac{U^{mn}_{{\bf k}{\bf k}''}U^{nm''}_{{\bf k}'{\bf k}}(f^{n}_{{\bf k}''}-f^{m}_{{\bf k}}){\hat i}\cdot \nabla_{{\bf k}'} \delta({\bf k}''-{\bf k}') }{(\varepsilon^n_{{\bf k}'}-\varepsilon^{m}_{{\bf k}})+i\hbar\eta}\{\frac{e^{i\omega t}}{i\hbar\omega-i(\varepsilon^m_{{\bf k}}-\varepsilon^n_{{\bf k}'})-\eta}+\frac{e^{-i\omega t}}{-i\hbar\omega-i(\varepsilon^m_{{\bf k}}-\varepsilon^n_{{\bf k}'})-\eta}\}\nonumber\\
&&-(\frac{eE\cos \theta_p }{2} \rightarrow \frac{E\sin \theta_p }{2i} ~{\text {and}}~ ({\hat i} \rightarrow {\hat j}) ~{\text {and}}~(...+... )\rightarrow (...-...))\nonumber\end{eqnarray}

We can simplify the above expression by making use of some assumptions for which ${\cal R}^{13}={\cal R}^{14}=0$. Furthermore, terms contain $x\delta(x)=0$. 
Therefore, the diagonal part of $J_E$ is given by
\begin{eqnarray}
&&\langle 1,{\bf k} |J_E(\langle \rho \rangle) |1, {\bf k} \rangle=\frac{i}{\hbar}\times\\
%m=1 m''=1 n=1
&-&\frac{eE\cos \theta_p }{2}\sum_{m' {\bf k}' \in I} \frac{U^{11}_{{\bf k}{\bf k}'}U^{m'1}_{{\bf k}'{\bf k}}{\hat i}\cdot {\cal R}^{1m'}_{\bf k'} (f^{1}_{{\bf k}}-f^{m'}_{{\bf k}'})}{(\varepsilon^1_{{\bf k}}-\varepsilon^{m'}_{{\bf k}'})+i\hbar\eta}\{\frac{e^{i\omega t}}{\hbar\omega-(\varepsilon^{1}_{{\bf k}'}-\varepsilon^1_{{\bf k}})+i\hbar\eta}+\frac{e^{-i\omega t}}{-\hbar\omega-(\varepsilon^{1}_{{\bf k}'}-\varepsilon^1_{{\bf k}})+i\hbar\eta}\}\nonumber\\
&+&\frac{eE\cos \theta_p }{2}\sum_{m' {\bf k}'} \frac{U^{11}_{{\bf k}{\bf k}'}U^{1m'}_{{\bf k}'{\bf k}}{\hat i}\cdot {\cal R}^{m'1}_{{\bf k}' }(f^{m'}_{{\bf k}}-f^{1}_{{\bf k}'})}{(\varepsilon^1_{{\bf k}}-\varepsilon^{1}_{{\bf k}'})+i\hbar\eta}\{\frac{e^{i\omega t}}{\hbar\omega-(\varepsilon^{1}_{{\bf k}'}-\varepsilon^1_{{\bf k}})+i\hbar\eta}+\frac{e^{-i\omega t}}{-\hbar\omega-(\varepsilon^{1}_{{\bf k}'}-\varepsilon^1_{{\bf k}})+i\hbar\eta}\}\nonumber\\
&+&\frac{eE\cos \theta_p }{2}\sum_{{\bf k}' {\bf k}''} \frac{U^{11}_{{\bf k}{\bf k}'}U^{11}_{{\bf k}''{\bf k}}(f^{1}_{{\bf k}}-f^{1}_{{\bf k}''}){\hat i}\cdot \nabla_{\bf k}' \delta({\bf k}''-{\bf k}') }{(\varepsilon^n_{{\bf k}}-\varepsilon^{1}_{{\bf k}''})+i\hbar\eta}\{\frac{e^{i\omega t}}{i\hbar\omega-i(\varepsilon^{1}_{{\bf k}'}-\varepsilon^1_{{\bf k}})-\eta}+\frac{e^{-i\omega t}}{-i\hbar\omega-i(\varepsilon^{1}_{{\bf k}'}-\varepsilon^1_{{\bf k}})-\eta}\}\nonumber\\
&-&\frac{eE\cos \theta_p }{2}\sum_{{\bf k}' {\bf k}''} \frac{U^{11}_{{\bf k}{\bf k}'}U^{11}_{{\bf k}'{\bf k}''}(f^{1}_{{\bf k}''}-f^{1}_{{\bf k}'}){\hat i}\cdot \nabla_{{\bf k}} \delta({\bf k}''-{\bf k}) }{(\varepsilon^1_{{\bf k}}-\varepsilon^{1}_{{\bf k}'})+i\hbar\eta}\{\frac{e^{i\omega t}}{i\hbar\omega-i(\varepsilon^{1}_{{\bf k}'}-\varepsilon^1_{{\bf k}})-\eta}+\frac{e^{-i\omega t}}{-i\hbar\omega-i(\varepsilon^{1}_{{\bf k}'}-\varepsilon^1_{{\bf k}})-\eta}\}\nonumber\\
%m=1 n=1 m''=1
&+&\frac{eE\cos \theta_p }{2}\sum_{ {\bf k}'} \frac{U^{21}_{{\bf k}{\bf k}'}U^{11}_{{\bf k}'{\bf k}}{\hat i}\cdot {\cal R}^{12}_{\bf k} (f^{1}_{{\bf k}'}-f^{2}_{{\bf k}})}{(\varepsilon^1_{{\bf k}'}-\varepsilon^{2}_{{\bf k}})+i\hbar\eta}\{\frac{e^{i\omega t}}{\hbar\omega-(\varepsilon^1_{{\bf k}}-\varepsilon^1_{{\bf k}'})+i\hbar\eta}+\frac{e^{-i\omega t}}{-\hbar\omega-(\varepsilon^1_{{\bf k}}-\varepsilon^1_{{\bf k}'})+i\hbar\eta}\}\nonumber\\
&-&\frac{eE\cos \theta_p }{2}\sum_{{\bf k}'} \frac{U^{12}_{{\bf k}{\bf k}'}U^{11}_{{\bf k}'{\bf k}}{\hat i}\cdot {\cal R}^{21}_{\bf k} (f^{2}_{{\bf k}'}-f^{1}_{{\bf k}})}{(\varepsilon^1_{{\bf k}'}-\varepsilon^{1}_{{\bf k}})+i\hbar\eta}\{\frac{e^{i\omega t}}{\hbar\omega-(\varepsilon^1_{{\bf k}}-\varepsilon^1_{{\bf k}'})+i\hbar\eta}+\frac{e^{-i\omega t}}{-\hbar\omega-(\varepsilon^1_{{\bf k}}-\varepsilon^1_{{\bf k}'})+i\hbar\eta}\}\nonumber\\
&-&\frac{eE\cos \theta_p }{2}\sum_{ {\bf k}'{\bf k}''} \frac{U^{11}_{{\bf k}''{\bf k}'}U^{11}_{{\bf k}'{\bf k}}(f^{1}_{{\bf k}'}-f^{1}_{{\bf k}''}){\hat i}\cdot \nabla_{\bf k} \delta({\bf k}''-{\bf k}) }{(\varepsilon^1_{{\bf k}'}-\varepsilon^{1}_{{\bf k}''})+i\hbar\eta}\{\frac{e^{i\omega t}}{i\hbar\omega-i(\varepsilon^1_{{\bf k}}-\varepsilon^1_{{\bf k}'})-\eta}+\frac{e^{-i\omega t}}{-i\hbar\omega-i(\varepsilon^1_{{\bf k}}-\varepsilon^1_{{\bf k}'})-\eta}\}\nonumber\\
&+&\frac{eE\cos \theta_p }{2}\sum_{{\bf k}' {\bf k}''} \frac{U^{11}_{{\bf k}{\bf k}''}U^{11}_{{\bf k}'{\bf k}}(f^{1}_{{\bf k}''}-f^{1}_{{\bf k}}){\hat i}\cdot \nabla_{{\bf k}'} \delta({\bf k}''-{\bf k}') }{(\varepsilon^n_{{\bf k}'}-\varepsilon^{1}_{{\bf k}})+i\hbar\eta}\{\frac{e^{i\omega t}}{i\hbar\omega-i(\varepsilon^1_{{\bf k}}-\varepsilon^1_{{\bf k}'})-\eta}+\frac{e^{-i\omega t}}{-i\hbar\omega-i(\varepsilon^1_{{\bf k}}-\varepsilon^1_{{\bf k}'})-\eta}\}\nonumber\\
%%%m=1 n=1 m''=2
&-&\frac{eE\cos \theta_p }{2}\sum_{m' {\bf k}'} \frac{U^{12}_{{\bf k}{\bf k}'}U^{m'1}_{{\bf k}'{\bf k}}{\hat i}\cdot {\cal R}^{2m'}_{\bf k'} (f^{1}_{{\bf k}}-f^{m'}_{{\bf k}'})}{(\varepsilon^1_{{\bf k}}-\varepsilon^{m'}_{{\bf k}'})+i\hbar\eta}\{\frac{e^{i\omega t}}{\hbar\omega-(\varepsilon^{2}_{{\bf k}'}-\varepsilon^1_{{\bf k}})+i\hbar\eta}+\frac{e^{-i\omega t}}{-\hbar\omega-(\varepsilon^{2}_{{\bf k}'}-\varepsilon^1_{{\bf k}})+i\hbar\eta}\}\nonumber\\
&+&\frac{eE\cos \theta_p }{2}\sum_{m' {\bf k}'} \frac{U^{12}_{{\bf k}{\bf k}'}U^{2m'}_{{\bf k}'{\bf k}}{\hat i}\cdot {\cal R}^{m'1}_{{\bf k}' }(f^{m'}_{{\bf k}}-f^{2}_{{\bf k}'})}{(\varepsilon^1_{{\bf k}}-\varepsilon^{2}_{{\bf k}'})+i\hbar\eta}\{\frac{e^{i\omega t}}{\hbar\omega-(\varepsilon^{2}_{{\bf k}'}-\varepsilon^1_{{\bf k}})+i\hbar\eta}+\frac{e^{-i\omega t}}{-\hbar\omega-(\varepsilon^{2}_{{\bf k}'}-\varepsilon^1_{{\bf k}})+i\hbar\eta}\}\nonumber\\
&+&\frac{eE\cos \theta_p }{2}\sum_{{\bf k}' {\bf k}''} \frac{U^{12}_{{\bf k}{\bf k}'}U^{21}_{{\bf k}''{\bf k}}(f^{1}_{{\bf k}}-f^{2}_{{\bf k}''}){\hat i}\cdot \nabla_{\bf k}' \delta({\bf k}''-{\bf k}') }{(\varepsilon^1_{{\bf k}}-\varepsilon^{2}_{{\bf k}''})+i\hbar\eta}\{\frac{e^{i\omega t}}{i\hbar\omega-i(\varepsilon^{2}_{{\bf k}'}-\varepsilon^1_{{\bf k}})-\eta}+\frac{e^{-i\omega t}}{-i\hbar\omega-i(\varepsilon^{2}_{{\bf k}'}-\varepsilon^1_{{\bf k}})-\eta}\}\nonumber\\
&-&\frac{eE\cos \theta_p }{2}\sum_{{\bf k}' {\bf k}''} \frac{U^{12}_{{\bf k}{\bf k}'}U^{21}_{{\bf k}'{\bf k}''}(f^{1}_{{\bf k}''}-f^{2}_{{\bf k}'}){\hat i}\cdot \nabla_{{\bf k}} \delta({\bf k}''-{\bf k}) }{(\varepsilon^1_{{\bf k}}-\varepsilon^{2}_{{\bf k}'})+i\hbar\eta}\{\frac{e^{i\omega t}}{i\hbar\omega-i(\varepsilon^{2}_{{\bf k}'}-\varepsilon^1_{{\bf k}})-\eta}+\frac{e^{-i\omega t}}{-i\hbar\omega-i(\varepsilon^{2}_{{\bf k}'}-\varepsilon^1_{{\bf k}})-\eta}\}\nonumber
\end{eqnarray}
\begin{eqnarray}
%%%m=1  n=2 m''=1
&+&\frac{eE\cos \theta_p }{2}\sum_{ {\bf k}'} \frac{U^{12}_{{\bf k}{\bf k}'}U^{21}_{{\bf k}'{\bf k}}{\hat i}\cdot {\cal R}^{11}_{\bf k} (f^{2}_{{\bf k}'}-f^{1}_{{\bf k}})}{(\varepsilon^2_{{\bf k}'}-\varepsilon^{1}_{{\bf k}})+i\hbar\eta}\{\frac{e^{i\omega t}}{\hbar\omega-(\varepsilon^1_{{\bf k}}-\varepsilon^2_{{\bf k}'})+i\hbar\eta}+\frac{e^{-i\omega t}}{-\hbar\omega-(\varepsilon^1_{{\bf k}}-\varepsilon^2_{{\bf k}'})+i\hbar\eta}\}\nonumber\\
&-&\frac{eE\cos \theta_p }{2}\sum_{m'=1, {\bf k}'} \frac{U^{1m'}_{{\bf k}{\bf k}'}U^{21}_{{\bf k}'{\bf k}}{\hat i}\cdot {\cal R}^{m'2}_{\bf k} (f^{m'}_{{\bf k}'}-f^{1}_{{\bf k}})}{(\varepsilon^2_{{\bf k}'}-\varepsilon^{1}_{{\bf k}})+i\hbar\eta}\{\frac{e^{i\omega t}}{\hbar\omega-(\varepsilon^1_{{\bf k}}-\varepsilon^2_{{\bf k}'})+i\hbar\eta}+\frac{e^{-i\omega t}}{-\hbar\omega-(\varepsilon^1_{{\bf k}}-\varepsilon^2_{{\bf k}'})+i\hbar\eta}\}\nonumber\\
&-&\frac{eE\cos \theta_p }{2}\sum_{ {\bf k}'{\bf k}''} \frac{U^{12}_{{\bf k}''{\bf k}'}U^{21}_{{\bf k}'{\bf k}}(f^{2}_{{\bf k}'}-f^{1}_{{\bf k}''}){\hat i}\cdot \nabla_{\bf k} \delta({\bf k}''-{\bf k}) }{(\varepsilon^2_{{\bf k}'}-\varepsilon^{1}_{{\bf k}''})+i\hbar\eta}\{\frac{e^{i\omega t}}{i\hbar\omega-i(\varepsilon^1_{{\bf k}}-\varepsilon^2_{{\bf k}'})-\eta}+\frac{e^{-i\omega t}}{-i\hbar\omega-i(\varepsilon^1_{{\bf k}}-\varepsilon^2_{{\bf k}'})-\eta}\}\nonumber\\
&+&\frac{eE\cos \theta_p }{2}\sum_{{\bf k}' {\bf k}''} \frac{U^{12}_{{\bf k}{\bf k}''}U^{21}_{{\bf k}'{\bf k}}(f^{2}_{{\bf k}''}-f^{1}_{{\bf k}}){\hat i}\cdot \nabla_{{\bf k}'} \delta({\bf k}''-{\bf k}') }{(\varepsilon^2_{{\bf k}'}-\varepsilon^{1}_{{\bf k}})+i\hbar\eta}\{\frac{e^{i\omega t}}{i\hbar\omega-i(\varepsilon^1_{{\bf k}}-\varepsilon^2_{{\bf k}'})-\eta}+\frac{e^{-i\omega t}}{-i\hbar\omega-i(\varepsilon^1_{{\bf k}}-\varepsilon^2_{{\bf k}'})-\eta}\}\nonumber\\
%m=1 m''=3 n=1
&-&\frac{eE\cos \theta_p }{2}\sum_{m' {\bf k}'} \frac{U^{13}_{{\bf k}{\bf k}'}U^{m'1}_{{\bf k}'{\bf k}}{\hat i}\cdot {\cal R}^{3m'}_{\bf k'} (f^{1}_{{\bf k}}-f^{m'}_{{\bf k}'})}{(\varepsilon^1_{{\bf k}}-\varepsilon^{m'}_{{\bf k}'})+i\hbar\eta}\{\frac{e^{i\omega t}}{\hbar\omega-(\varepsilon^{3}_{{\bf k}'}-\varepsilon^1_{{\bf k}})+i\hbar\eta}+\frac{e^{-i\omega t}}{-\hbar\omega-(\varepsilon^{3}_{{\bf k}'}-\varepsilon^1_{{\bf k}})+i\hbar\eta}\}\nonumber\\
&+&\frac{eE\cos \theta_p }{2}\sum_{m' {\bf k}'} \frac{U^{13}_{{\bf k}{\bf k}'}U^{3m'}_{{\bf k}'{\bf k}}{\hat i}\cdot {\cal R}^{m'1}_{{\bf k}' }(f^{m'}_{{\bf k}}-f^{3}_{{\bf k}'})}{(\varepsilon^1_{{\bf k}}-\varepsilon^{3}_{{\bf k}'})+i\hbar\eta}\{\frac{e^{i\omega t}}{\hbar\omega-(\varepsilon^{3}_{{\bf k}'}-\varepsilon^1_{{\bf k}})+i\hbar\eta}+\frac{e^{-i\omega t}}{-\hbar\omega-(\varepsilon^{3}_{{\bf k}'}-\varepsilon^1_{{\bf k}})+i\hbar\eta}\}\nonumber\\
&+&\frac{eE\cos \theta_p }{2}\sum_{{\bf k}' {\bf k}''} \frac{U^{13}_{{\bf k}{\bf k}'}U^{3n}_{{\bf k}''{\bf k}}(f^{1}_{{\bf k}}-f^{3}_{{\bf k}''}){\hat i}\cdot \nabla_{\bf k}' \delta({\bf k}''-{\bf k}') }{(\varepsilon^1_{{\bf k}}-\varepsilon^{3}_{{\bf k}''})+i\hbar\eta}\{\frac{e^{i\omega t}}{i\hbar\omega-i(\varepsilon^{3}_{{\bf k}'}-\varepsilon^1_{{\bf k}})-\eta}+\frac{e^{-i\omega t}}{-i\hbar\omega-i(\varepsilon^{3}_{{\bf k}'}-\varepsilon^1_{{\bf k}})-\eta}\}\nonumber\\
&-&\frac{eE\cos \theta_p }{2}\sum_{{\bf k}' {\bf k}''} \frac{U^{13}_{{\bf k}{\bf k}'}U^{31}_{{\bf k}'{\bf k}''}(f^{1}_{{\bf k}''}-f^{3}_{{\bf k}'}){\hat i}\cdot \nabla_{{\bf k}} \delta({\bf k}''-{\bf k}) }{(\varepsilon^1_{{\bf k}}-\varepsilon^{3}_{{\bf k}'})+i\hbar\eta}\{\frac{e^{i\omega t}}{i\hbar\omega-i(\varepsilon^{3}_{{\bf k}'}-\varepsilon^1_{{\bf k}})-\eta}+\frac{e^{-i\omega t}}{-i\hbar\omega-i(\varepsilon^{3}_{{\bf k}'}-\varepsilon^1_{{\bf k}})-\eta}\}\nonumber\\
% m=1  n=3 m''=1
&+&\frac{eE\cos \theta_p }{2}\sum_{{\bf k}'} \frac{U^{23}_{{\bf k}{\bf k}'}U^{31}_{{\bf k}'{\bf k}}{\hat i}\cdot {\cal R}^{12}_{\bf k} (f^{3}_{{\bf k}'}-f^{2}_{{\bf k}})}{(\varepsilon^3_{{\bf k}'}-\varepsilon^{2}_{{\bf k}})+i\hbar\eta}\{\frac{e^{i\omega t}}{\hbar\omega-(\varepsilon^1_{{\bf k}}-\varepsilon^3_{{\bf k}'})+i\hbar\eta}+\frac{e^{-i\omega t}}{-\hbar\omega-(\varepsilon^1_{{\bf k}}-\varepsilon^3_{{\bf k}'})+i\hbar\eta}\}\nonumber\\
&-&\frac{eE\cos \theta_p }{2}\sum_{m'=4, {\bf k}'} \frac{U^{1m'}_{{\bf k}{\bf k}'}U^{31}_{{\bf k}'{\bf k}}{\hat i}\cdot {\cal R}^{m'3}_{\bf k} (f^{m'}_{{\bf k}'}-f^{1}_{{\bf k}})}{(\varepsilon^1_{{\bf k}'}-\varepsilon^{1}_{{\bf k}})+i\hbar\eta}\{\frac{e^{i\omega t}}{\hbar\omega-(\varepsilon^1_{{\bf k}}-\varepsilon^3_{{\bf k}'})+i\hbar\eta}+\frac{e^{-i\omega t}}{-\hbar\omega-(\varepsilon^1_{{\bf k}}-\varepsilon^3_{{\bf k}'})+i\hbar\eta}\}\nonumber\\
&-&\frac{eE\cos \theta_p }{2}\sum_{ {\bf k}'{\bf k}''} \frac{U^{13}_{{\bf k}''{\bf k}'}U^{31}_{{\bf k}'{\bf k}}(f^{3}_{{\bf k}'}-f^{1}_{{\bf k}''}){\hat i}\cdot \nabla_{\bf k} \delta({\bf k}''-{\bf k}) }{(\varepsilon^3_{{\bf k}'}-\varepsilon^{1}_{{\bf k}''})+i\hbar\eta}\{\frac{e^{i\omega t}}{i\hbar\omega-i(\varepsilon^1_{{\bf k}}-\varepsilon^3_{{\bf k}'})-\eta}+\frac{e^{-i\omega t}}{-i\hbar\omega-i(\varepsilon^1_{{\bf k}}-\varepsilon^3_{{\bf k}'})-\eta}\}\nonumber\\
&+&\frac{eE\cos \theta_p }{2}\sum_{{\bf k}' {\bf k}''} \frac{U^{13}_{{\bf k}{\bf k}''}U^{31}_{{\bf k}'{\bf k}}(f^{3}_{{\bf k}''}-f^{1}_{{\bf k}}){\hat i}\cdot \nabla_{{\bf k}'} \delta({\bf k}''-{\bf k}') }{(\varepsilon^3_{{\bf k}'}-\varepsilon^{1}_{{\bf k}})+i\hbar\eta}\{\frac{e^{i\omega t}}{i\hbar\omega-i(\varepsilon^3_{{\bf k}}-\varepsilon^1_{{\bf k}'})-\eta}+\frac{e^{-i\omega t}}{-i\hbar\omega-i(\varepsilon^1_{{\bf k}}-\varepsilon^3_{{\bf k}'})-\eta}\}\nonumber\\
%%m=1  m''=4  n=1
&-&\frac{eE\cos \theta_p }{2}\sum_{m' {\bf k}'} \frac{U^{14}_{{\bf k}{\bf k}'}U^{m'1}_{{\bf k}'{\bf k}}{\hat i}\cdot {\cal R}^{4m'}_{\bf k'} (f^{1}_{{\bf k}}-f^{m'}_{{\bf k}'})}{(\varepsilon^1_{{\bf k}}-\varepsilon^{m'}_{{\bf k}'})+i\hbar\eta}\{\frac{e^{i\omega t}}{\hbar\omega-(\varepsilon^{4}_{{\bf k}'}-\varepsilon^1_{{\bf k}})+i\hbar\eta}+\frac{e^{-i\omega t}}{-\hbar\omega-(\varepsilon^{4}_{{\bf k}'}-\varepsilon^1_{{\bf k}})+i\hbar\eta}\}\nonumber\\
&+&\frac{eE\cos \theta_p }{2}\sum_{m' {\bf k}'} \frac{U^{14}_{{\bf k}{\bf k}'}U^{4m'}_{{\bf k}'{\bf k}}{\hat i}\cdot {\cal R}^{m'1}_{{\bf k}' }(f^{m'}_{{\bf k}}-f^{4}_{{\bf k}'})}{(\varepsilon^1_{{\bf k}}-\varepsilon^{4}_{{\bf k}'})+i\hbar\eta}\{\frac{e^{i\omega t}}{\hbar\omega-(\varepsilon^{4}_{{\bf k}'}-\varepsilon^1_{{\bf k}})+i\hbar\eta}+\frac{e^{-i\omega t}}{-\hbar\omega-(\varepsilon^{4}_{{\bf k}'}-\varepsilon^1_{{\bf k}})+i\hbar\eta}\}\nonumber\\
&+&\frac{eE\cos \theta_p }{2}\sum_{{\bf k}' {\bf k}''} \frac{U^{14}_{{\bf k}{\bf k}'}U^{41}_{{\bf k}''{\bf k}}(f^{1}_{{\bf k}}-f^{4}_{{\bf k}''}){\hat i}\cdot \nabla_{\bf k}' \delta({\bf k}''-{\bf k}') }{(\varepsilon^1_{{\bf k}}-\varepsilon^{4}_{{\bf k}''})+i\hbar\eta}\{\frac{e^{i\omega t}}{i\hbar\omega-i(\varepsilon^{4}_{{\bf k}'}-\varepsilon^1_{{\bf k}})-\eta}+\frac{e^{-i\omega t}}{-i\hbar\omega-i(\varepsilon^{4}_{{\bf k}'}-\varepsilon^1_{{\bf k}})-\eta}\}\nonumber\\
&-&\frac{eE\cos \theta_p }{2}\sum_{{\bf k}' {\bf k}''} \frac{U^{14}_{{\bf k}{\bf k}'}U^{41}_{{\bf k}'{\bf k}''}(f^{1}_{{\bf k}''}-f^{4}_{{\bf k}'}){\hat i}\cdot \nabla_{{\bf k}} \delta({\bf k}''-{\bf k}) }{(\varepsilon^1_{{\bf k}}-\varepsilon^{4}_{{\bf k}'})+i\hbar\eta}\{\frac{e^{i\omega t}}{i\hbar\omega-i(\varepsilon^{4}_{{\bf k}'}-\varepsilon^1_{{\bf k}})-\eta}+\frac{e^{-i\omega t}}{-i\hbar\omega-i(\varepsilon^{4}_{{\bf k}'}-\varepsilon^1_{{\bf k}})-\eta}\}\nonumber\\
%m=1 n=4 m''=1
&+&\frac{eE\cos \theta_p }{2}\sum_{m'=1,2, {\bf k}'} \frac{U^{m'4}_{{\bf k}{\bf k}'}U^{41}_{{\bf k}'{\bf k}}{\hat i}\cdot {\cal R}^{1m'}_{\bf k} (f^{4}_{{\bf k}'}-f^{m'}_{{\bf k}})}{(\varepsilon^4_{{\bf k}'}-\varepsilon^{m'}_{{\bf k}})+i\hbar\eta}\{\frac{e^{i\omega t}}{\hbar\omega-(\varepsilon^1_{{\bf k}}-\varepsilon^4_{{\bf k}'})+i\hbar\eta}+\frac{e^{-i\omega t}}{-\hbar\omega-(\varepsilon^1_{{\bf k}}-\varepsilon^4_{{\bf k}'})+i\hbar\eta}\}\nonumber\\
&-&\frac{eE\cos \theta_p }{2}\sum_{m'=3,4, {\bf k}'} \frac{U^{1m'}_{{\bf k}{\bf k}'}U^{4m''}_{{\bf k}'{\bf k}}{\hat i}\cdot {\cal R}^{m'4}_{\bf k} (f^{m'}_{{\bf k}'}-f^{1}_{{\bf k}})}{(\varepsilon^4_{{\bf k}'}-\varepsilon^{1}_{{\bf k}})+i\hbar\eta}\{\frac{e^{i\omega t}}{\hbar\omega-(\varepsilon^1_{{\bf k}}-\varepsilon^4_{{\bf k}'})+i\hbar\eta}+\frac{e^{-i\omega t}}{-\hbar\omega-(\varepsilon^1_{{\bf k}}-\varepsilon^4_{{\bf k}'})+i\hbar\eta}\}\nonumber\\
&-&\frac{eE\cos \theta_p }{2}\sum_{ {\bf k}'{\bf k}''} \frac{U^{14}_{{\bf k}''{\bf k}'}U^{41}_{{\bf k}'{\bf k}}(f^{4}_{{\bf k}'}-f^{1}_{{\bf k}''}){\hat i}\cdot \nabla_{\bf k} \delta({\bf k}''-{\bf k}) }{(\varepsilon^1_{{\bf k}'}-\varepsilon^{1}_{{\bf k}''})+i\hbar\eta}\{\frac{e^{i\omega t}}{i\hbar\omega-i(\varepsilon^1_{{\bf k}}-\varepsilon^4_{{\bf k}'})-\eta}+\frac{e^{-i\omega t}}{-i\hbar\omega-i(\varepsilon^1_{{\bf k}}-\varepsilon^4_{{\bf k}'})-\eta}\}\nonumber\\
&+&\frac{eE\cos \theta_p }{2}\sum_{{\bf k}' {\bf k}''} \frac{U^{14}_{{\bf k}{\bf k}''}U^{41}_{{\bf k}'{\bf k}}(f^{4}_{{\bf k}''}-f^{1}_{{\bf k}}){\hat i}\cdot \nabla_{{\bf k}'} \delta({\bf k}''-{\bf k}') }{(\varepsilon^4_{{\bf k}'}-\varepsilon^{1}_{{\bf k}})+i\hbar\eta}\{\frac{e^{i\omega t}}{i\hbar\omega-i(\varepsilon^1_{{\bf k}}-\varepsilon^4_{{\bf k}'})-\eta}+\frac{e^{-i\omega t}}{-i\hbar\omega-i(\varepsilon^1_{{\bf k}}-\varepsilon^4_{{\bf k}'})-\eta}\}\nonumber\\
&&+ {\text{other 32 terms by making use of }} (\frac{E\cos \theta_p }{2} \rightarrow \frac{E\sin \theta_p }{2i} ~{\text {and}}~ ({\hat i} \rightarrow {\hat j}) ~{\text {and}}~(...+... )\rightarrow (...-...))\nonumber
\end{eqnarray}
%%%%%%%%%%%%%%%%%%%%%%%%%%%%%

In the same manner we can calculate the off-diagonal part
\begin{eqnarray}
&&\langle 1,{\bf k} |J_E(\langle \rho \rangle) |2, {\bf k} \rangle=\frac{i}{\hbar}\times\\
%m=1 m''=1 n=2
&-&\frac{eE\cos \theta_p }{2}\sum_{m' {\bf k}'} \frac{U^{11}_{{\bf k}{\bf k}'}U^{m'2}_{{\bf k}'{\bf k}}{\hat i}\cdot {\cal R}^{1m'}_{\bf k'} (f^{2}_{{\bf k}}-f^{m'}_{{\bf k}'})}{(\varepsilon^2_{{\bf k}}-\varepsilon^{m'}_{{\bf k}'})+i\hbar\eta}\{\frac{e^{i\omega t}}{\hbar\omega-(\varepsilon^{1}_{{\bf k}'}-\varepsilon^2_{{\bf k}})+i\hbar\eta}+\frac{e^{-i\omega t}}{-\hbar\omega-(\varepsilon^{1}_{{\bf k}'}-\varepsilon^2_{{\bf k}})+i\hbar\eta}\}\nonumber\\
&+&\frac{eE\cos \theta_p }{2}\sum_{m' {\bf k}'} \frac{U^{11}_{{\bf k}{\bf k}'}U^{1m'}_{{\bf k}'{\bf k}}{\hat i}\cdot {\cal R}^{m'2}_{{\bf k}' }(f^{m'}_{{\bf k}}-f^{1}_{{\bf k}'})}{(\varepsilon^2_{{\bf k}}-\varepsilon^{1}_{{\bf k}'})+i\hbar\eta}\{\frac{e^{i\omega t}}{\hbar\omega-(\varepsilon^{1}_{{\bf k}'}-\varepsilon^2_{{\bf k}})+i\hbar\eta}+\frac{e^{-i\omega t}}{-\hbar\omega-(\varepsilon^{1}_{{\bf k}'}-\varepsilon^2_{{\bf k}})+i\hbar\eta}\}\nonumber\\
&+&\frac{eE\cos \theta_p }{2}\sum_{{\bf k}' {\bf k}''} \frac{U^{m1}_{{\bf k}{\bf k}'}U^{12}_{{\bf k}''{\bf k}}(f^{2}_{{\bf k}}-f^{1}_{{\bf k}''}){\hat i}\cdot \nabla_{\bf k}' \delta({\bf k}''-{\bf k}') }{(\varepsilon^2_{{\bf k}}-\varepsilon^{1}_{{\bf k}''})+i\hbar\eta}\{\frac{e^{i\omega t}}{i\hbar\omega-i(\varepsilon^{1}_{{\bf k}'}-\varepsilon^2_{{\bf k}})-\eta}+\frac{e^{-i\omega t}}{-i\hbar\omega-i(\varepsilon^{1}_{{\bf k}'}-\varepsilon^2_{{\bf k}})-\eta}\}\nonumber\\
&-&\frac{eE\cos \theta_p }{2}\sum_{{\bf k}' {\bf k}''} \frac{U^{11}_{{\bf k}{\bf k}'}U^{12}_{{\bf k}'{\bf k}''}(f^{n}_{{\bf k}''}-f^{1}_{{\bf k}'}){\hat i}\cdot \nabla_{{\bf k}} \delta({\bf k}''-{\bf k}) }{(\varepsilon^2_{{\bf k}}-\varepsilon^{1}_{{\bf k}'})+i\hbar\eta}\{\frac{e^{i\omega t}}{i\hbar\omega-i(\varepsilon^{1}_{{\bf k}'}-\varepsilon^2_{{\bf k}})-\eta}+\frac{e^{-i\omega t}}{-i\hbar\omega-i(\varepsilon^{1}_{{\bf k}'}-\varepsilon^2_{{\bf k}})-\eta}\}\nonumber\\
%m=1, n=1 m''=2
&+&\frac{eE\cos \theta_p }{2}\sum_{m' {\bf k}'} \frac{U^{m'1}_{{\bf k}{\bf k}'}U^{12}_{{\bf k}'{\bf k}}{\hat i}\cdot {\cal R}^{1m'}_{\bf k} (f^{1}_{{\bf k}'}-f^{m'}_{{\bf k}})}{(\varepsilon^1_{{\bf k}'}-\varepsilon^{m'}_{{\bf k}})+i\hbar\eta}\{\frac{e^{i\omega t}}{\hbar\omega-(\varepsilon^1_{{\bf k}}-\varepsilon^1_{{\bf k}'})+i\hbar\eta}+\frac{e^{-i\omega t}}{-\hbar\omega-(\varepsilon^1_{{\bf k}}-\varepsilon^1_{{\bf k}'})+i\hbar\eta}\}\nonumber\\
&-&\frac{eE\cos \theta_p }{2}\sum_{ {\bf k}'} \frac{U^{11}_{{\bf k}{\bf k}'}U^{12}_{{\bf k}'{\bf k}}{\hat i}\cdot {\cal R}^{21}_{\bf k} (f^{1}_{{\bf k}'}-f^{1}_{{\bf k}})}{(\varepsilon^1_{{\bf k}'}-\varepsilon^{1}_{{\bf k}})+i\hbar\eta}\{\frac{e^{i\omega t}}{\hbar\omega-(\varepsilon^1_{{\bf k}}-\varepsilon^1_{{\bf k}'})+i\hbar\eta}+\frac{e^{-i\omega t}}{-\hbar\omega-(\varepsilon^1_{{\bf k}}-\varepsilon^1_{{\bf k}'})+i\hbar\eta}\}\nonumber\\
&-&\frac{eE\cos \theta_p }{2}\sum_{ {\bf k}'{\bf k}''} \frac{U^{11}_{{\bf k}''{\bf k}'}U^{12}_{{\bf k}'{\bf k}}(f^{1}_{{\bf k}'}-f^{1}_{{\bf k}''}){\hat i}\cdot \nabla_{\bf k} \delta({\bf k}''-{\bf k}) }{(\varepsilon^1_{{\bf k}'}-\varepsilon^{1}_{{\bf k}''})+i\hbar\eta}\{\frac{e^{i\omega t}}{i\hbar\omega-i(\varepsilon^1_{{\bf k}}-\varepsilon^1_{{\bf k}'})-\eta}+\frac{e^{-i\omega t}}{-i\hbar\omega-i(\varepsilon^1_{{\bf k}}-\varepsilon^1_{{\bf k}'})-\eta}\}\nonumber\\
&+&\frac{eE\cos \theta_p }{2}\sum_{{\bf k}' {\bf k}''} \frac{U^{11}_{{\bf k}{\bf k}''}U^{12}_{{\bf k}'{\bf k}}(f^{2}_{{\bf k}''}-f^{1}_{{\bf k}}){\hat i}\cdot \nabla_{{\bf k}'} \delta({\bf k}''-{\bf k}') }{(\varepsilon^2_{{\bf k}'}-\varepsilon^{1}_{{\bf k}})+i\hbar\eta}\{\frac{e^{i\omega t}}{i\hbar\omega-i(\varepsilon^1_{{\bf k}}-\varepsilon^2_{{\bf k}'})-\eta}+\frac{e^{-i\omega t}}{-i\hbar\omega-i(\varepsilon^1_{{\bf k}}-\varepsilon^2_{{\bf k}'})-\eta}\}\nonumber\\
%m=1 m''=2 n=2
&-&\frac{eE\cos \theta_p }{2}\sum_{m' {\bf k}'} \frac{U^{12}_{{\bf k}{\bf k}'}U^{m'2}_{{\bf k}'{\bf k}}{\hat i}\cdot {\cal R}^{2m'}_{\bf k'} (f^{2}_{{\bf k}}-f^{m'}_{{\bf k}'})}{(\varepsilon^2_{{\bf k}}-\varepsilon^{m'}_{{\bf k}'})+i\hbar\eta}\{\frac{e^{i\omega t}}{\hbar\omega-(\varepsilon^{2}_{{\bf k}'}-\varepsilon^2_{{\bf k}})+i\hbar\eta}+\frac{e^{-i\omega t}}{-\hbar\omega-(\varepsilon^{2}_{{\bf k}'}-\varepsilon^2_{{\bf k}})+i\hbar\eta}\}\nonumber\\
&+&\frac{eE\cos \theta_p }{2}\sum_{m' {\bf k}'} \frac{U^{12}_{{\bf k}{\bf k}'}U^{2m'}_{{\bf k}'{\bf k}}{\hat i}\cdot {\cal R}^{m'2}_{{\bf k}' }(f^{m'}_{{\bf k}}-f^{2}_{{\bf k}'})}{(\varepsilon^2_{{\bf k}}-\varepsilon^{2}_{{\bf k}'})+i\hbar\eta}\{\frac{e^{i\omega t}}{\hbar\omega-(\varepsilon^{2}_{{\bf k}'}-\varepsilon^2_{{\bf k}})+i\hbar\eta}+\frac{e^{-i\omega t}}{-\hbar\omega-(\varepsilon^{2}_{{\bf k}'}-\varepsilon^2_{{\bf k}})+i\hbar\eta}\}\nonumber\\
&+&\frac{eE\cos \theta_p }{2}\sum_{{\bf k}' {\bf k}''} \frac{U^{12}_{{\bf k}{\bf k}'}U^{22}_{{\bf k}''{\bf k}}(f^{2}_{{\bf k}}-f^{2}_{{\bf k}''}){\hat i}\cdot \nabla_{\bf k}' \delta({\bf k}''-{\bf k}') }{(\varepsilon^2_{{\bf k}}-\varepsilon^{2}_{{\bf k}''})+i\hbar\eta}\{\frac{e^{i\omega t}}{i\hbar\omega-i(\varepsilon^{2}_{{\bf k}'}-\varepsilon^2_{{\bf k}})-\eta}+\frac{e^{-i\omega t}}{-i\hbar\omega-i(\varepsilon^{2}_{{\bf k}'}-\varepsilon^2_{{\bf k}})-\eta}\}\nonumber\\
&-&\frac{eE\cos \theta_p }{2}\sum_{{\bf k}' {\bf k}''} \frac{U^{12}_{{\bf k}{\bf k}'}U^{22}_{{\bf k}'{\bf k}''}(f^{2}_{{\bf k}''}-f^{2}_{{\bf k}'}){\hat i}\cdot \nabla_{{\bf k}} \delta({\bf k}''-{\bf k}) }{(\varepsilon^2_{{\bf k}}-\varepsilon^{2}_{{\bf k}'})+i\hbar\eta}\{\frac{e^{i\omega t}}{i\hbar\omega-i(\varepsilon^{2}_{{\bf k}'}-\varepsilon^2_{{\bf k}})-\eta}+\frac{e^{-i\omega t}}{-i\hbar\omega-i(\varepsilon^{2}_{{\bf k}'}-\varepsilon^2_{{\bf k}})-\eta}\}\nonumber\\
%m=1 n=2 m''=2
&+&\frac{eE\cos \theta_p }{2}\sum_{ {\bf k}'} \frac{U^{12}_{{\bf k}{\bf k}'}U^{22}_{{\bf k}'{\bf k}}{\hat i}\cdot {\cal R}^{11}_{\bf k} (f^{2}_{{\bf k}'}-f^{1}_{{\bf k}})}{(\varepsilon^2_{{\bf k}'}-\varepsilon^{1}_{{\bf k}})+i\hbar\eta}\{\frac{e^{i\omega t}}{\hbar\omega-(\varepsilon^1_{{\bf k}}-\varepsilon^2_{{\bf k}'})+i\hbar\eta}+\frac{e^{-i\omega t}}{-\hbar\omega-(\varepsilon^1_{{\bf k}}-\varepsilon^2_{{\bf k}'})+i\hbar\eta}\}\nonumber\\
&-&\frac{eE\cos \theta_p }{2}\sum_{m'=1 {\bf k}'} \frac{U^{1m'}_{{\bf k}{\bf k}'}U^{2m''}_{{\bf k}'{\bf k}}{\hat i}\cdot {\cal R}^{m'2}_{\bf k} (f^{m'}_{{\bf k}'}-f^{1}_{{\bf k}})}{(\varepsilon^2_{{\bf k}'}-\varepsilon^{1}_{{\bf k}})+i\hbar\eta}\{\frac{e^{i\omega t}}{\hbar\omega-(\varepsilon^1_{{\bf k}}-\varepsilon^2_{{\bf k}'})+i\hbar\eta}+\frac{e^{-i\omega t}}{-\hbar\omega-(\varepsilon^1_{{\bf k}}-\varepsilon^2_{{\bf k}'})+i\hbar\eta}\}\nonumber\\
&-&\frac{eE\cos \theta_p }{2}\sum_{ {\bf k}'{\bf k}''} \frac{U^{12}_{{\bf k}''{\bf k}'}U^{22}_{{\bf k}'{\bf k}}(f^{2}_{{\bf k}'}-f^{1}_{{\bf k}''}){\hat i}\cdot \nabla_{\bf k} \delta({\bf k}''-{\bf k}) }{(\varepsilon^2_{{\bf k}'}-\varepsilon^{1}_{{\bf k}''})+i\hbar\eta}\{\frac{e^{i\omega t}}{i\hbar\omega-i(\varepsilon^1_{{\bf k}}-\varepsilon^2_{{\bf k}'})-\eta}+\frac{e^{-i\omega t}}{-i\hbar\omega-i(\varepsilon^1_{{\bf k}}-\varepsilon^2_{{\bf k}'})-\eta}\}\nonumber\\
&+&\frac{eE\cos \theta_p }{2}\sum_{{\bf k}' {\bf k}''} \frac{U^{12}_{{\bf k}{\bf k}''}U^{22}_{{\bf k}'{\bf k}}(f^{2}_{{\bf k}''}-f^{1}_{{\bf k}}){\hat i}\cdot \nabla_{{\bf k}'} \delta({\bf k}''-{\bf k}') }{(\varepsilon^2_{{\bf k}'}-\varepsilon^{1}_{{\bf k}})+i\hbar\eta}\{\frac{e^{i\omega t}}{i\hbar\omega-i(\varepsilon^1_{{\bf k}}-\varepsilon^2_{{\bf k}'})-\eta}+\frac{e^{-i\omega t}}{-i\hbar\omega-i(\varepsilon^1_{{\bf k}}-\varepsilon^2_{{\bf k}'})-\eta}\}\nonumber\\
%m=1 m''=3 n=2
&-&\frac{eE\cos \theta_p }{2}\sum_{m' {\bf k}'} \frac{U^{13}_{{\bf k}{\bf k}'}U^{m'2}_{{\bf k}'{\bf k}}{\hat i}\cdot {\cal R}^{3m'}_{\bf k'} (f^{2}_{{\bf k}}-f^{m'}_{{\bf k}'})}{(\varepsilon^2_{{\bf k}}-\varepsilon^{m'}_{{\bf k}'})+i\hbar\eta}\{\frac{e^{i\omega t}}{\hbar\omega-(\varepsilon^{3}_{{\bf k}'}-\varepsilon^2_{{\bf k}})+i\hbar\eta}+\frac{e^{-i\omega t}}{-\hbar\omega-(\varepsilon^{3}_{{\bf k}'}-\varepsilon^2_{{\bf k}})+i\hbar\eta}\}\nonumber\\
&+&\frac{eE\cos \theta_p }{2}\sum_{m' {\bf k}'} \frac{U^{13}_{{\bf k}{\bf k}'}U^{3m'}_{{\bf k}'{\bf k}}{\hat i}\cdot {\cal R}^{m'2}_{{\bf k}' }(f^{m'}_{{\bf k}}-f^{3}_{{\bf k}'})}{(\varepsilon^2_{{\bf k}}-\varepsilon^{3}_{{\bf k}'})+i\hbar\eta}\{\frac{e^{i\omega t}}{\hbar\omega-(\varepsilon^{3}_{{\bf k}'}-\varepsilon^2_{{\bf k}})+i\hbar\eta}+\frac{e^{-i\omega t}}{-\hbar\omega-(\varepsilon^{3}_{{\bf k}'}-\varepsilon^2_{{\bf k}})+i\hbar\eta}\}\nonumber\\
&+&\frac{eE\cos \theta_p }{2}\sum_{{\bf k}' {\bf k}''} \frac{U^{13}_{{\bf k}{\bf k}'}U^{32}_{{\bf k}''{\bf k}}(f^{2}_{{\bf k}}-f^{3}_{{\bf k}''}){\hat i}\cdot \nabla_{\bf k}' \delta({\bf k}''-{\bf k}') }{(\varepsilon^2_{{\bf k}}-\varepsilon^{3}_{{\bf k}''})+i\hbar\eta}\{\frac{e^{i\omega t}}{i\hbar\omega-i(\varepsilon^{3}_{{\bf k}'}-\varepsilon^2_{{\bf k}})-\eta}+\frac{e^{-i\omega t}}{-i\hbar\omega-i(\varepsilon^{3}_{{\bf k}'}-\varepsilon^2_{{\bf k}})-\eta}\}\nonumber\\
&-&\frac{eE\cos \theta_p }{2}\sum_{{\bf k}' {\bf k}''} \frac{U^{13}_{{\bf k}{\bf k}'}U^{32}_{{\bf k}'{\bf k}''}(f^{2}_{{\bf k}''}-f^{3}_{{\bf k}'}){\hat i}\cdot \nabla_{{\bf k}} \delta({\bf k}''-{\bf k}) }{(\varepsilon^2_{{\bf k}}-\varepsilon^{3}_{{\bf k}'})+i\hbar\eta}\{\frac{e^{i\omega t}}{i\hbar\omega-i(\varepsilon^{3}_{{\bf k}'}-\varepsilon^2_{{\bf k}})-\eta}+\frac{e^{-i\omega t}}{-i\hbar\omega-i(\varepsilon^{3}_{{\bf k}'}-\varepsilon^2_{{\bf k}})-\eta}\}\nonumber
\end{eqnarray}
\begin{eqnarray}
%m=1 n=3 m''=2
&+&\frac{eE\cos \theta_p }{2}\sum_{m'=1,2, {\bf k}'} \frac{U^{m'3}_{{\bf k}{\bf k}'}U^{32}_{{\bf k}'{\bf k}}{\hat i}\cdot {\cal R}^{1m'}_{\bf k} (f^{3}_{{\bf k}'}-f^{m'}_{{\bf k}})}{(\varepsilon^3_{{\bf k}'}-\varepsilon^{m'}_{{\bf k}})+i\hbar\eta}\{\frac{e^{i\omega t}}{\hbar\omega-(\varepsilon^1_{{\bf k}}-\varepsilon^3_{{\bf k}'})+i\hbar\eta}+\frac{e^{-i\omega t}}{-\hbar\omega-(\varepsilon^1_{{\bf k}}-\varepsilon^3_{{\bf k}'})+i\hbar\eta}\}\nonumber\\
&-&\frac{eE\cos \theta_p }{2}\sum_{m'=3,4, {\bf k}'} \frac{U^{1m'}_{{\bf k}{\bf k}'}U^{32}_{{\bf k}'{\bf k}}{\hat i}\cdot {\cal R}^{m'3}_{\bf k} (f^{m'}_{{\bf k}'}-f^{1}_{{\bf k}})}{(\varepsilon^3_{{\bf k}'}-\varepsilon^{1}_{{\bf k}})+i\hbar\eta}\{\frac{e^{i\omega t}}{\hbar\omega-(\varepsilon^1_{{\bf k}}-\varepsilon^3_{{\bf k}'})+i\hbar\eta}+\frac{e^{-i\omega t}}{-\hbar\omega-(\varepsilon^1_{{\bf k}}-\varepsilon^3_{{\bf k}'})+i\hbar\eta}\}\nonumber\\
&-&\frac{eE\cos \theta_p }{2}\sum_{ {\bf k}'{\bf k}''} \frac{U^{13}_{{\bf k}''{\bf k}'}U^{32}_{{\bf k}'{\bf k}}(f^{3}_{{\bf k}'}-f^{1}_{{\bf k}''}){\hat i}\cdot \nabla_{\bf k} \delta({\bf k}''-{\bf k}) }{(\varepsilon^3_{{\bf k}'}-\varepsilon^{1}_{{\bf k}''})+i\hbar\eta}\{\frac{e^{i\omega t}}{i\hbar\omega-i(\varepsilon^1_{{\bf k}}-\varepsilon^3_{{\bf k}'})-\eta}+\frac{e^{-i\omega t}}{-i\hbar\omega-i(\varepsilon^1_{{\bf k}}-\varepsilon^3_{{\bf k}'})-\eta}\}\nonumber\\
&+&\frac{eE\cos \theta_p }{2}\sum_{{\bf k}' {\bf k}''} \frac{U^{13}_{{\bf k}{\bf k}''}U^{32}_{{\bf k}'{\bf k}}(f^{3}_{{\bf k}''}-f^{1}_{{\bf k}}){\hat i}\cdot \nabla_{{\bf k}'} \delta({\bf k}''-{\bf k}') }{(\varepsilon^3_{{\bf k}'}-\varepsilon^{1}_{{\bf k}})+i\hbar\eta}\{\frac{e^{i\omega t}}{i\hbar\omega-i(\varepsilon^1_{{\bf k}}-\varepsilon^3_{{\bf k}'})-\eta}+\frac{e^{-i\omega t}}{-i\hbar\omega-i(\varepsilon^1_{{\bf k}}-\varepsilon^3_{{\bf k}'})-\eta}\}\nonumber\\
%m=1 m''=4 n=2
&-&\frac{eE\cos \theta_p }{2}\sum_{m' {\bf k}'} \frac{U^{14}_{{\bf k}{\bf k}'}U^{m'2}_{{\bf k}'{\bf k}}{\hat i}\cdot {\cal R}^{4m'}_{\bf k'} (f^{2}_{{\bf k}}-f^{m'}_{{\bf k}'})}{(\varepsilon^2_{{\bf k}}-\varepsilon^{m'}_{{\bf k}'})+i\hbar\eta}\{\frac{e^{i\omega t}}{\hbar\omega-(\varepsilon^{4}_{{\bf k}'}-\varepsilon^2_{{\bf k}})+i\hbar\eta}+\frac{e^{-i\omega t}}{-\hbar\omega-(\varepsilon^{m''}_{{\bf k}'}-\varepsilon^2_{{\bf k}})+i\hbar\eta}\}\nonumber\\
&+&\frac{eE\cos \theta_p }{2}\sum_{m' {\bf k}'} \frac{U^{14}_{{\bf k}{\bf k}'}U^{4m'}_{{\bf k}'{\bf k}}{\hat i}\cdot {\cal R}^{m'2}_{{\bf k}' }(f^{m'}_{{\bf k}}-f^{4}_{{\bf k}'})}{(\varepsilon^2_{{\bf k}}-\varepsilon^{4}_{{\bf k}'})+i\hbar\eta}\{\frac{e^{i\omega t}}{\hbar\omega-(\varepsilon^{4}_{{\bf k}'}-\varepsilon^2_{{\bf k}})+i\hbar\eta}+\frac{e^{-i\omega t}}{-\hbar\omega-(\varepsilon^{4}_{{\bf k}'}-\varepsilon^2_{{\bf k}})+i\hbar\eta}\}\nonumber\\
&+&\frac{eE\cos \theta_p }{2}\sum_{{\bf k}' {\bf k}''} \frac{U^{14}_{{\bf k}{\bf k}'}U^{42}_{{\bf k}''{\bf k}}(f^{2}_{{\bf k}}-f^{4}_{{\bf k}''}){\hat i}\cdot \nabla_{\bf k}' \delta({\bf k}''-{\bf k}') }{(\varepsilon^2_{{\bf k}}-\varepsilon^{4}_{{\bf k}''})+i\hbar\eta}\{\frac{e^{i\omega t}}{i\hbar\omega-i(\varepsilon^{4}_{{\bf k}'}-\varepsilon^2_{{\bf k}})-\eta}+\frac{e^{-i\omega t}}{-i\hbar\omega-i(\varepsilon^{4}_{{\bf k}'}-\varepsilon^2_{{\bf k}})-\eta}\}\nonumber\\
&-&\frac{eE\cos \theta_p }{2}\sum_{{\bf k}' {\bf k}''} \frac{U^{14}_{{\bf k}{\bf k}'}U^{42}_{{\bf k}'{\bf k}''}(f^{2}_{{\bf k}''}-f^{4}_{{\bf k}'}){\hat i}\cdot \nabla_{{\bf k}} \delta({\bf k}''-{\bf k}) }{(\varepsilon^2_{{\bf k}}-\varepsilon^{4}_{{\bf k}'})+i\hbar\eta}\{\frac{e^{i\omega t}}{i\hbar\omega-i(\varepsilon^{4}_{{\bf k}'}-\varepsilon^2_{{\bf k}})-\eta}+\frac{e^{-i\omega t}}{-i\hbar\omega-i(\varepsilon^{4}_{{\bf k}'}-\varepsilon^2_{{\bf k}})-\eta}\}\nonumber\\
%m=1 n=4 m''=2
&+&\frac{eE\cos \theta_p }{2}\sum_{m'=1,2, {\bf k}'} \frac{U^{m'4}_{{\bf k}{\bf k}'}U^{42}_{{\bf k}'{\bf k}}{\hat i}\cdot {\cal R}^{1m'}_{\bf k} (f^{4}_{{\bf k}'}-f^{m'}_{{\bf k}})}{(\varepsilon^4_{{\bf k}'}-\varepsilon^{m'}_{{\bf k}})+i\hbar\eta}\{\frac{e^{i\omega t}}{\hbar\omega-(\varepsilon^1_{{\bf k}}-\varepsilon^4_{{\bf k}'})+i\hbar\eta}+\frac{e^{-i\omega t}}{-\hbar\omega-(\varepsilon^1_{{\bf k}}-\varepsilon^4_{{\bf k}'})+i\hbar\eta}\}\nonumber\\
&-&\frac{eE\cos \theta_p }{2}\sum_{m'=4,3, {\bf k}'} \frac{U^{1m'}_{{\bf k}{\bf k}'}U^{42}_{{\bf k}'{\bf k}}{\hat i}\cdot {\cal R}^{m'4}_{\bf k} (f^{m'}_{{\bf k}'}-f^{1}_{{\bf k}})}{(\varepsilon^4_{{\bf k}'}-\varepsilon^{1}_{{\bf k}})+i\hbar\eta}\{\frac{e^{i\omega t}}{\hbar\omega-(\varepsilon^1_{{\bf k}}-\varepsilon^4_{{\bf k}'})+i\hbar\eta}+\frac{e^{-i\omega t}}{-\hbar\omega-(\varepsilon^1_{{\bf k}}-\varepsilon^4_{{\bf k}'})+i\hbar\eta}\}\nonumber\\
&-&\frac{eE\cos \theta_p }{2}\sum_{ {\bf k}'{\bf k}''} \frac{U^{14}_{{\bf k}''{\bf k}'}U^{42}_{{\bf k}'{\bf k}}(f^{4}_{{\bf k}'}-f^{1}_{{\bf k}''}){\hat i}\cdot \nabla_{\bf k} \delta({\bf k}''-{\bf k}) }{(\varepsilon^4_{{\bf k}'}-\varepsilon^{1}_{{\bf k}''})+i\hbar\eta}\{\frac{e^{i\omega t}}{i\hbar\omega-i(\varepsilon^1_{{\bf k}}-\varepsilon^4_{{\bf k}'})-\eta}+\frac{e^{-i\omega t}}{-i\hbar\omega-i(\varepsilon^1_{{\bf k}}-\varepsilon^4_{{\bf k}'})-\eta}\}\nonumber\\
&+&\frac{eE\cos \theta_p }{2}\sum_{{\bf k}' {\bf k}''} \frac{U^{14}_{{\bf k}{\bf k}''}U^{42}_{{\bf k}'{\bf k}}(f^{4}_{{\bf k}''}-f^{1}_{{\bf k}}){\hat i}\cdot \nabla_{{\bf k}'} \delta({\bf k}''-{\bf k}') }{(\varepsilon^4_{{\bf k}'}-\varepsilon^{1}_{{\bf k}})+i\hbar\eta}\{\frac{e^{i\omega t}}{i\hbar\omega-i(\varepsilon^1_{{\bf k}}-\varepsilon^4_{{\bf k}'})-\eta}+\frac{e^{-i\omega t}}{-i\hbar\omega-i(\varepsilon^1_{{\bf k}}-\varepsilon^4_{{\bf k}'})-\eta}\}\nonumber\\
&&+ {\text{other 32 terms by making use of }} (\frac{E\cos \theta_p }{2} \rightarrow \frac{E\sin \theta_p }{2i} ~{\text {and}}~ ({\hat i} \rightarrow {\hat j}) ~{\text {and}}~(...+... )\rightarrow (...-...))\nonumber
\end{eqnarray}
The density matrix for two valleys are written as  
\begin{eqnarray}
&&\frac{\partial \langle \rho_1 \rangle}{\partial t}+\frac{i}{\hbar}[{\mathcal H}_0+{\mathcal H}_E, \langle \rho_1 \rangle ]=-J_0(\langle \rho \rangle)-J_E(\langle \rho \rangle)\\
&&\frac{\partial \langle \rho_2 \rangle}{\partial t}+\frac{i}{\hbar}[{\mathcal H}_0+{\mathcal H}_E, \langle \rho_2 \rangle ]=-J_0(\langle \rho \rangle)-J_E(\langle \rho \rangle)\nonumber
\end{eqnarray}

Let us begin with considering one valley with all in- and out-scattering processes to that valley. 
The density matrix can be expanded in the powers of the electric field, and thus the quantum kinetic equation can be simplified as

\begin{eqnarray}
&&\frac{\partial f^n_d}{\partial t}+\frac{f^n_d}{\tau_1}+J^d_{E}(f^{(n-1)})=-\frac{i}{\hbar}\langle[{\mathcal H}_E,f^{(n-1)}_d]\rangle-J^{'}_0(\langle \rho \rangle)\\
&&\frac{\partial f^n_{od}}{\partial t}+\frac{i}{\hbar}\langle |[{\mathcal H}_0,f^n_{od}]|\rangle+\frac{f^n_{od}}{\tau_1}+J^{od}_{E}(f^{(n-1)})=-\frac{i}{\hbar}\langle |[{\mathcal H}_E,f^{(n-1)}_{od}]| \rangle
-J^{''}_{0}(\langle \rho \rangle) \nonumber
 \end{eqnarray}
where relaxation-times are given by Eq. (\ref{tau}) and $J^{'}_0=J_0(\langle \rho \rangle)-f^n_d/\tau_1$ (see Eq. (\ref{Jtau1})), and equally $J^{''}_0=J_0(\langle \rho \rangle)-f^n_{od}/\tau_1$, Eq. (\ref{Jtau2}).  Meanwhile we can also use the covariant derivative where
\begin{equation}
-\frac{i}{\hbar} \langle [H_{E}, f^{(n-1)}]\rangle=\frac{e{\bf E}}{\hbar}\frac{D f^{(n-1)}}{D{\bf k}}=\frac{e{\bf E}}{\hbar}\cdot\{ {\nabla f^{(n-1)}}-i\langle[{\cal R}_{\bf k}, f^{(n-1)}]\rangle\}
\end{equation}

Note that $f^{(0)}_d=f^{(0)}(\varepsilon_{k})$ and $f^{(0)}_{0d}=0$. In this stage we follow the perturbation recipe to calculate first order density matrices, $f^{(1)}_{od,{\bf k}}(t)$ and $f^{(1)}_{d,{\bf k}}(t)$. They are given by 
\begin{equation}
        f^{(1)}_{d,{\bf k}}(t)=\int_{-\infty}^t ~dt' e^{-\frac{t-t'}{\tau}} \{ \frac{e}{\hbar}{\bf E}(t')\cdot \nabla_{\bf k}f^0(\varepsilon^s(k))-J^d_{E}(f^0(\varepsilon(k))-J^{'}_0(t')\}
\end{equation}
and
\begin{equation}\label{f1od}
f^{(1)}_{od,{\bf k}}(t)=\int_{-\infty}^t ~dt' e^{-\frac{t-t'}{\tau}} e^{-{i\varepsilon^c(k)(t-t')/\hbar}}\{ \frac{e}{\hbar}{\bf E}(t')\cdot [i{\cal R}^{12}_{\bf k} [f_0(\varepsilon^c(k))-f_0(\varepsilon^{v}(k))]-J_{E}(f^{0}_{\bf k})-J^{''}_0(t') \}e^{i\varepsilon^{v}(k)(t-t')/\hbar}
\end{equation}
where $f_0(\varepsilon^s(k))$ is an equilibrium Fermi-Dirac distribution function. 
Note that we first solve the kinetic equation for first order in ${\bf E}$;
\begin{equation}
\frac{f^{(1)}_{d,{\bf k}_1}(t)}{\tau}=\frac{e}{\hbar}{\bf E}\cdot \nabla_{{\bf k}_1}f^{0}_{{\bf k}_1},\hspace{1.5cm} 
\frac{f^{(1)}_{d II,{\bf k}_2}(t)}{\tau}=\frac{e}{\hbar}{\bf E}\cdot \nabla_{{\bf k}_2}f^{0}_{{\bf k}_2}
\end{equation} 
where ${\bf k}_1$ and ${\bf k}_2$ refer to valley I and II, respectively. This solution of $f^{(1)}_{d,{\bf k}}(t)$ appears in $J^{''}_0(t')$ in Eq. (\ref{f1od}).
To do so, we start by looking at more dominate terms in $J^{''}_0(t')$ and $J_{E}(f^{0}_{\bf k})$ and use them selectively.

\begin{eqnarray}
J^{''}_{0,{\bf k}}(\langle \rho \rangle)=
&&\sum_{{{\bf k}'_1}\in I}[ \frac{f^{(1)}_{d,{\bf k}}}{-\varepsilon^{1}_{{\bf k}}+\varepsilon^{2}_{{\bf k}'_1}-i\eta} U^{12}_{{\bf k}{\bf k}'_1}U^{22}_{{\bf k}'_1{\bf k}}-U^{11}_{{\bf k}'_1{\bf k}}\frac{f^{(1)}_{d,{\bf k}}}{-\varepsilon^{1}_{{\bf k}'_1}+\varepsilon^{2}_{{\bf k}}-i\eta} U^{12}_{{\bf k}'_1{\bf k}}]\nonumber\\
%%%
&&+\sum_{{{\bf k}'_2}\in II}[ \frac{f^{(1)}_{d,{\bf k}}}{-\varepsilon^{1}_{{\bf k}}+\varepsilon^{3}_{{\bf k}'_2}-i\eta} U^{13}_{{\bf k}{\bf k}'_2}U^{32}_{{\bf k}'_2{\bf k}}+\frac{f^{(1)}_{d,{\bf k}}}{-\varepsilon^{1}_{{\bf k}}+\varepsilon^{4}_{{\bf k}'_2}-i\eta} U^{14}_{{\bf k}{\bf k}'_2}U^{42}_{{\bf k}'_2{\bf k}}]\nonumber\\
&&+\sum_{{{\bf k}'_2}\in II}[ \frac{f^{(1)}_{d,{\bf k}}}{-\varepsilon^{1}_{{\bf k}}+\varepsilon^{3}_{{\bf k}'_2}-i\eta} U^{13}_{{\bf k}{\bf k}'_2}U^{32}_{{\bf k}'_2{\bf k}}-U^{13}_{{\bf k}{\bf k}'_2}\frac{f^{(1)}_{d II,{\bf k}'_2}}{-\varepsilon^{1}_{{\bf k}}+\varepsilon^{3}_{{\bf k}'_2}-i\eta} U^{32}_{{\bf k}'_2{\bf k}}]\nonumber
\end{eqnarray}
and we also have
\begin{eqnarray} 
J^{'}_{0,{\bf k}}(\langle \rho \rangle)=
&&\sum_{{{\bf k}'_2}\in II}[ \frac{f^{(1)}_{d,{\bf k}}}{-\varepsilon^{1}_{{\bf k}}+\varepsilon^{3}_{{\bf k}'_2}-i\eta} U^{13}_{{\bf k}{\bf k}'_2}U^{32}_{{\bf k}'_2{\bf k}}-U^{13}_{{\bf k}{\bf k}'_2}\frac{f^{(1)}_{d II,{\bf k}'_2}}{-\varepsilon^{1}_{{\bf k}}+\varepsilon^{3}_{{\bf k}'_2}-i\eta} U^{32}_{{\bf k}'_2{\bf k}}]\nonumber
\end{eqnarray}

 In the case of $J^d_E$ we have
\begin{eqnarray}
\frac{\hbar}{i}J^d_{E, {\bf k}}(\langle \rho \rangle)=%%%m'=4
&+&\frac{eE\cos \theta_p }{2}\sum_{ {\bf k}'} \frac{U^{12}_{{\bf k}{\bf k}'}U^{21}_{{\bf k}'{\bf k}}{\hat i}\cdot {\cal R}^{11}_{\bf k} (f^{2}_{{\bf k}'}-f^{1}_{{\bf k}})}{(\varepsilon^2_{{\bf k}'}-\varepsilon^{1}_{{\bf k}})+i\hbar\eta}\{\frac{e^{i\omega t}}{\hbar\omega-(\varepsilon^1_{{\bf k}}-\varepsilon^2_{{\bf k}'})+i\hbar\eta}+\frac{e^{-i\omega t}}{-\hbar\omega-(\varepsilon^1_{{\bf k}}-\varepsilon^2_{{\bf k}'})+i\hbar\eta}\}\\
%%%%
&-&\frac{eE\cos \theta_p }{2}\sum_{ {\bf k}'\in II} \frac{U^{13}_{{\bf k}{\bf k}'}U^{41}_{{\bf k}'{\bf k}}{\hat i}\cdot {\cal R}^{34}_{\bf k'} (f^{1}_{{\bf k}}-f^{4}_{{\bf k}'})}{(\varepsilon^1_{{\bf k}}-\varepsilon^{4}_{{\bf k}'})+i\hbar\eta}\{\frac{e^{i\omega t}}{\hbar\omega-(\varepsilon^{3}_{{\bf k}'}-\varepsilon^1_{{\bf k}})+i\hbar\eta}+\frac{e^{-i\omega t}}{-\hbar\omega-(\varepsilon^{3}_{{\bf k}'}-\varepsilon^1_{{\bf k}})+i\hbar\eta}\}\nonumber\\
%%m'=3
&-&\frac{eE\cos \theta_p }{2}\sum_{ {\bf k}'\in II} \frac{U^{14}_{{\bf k}{\bf k}'}U^{31}_{{\bf k}'{\bf k}}{\hat i}\cdot {\cal R}^{43}_{\bf k'} (f^{1}_{{\bf k}}-f^{3}_{{\bf k}'})}{(\varepsilon^1_{{\bf k}}-\varepsilon^{3}_{{\bf k}'})+i\hbar\eta}\{\frac{e^{i\omega t}}{\hbar\omega-(\varepsilon^{4}_{{\bf k}'}-\varepsilon^1_{{\bf k}})+i\hbar\eta}+\frac{e^{-i\omega t}}{-\hbar\omega-(\varepsilon^{4}_{{\bf k}'}-\varepsilon^1_{{\bf k}})+i\hbar\eta}\}\nonumber\\
%%m'=4
&-&\frac{eE\cos \theta_p }{2}\sum_{{\bf k}'\in II} \frac{U^{14}_{{\bf k}{\bf k}'}U^{41}_{{\bf k}'{\bf k}}{\hat i}\cdot {\cal R}^{44}_{\bf k'} (f^{1}_{{\bf k}}-f^{4}_{{\bf k}'})}{(\varepsilon^1_{{\bf k}}-\varepsilon^{4}_{{\bf k}'})+i\hbar\eta}\{\frac{e^{i\omega t}}{\hbar\omega-(\varepsilon^{4}_{{\bf k}'}-\varepsilon^1_{{\bf k}})+i\hbar\eta}+\frac{e^{-i\omega t}}{-\hbar\omega-(\varepsilon^{4}_{{\bf k}'}-\varepsilon^1_{{\bf k}})+i\hbar\eta}\}\nonumber\\
&&+{\text{likewise terms by making use of }} (\frac{E\cos \theta_p }{2} \rightarrow \frac{E\sin \theta_p }{2i} ~{\text {and}}~ ({\hat i} \rightarrow {\hat j}) ~{\text {and}}~(...+... )\rightarrow (...-...))\nonumber
\end{eqnarray}  
and for $J^{od}_E$ we will have
\begin{eqnarray}
\frac{\hbar}{i}J^{od}_{E, {\bf k}}(\langle \rho \rangle)=
%%%m'=1
&&\frac{eE\cos \theta_p }{2}\sum_{{\bf k}'\in I} \frac{U^{11}_{{\bf k}{\bf k}'}U^{12}_{{\bf k}'{\bf k}}{\hat i}\cdot {\cal R}^{22}_{{\bf k}' }(f^{2}_{{\bf k}}-f^{1}_{{\bf k}'})}{(\varepsilon^2_{{\bf k}}-\varepsilon^{1}_{{\bf k}'})+i\hbar\eta}\{\frac{e^{i\omega t}}{\hbar\omega-(\varepsilon^{1}_{{\bf k}'}-\varepsilon^2_{{\bf k}})+i\hbar\eta}+\frac{e^{-i\omega t}}{-\hbar\omega-(\varepsilon^{1}_{{\bf k}'}-\varepsilon^2_{{\bf k}})+i\hbar\eta}\}\\
%%m'=1
&&\frac{eE\cos \theta_p }{2}\sum_{{\bf k}'\in I} \frac{U^{13}_{{\bf k}{\bf k}'}U^{33}_{{\bf k}'{\bf k}}{\hat i}\cdot {\cal R}^{11}_{{\bf k}' }(f^{3}_{{\bf k}'}-f^{1}_{{\bf k}})}{(\varepsilon^3_{{\bf k}'}-\varepsilon^{1}_{{\bf k}})+i\hbar\eta}\{\frac{e^{i\omega t}}{\hbar\omega-(\varepsilon^{1}_{{\bf k}}-\varepsilon^3_{{\bf k}'})+i\hbar\eta}+\frac{e^{-i\omega t}}{-\hbar\omega-(\varepsilon^{1}_{{\bf k}}-\varepsilon^3_{{\bf k}'})+i\hbar\eta}\}\\
%%m'=1
&+&\frac{eE\cos \theta_p }{2}\sum_{ {\bf k}'\in II} \frac{U^{14}_{{\bf k}{\bf k}'}U^{42}_{{\bf k}'{\bf k}}{\hat i}\cdot {\cal R}^{11}_{\bf k} (f^{4}_{{\bf k}'}-f^{1}_{{\bf k}})}{(\varepsilon^4_{{\bf k}'}-\varepsilon^{1}_{{\bf k}})+i\hbar\eta}\{\frac{e^{i\omega t}}{\hbar\omega-(\varepsilon^1_{{\bf k}}-\varepsilon^4_{{\bf k}'})+i\hbar\eta}+\frac{e^{-i\omega t}}{-\hbar\omega-(\varepsilon^1_{{\bf k}}-\varepsilon^4_{{\bf k}'})+i\hbar\eta}\}\nonumber\\
%%m'=4
&-&\frac{eE\cos \theta_p }{2}\sum_{ {\bf k}'\in II} \frac{U^{14}_{{\bf k}{\bf k}'}U^{42}_{{\bf k}'{\bf k}}{\hat i}\cdot {\cal R}^{44}_{\bf k} (f^{4}_{{\bf k}'}-f^{1}_{{\bf k}})}{(\varepsilon^4_{{\bf k}'}-\varepsilon^{1}_{{\bf k}})+i\hbar\eta}\{\frac{e^{i\omega t}}{\hbar\omega-(\varepsilon^1_{{\bf k}}-\varepsilon^4_{{\bf k}'})+i\hbar\eta}+\frac{e^{-i\omega t}}{-\hbar\omega-(\varepsilon^1_{{\bf k}}-\varepsilon^4_{{\bf k}'})+i\hbar\eta}\}\nonumber\\
&&+{\text{likewise terms by making use of }} (\frac{E\cos \theta_p }{2} \rightarrow \frac{E\sin \theta_p }{2i} ~{\text {and}}~ ({\hat i} \rightarrow {\hat j}) ~{\text {and}}~(...+... )\rightarrow (...-...))\nonumber
\end{eqnarray}

Therefore, all terms can be given by
\begin{eqnarray}
&&  f^{(1)}_{d,{\bf k}}(t)=\int_{-\infty}^t ~dt' e^{-\frac{t-t'}{\tau}} \{ \frac{e}{\hbar}{\bf E}(t')\cdot \nabla_{\bf k}f^0(\varepsilon(k))-J^d_{E}(f^0(\varepsilon(k))-J^{'}_0(t')\}\nonumber\\
&&f^{(1)}_{od,{\bf k}}(t)=\int_{-\infty}^t ~dt' e^{-\frac{t-t'}{\tau}} e^{-{i\varepsilon^1_{\bf k}(t-t')/\hbar}}\{ \frac{e}{\hbar}{\bf E}(t')\cdot [i{\cal R}^{12}_{\bf k} [f_0(\varepsilon^c(k))-f_0(\varepsilon^{v}(k))]-J^{od}_{E}(f^{0}_{\bf k})-J^{''}_0(t') \}e^{i\varepsilon^{2}_{\bf k}(t-t')/\hbar}
\nonumber\\
&&  f^{(1)}_{d II,{\bf k}}(t)=\int_{-\infty}^t ~dt' e^{-\frac{t-t'}{\tau}} \{ \frac{e}{\hbar}{\bf E}(t')\cdot \nabla_{\bf k}f^0(\varepsilon(k))-J^{d II}_{E}(f^0(\varepsilon(k))-J^{' II}_0(t')\}\nonumber\\
&&f^{(2)}_{d,{\bf k}}(t)=\int_{-\infty}^t ~dt' e^{-\frac{t-t'}{\tau}} [\frac{e}{\hbar}{\bf E}(t')\cdot \nabla_{\bf k}f^{(1)}_{d,{\bf k}}(t')-J^d_{E}(f^{(1)}_{{\bf k}}(t'))-J^{'}_0(t')]\nonumber\\
&&f^{(2)}_{od,{\bf k}}(t)=\int_{-\infty}^t ~dt' e^{-\frac{t-t'}{\tau}} e^{-{i\varepsilon^1_{\bf k}(t-t')/\hbar}}[\frac{e}{\hbar}{\bf E}(t')\cdot \{\nabla_{\bf k}f^{(1)}_{od,{\bf k}}(t')-if^{(1)}_{od,{\bf k}}(t')[{\cal R}^{11}_{\bf k}-{\cal R}^{22}_{\bf k}]\}
-J^{od}_{E}(f^{(1)}_{{\bf k}}(t')) -J^{''}_0(t')]e^{i\varepsilon^{2}_{\bf k}(t-t')/\hbar}\nonumber\\
&&f^{(2)}_{d II,{\bf k}}(t)=\int_{-\infty}^t ~dt' e^{-\frac{t-t'}{\tau}} [\frac{e}{\hbar}{\bf E}(t')\cdot \nabla_{\bf k}f^{(1)}_{d II,{\bf k}}(t')-J^{d II}_{E}(f^{(1)}_{{\bf k}}(t'))-J^{' II}_0(t')]\nonumber
\end{eqnarray}

By substituting all quantities in integrands, we would calculate the $f^{(2)}_{d,{\bf k}}(t)$ and $f^{(2)}_{od,{\bf k}}(t)$.
\begin{equation}
f^{(1)}_{d,{\bf k}}=f^{(1)}_{1d,{\bf k}}+f^{(1)}_{2d,{\bf k}}+f^{(1)}_{3d,{\bf k}}+O(\sin \theta_p)
\end{equation}
where
\begin{eqnarray}
&&f^{(1)}_{1d,{\bf k}}=\frac{eE\cos \theta_p }{2\hbar}\{\frac{e^{i\omega t}}{i\omega+\frac{1}{\tau}}+\frac{e^{-i\omega t}}{-i\omega+\frac{1}{\tau}}\}{\hat i}\cdot\nabla_{\bf k}f^0=\frac{eE\cos \theta_p }{2\hbar}\{F_0(\omega)+F_0(-\omega)\}{\hat i}\cdot\nabla_{\bf k}f^0\\
&&f^{(1)}_{2d,{\bf k}}=\frac{ieE\cos \theta_p }{2\hbar}\sum_{ {\bf k}'\in II} \frac{U^{13}_{{\bf k}{\bf k}'}U^{41}_{{\bf k}'{\bf k}}{\hat i}\cdot {\cal R}^{34}_{\bf k'} (f^{1}_{{\bf k}}-f^{4}_{{\bf k}'})}{(\varepsilon^1_{{\bf k}}-\varepsilon^{4}_{{\bf k}'})+i\hbar\eta}\{\frac{F_0(\omega)}{\hbar\omega-(\varepsilon^{3}_{{\bf k}'}-\varepsilon^1_{{\bf k}})+i\hbar\eta}+\frac{F_0(-\omega)}{-\hbar\omega-(\varepsilon^{3}_{{\bf k}'}-\varepsilon^1_{{\bf k}})+i\hbar\eta}\}\nonumber\\
%%m'=3
&&{\hspace {0.8 cm}}+\frac{ieE\cos \theta_p }{2\hbar}\sum_{ {\bf k}'\in II} \frac{U^{14}_{{\bf k}{\bf k}'}U^{31}_{{\bf k}'{\bf k}}{\hat i}\cdot {\cal R}^{43}_{\bf k'} (f^{1}_{{\bf k}}-f^{3}_{{\bf k}'})}{(\varepsilon^1_{{\bf k}}-\varepsilon^{3}_{{\bf k}'})+i\hbar\eta}\{\frac{F_0(\omega)}{\hbar\omega-(\varepsilon^{4}_{{\bf k}'}-\varepsilon^1_{{\bf k}})+i\hbar\eta}+\frac{F_0(-\omega)}{-\hbar\omega-(\varepsilon^{4}_{{\bf k}'}-\varepsilon^1_{{\bf k}})+i\hbar\eta}\}\nonumber\\
%%m'=4
&&{\hspace {0.8 cm}}+\frac{ieE\cos \theta_p }{2\hbar}\sum_{{\bf k}'\in II} \frac{U^{14}_{{\bf k}{\bf k}'}U^{41}_{{\bf k}'{\bf k}}{\hat i}\cdot {\cal R}^{44}_{\bf k'} (f^{1}_{{\bf k}}-f^{4}_{{\bf k}'})}{(\varepsilon^1_{{\bf k}}-\varepsilon^{4}_{{\bf k}'})+i\hbar\eta}\{\frac{F_0(\omega)}{\hbar\omega-(\varepsilon^{4}_{{\bf k}'}-\varepsilon^1_{{\bf k}})+i\hbar\eta}+\frac{F_0(-\omega)}{-\hbar\omega-(\varepsilon^{4}_{{\bf k}'}-\varepsilon^1_{{\bf k}})+i\hbar\eta}\}\nonumber\\
&&f^{(1)}_{3d,{\bf k}}=-\frac{e\tau E\cos \theta_p }{2\hbar}\sum_{{{\bf k}'_2}\in II}[\frac{ {\hat i}\cdot\nabla_{\bf k}f^0}{-\varepsilon^{1}_{{\bf k}}+\varepsilon^{3}_{{\bf k}'_2}-i\eta} U^{13}_{{\bf k}{\bf k}'_2}U^{32}_{{\bf k}'_2{\bf k}}-U^{13}_{{\bf k}{\bf k}'_2}\frac{{\hat i}\cdot\nabla_{{\bf k}'_2}f^0}{-\varepsilon^{1}_{{\bf k}}+\varepsilon^{3}_{{\bf k}'_2}-i\eta} U^{32}_{{\bf k}'_2{\bf k}}][F_0(\omega)+F_0(-\omega)]\nonumber
\end{eqnarray}
where we define $F_g(\omega)=e^{i\omega t}/(i\omega+ig(\varepsilon^{1}_{{\bf k}}-\varepsilon^2_{{\bf k}}))/\hbar+1/\tau)$.
\begin{equation}
f^{(1)}_{od,{\bf k}}=f^{(1)}_{1od,{\bf k}}+f^{(1)}_{2od,{\bf k}}+f^{(1)}_{3od,{\bf k}}+O(\sin \theta_p)
\end{equation}
\begin{eqnarray}
&&f^{(1)}_{1od,{\bf k}}=\frac{ieE\cos \theta_p }{2\hbar}(f^0(\varepsilon^1_{\bf k})-f^0(\varepsilon^2_{\bf k}))\{F_1(\omega)+F_1(-\omega)\}{\hat i}\cdot{\cal R}^{12}\\
&&f^{(1)}_{2od,{\bf k}}=-\frac{ieE\cos \theta_p }{2\hbar}\sum_{{\bf k}'\in I} \frac{U^{11}_{{\bf k}{\bf k}'}U^{12}_{{\bf k}'{\bf k}}{\hat i}\cdot {\cal R}^{22}_{{\bf k}' }(f^{2}_{{\bf k}}-f^{1}_{{\bf k}'})}{(\varepsilon^2_{{\bf k}}-\varepsilon^{1}_{{\bf k}'})+i\hbar\eta}\{\frac{F_1(\omega)}{\hbar\omega-(\varepsilon^{1}_{{\bf k}'}-\varepsilon^2_{{\bf k}})+i\hbar\eta}+\frac{F_1(-\omega)}{-\hbar\omega-(\varepsilon^{2}_{{\bf k}'}-\varepsilon^2_{{\bf k}})+i\hbar\eta}\}\nonumber\\
&&f^{(1)}_{22od,{\bf k}}=-\frac{ieE\cos \theta_p }{2\hbar}\sum_{{\bf k}'\in I} \frac{U^{33}_{{\bf k}{\bf k}'}U^{32}_{{\bf k}'{\bf k}}{\hat i}\cdot {\cal R}^{11}_{{\bf k}' }(f^{3}_{{\bf k}'}-f^{1}_{{\bf k}})}{(\varepsilon^3_{{\bf k}'}-\varepsilon^{1}_{{\bf k}})+i\hbar\eta}\{\frac{F_1(\omega)}{\hbar\omega-(\varepsilon^{1}_{{\bf k}}-\varepsilon^3_{{\bf k}'})+i\hbar\eta}+\frac{F_1(-\omega)}{-\hbar\omega-(\varepsilon^{1}_{{\bf k}}-\varepsilon^3_{{\bf k}'})+i\hbar\eta}\}\nonumber\\
&&f^{(1)}_{3od,{\bf k}}=-\frac{e\tau E\cos \theta_p }{2\hbar}\sum_{{{\bf k}'_1}\in I}[ \frac{{\hat i}\cdot\nabla_{\bf k}f^0}{-\varepsilon^{1}_{{\bf k}}+\varepsilon^{2}_{{\bf k}'_1}-i\eta} U^{12}_{{\bf k}{\bf k}'_1}U^{22}_{{\bf k}'_1{\bf k}}-U^{11}_{{\bf k}'_1{\bf k}}\frac{{\hat i}\cdot\nabla_{\bf k}f^0}{-\varepsilon^{1}_{{\bf k}'_1}+\varepsilon^{2}_{{\bf k}}-i\eta} U^{12}_{{\bf k}'_1{\bf k}}][F_1(\omega)+F_1(-\omega)]\nonumber\\
%%%
&&{\hspace {0.8 cm}}-\frac{e\tau E\cos \theta_p }{2\hbar}\sum_{{{\bf k}'_2}\in II}[ \frac{{\hat i}\cdot\nabla_{\bf k}f^0}{-\varepsilon^{1}_{{\bf k}}+\varepsilon^{3}_{{\bf k}'_2}-i\eta} U^{13}_{{\bf k}{\bf k}'_2}U^{32}_{{\bf k}'_2{\bf k}}+\frac{{\hat i}\cdot\nabla_{\bf k}f^0}{-\varepsilon^{1}_{{\bf k}}+\varepsilon^{4}_{{\bf k}'_2}-i\eta} U^{14}_{{\bf k}{\bf k}'_2}U^{42}_{{\bf k}'_2{\bf k}}][F_1(\omega)+F_1(-\omega)]\nonumber
\end{eqnarray}
Therefore, the time-independent contributions of second order density matrix would be given by
 \begin{equation}
f^{(2)}_{d,{\bf k}}=f^{(2)}_{1d,{\bf k}}+f^{(2)}_{2d,{\bf k}}+f^{(2)}_{3d,{\bf k}}-\int_{-\infty}^t ~dt' e^{-\frac{t-t'}{\tau}} [J^d_{E}(f^{(1)}_{{\bf k}}(t'))+J^{'}_0(t')]
\end{equation}
where
\begin{eqnarray}
&&f^{(2)}_{1d,{\bf k}}=\frac{e^2\tau E^2\cos^2 \theta_p }{2 \times 2\hbar^2}\{F^{'}_0(\omega)+F^{'}_0(-\omega)\}({\hat i}\cdot\nabla_{\bf k})({\hat i}\cdot\nabla_{\bf k}f^0)\\
&&{\hspace {0.8 cm}}+\frac{e^2\tau E^2\cos \theta_p \sin \theta_p}{2 \times 2i\hbar^2}\{F^{'}_0(\omega)-F^{'}_0(-\omega)\}({\hat i}\cdot\nabla_{\bf k})({\hat j}\cdot\nabla_{\bf k}f^0)\nonumber\\
&&{\hspace {0.8 cm}}+\frac{e^2\tau E^2\cos \theta_p \sin \theta_p}{2 \times 2i\hbar^2}\{-F^{'}_0(\omega)+F^{'}_0(-\omega)\}({\hat j}\cdot\nabla_{\bf k})({\hat i}\cdot\nabla_{\bf k}f^0)\nonumber\\
&&{\hspace {0.8 cm}}+\frac{e^2\tau E^2\sin \theta_p \sin \theta_p}{2i \times 2i\hbar^2}\{-F^{'}_0(\omega)-F^{'}_0(-\omega)\}({\hat j}\cdot\nabla_{\bf k})({\hat j}\cdot\nabla_{\bf k}f^0)\nonumber
\end{eqnarray}
where we define $F^{'}_g(\omega)=1/(i\omega+ig(\varepsilon^{1}_{{\bf k}}-\varepsilon^2_{{\bf k}}))/\hbar+1/\tau)$. Meanwhile, to calculate $f^{(2)}_{2d,{\bf k}}$ and $f^{(2)}_{od,{\bf k}}$ we have to operate $\nabla_{\bf k}$ on all quantities and also include terms contain $\sin \theta_p \cos \theta_p$. All details are carefully considered in numerics although I drop some terms in the note for simplicity.
\begin{eqnarray}\label{eq:f22d}
&&f^{(2)}_{2d,{\bf k}}=\frac{ie^2\tau E^2\cos^2 \theta_p }{2\times2\hbar^2 } \sum_{ {\bf k}'\in II} {\hat i}\cdot{\nabla}_{\bf k}(\frac{U^{13}_{{\bf k}{\bf k}'}U^{41}_{{\bf k}'{\bf k}}{\hat i}\cdot {\cal R}^{34}_{\bf k'} (f^{1}_{{\bf k}}-f^{4}_{{\bf k}'})}{(\varepsilon^1_{{\bf k}}-\varepsilon^{4}_{{\bf k}'})+i\hbar\eta})\{\frac{F^{'}_0(\omega)}{\hbar\omega-(\varepsilon^{3}_{{\bf k}'}-\varepsilon^1_{{\bf k}})+i\hbar\eta}+\frac{F^{'}_0(-\omega)}{-\hbar\omega-(\varepsilon^{3}_{{\bf k}'}-\varepsilon^1_{{\bf k}})+i\hbar\eta}\}\\
&&{\hspace {0.7 cm}}-\frac{ie^2\tau E^2\cos^2 \theta_p }{2\times2\hbar^2 } \sum_{ {\bf k}'\in II} \frac{U^{13}_{{\bf k}{\bf k}'}U^{41}_{{\bf k}'{\bf k}} {\hat i}\cdot {\cal R}^{34}_{\bf k'} (f^{1}_{{\bf k}}-f^{4}_{{\bf k}'})({\hat i}\cdot{\nabla}_{\bf k}\varepsilon^1_{{\bf k}})}{(\varepsilon^1_{{\bf k}}-\varepsilon^{4}_{{\bf k}'})+i\hbar\eta}\{\frac{F^{'}_0(\omega)}{[\hbar\omega-(\varepsilon^{3}_{{\bf k}'}-\varepsilon^1_{{\bf k}})+i\hbar\eta]^2}+\frac{F^{'}_0(-\omega)}{[-\hbar\omega-(\varepsilon^{3}_{{\bf k}'}-\varepsilon^1_{{\bf k}})+i\hbar\eta]^2}\}\nonumber\\
%%%
&&{\hspace {0.7 cm}}+\frac{ie^2\tau E^2\sin^2 \theta_p }{2i\times2i\hbar^2 } \sum_{ {\bf k}'\in II}{\hat j}\cdot{\nabla}_{\bf k}( \frac{U^{13}_{{\bf k}{\bf k}'}U^{41}_{{\bf k}'{\bf k}} {\hat j}\cdot {\cal R}^{34}_{\bf k'} (f^{1}_{{\bf k}}-f^{4}_{{\bf k}'})}{(\varepsilon^1_{{\bf k}}-\varepsilon^{4}_{{\bf k}'})+i\hbar\eta})\{\frac{-F^{'}_0(\omega)}{\hbar\omega-(\varepsilon^{3}_{{\bf k}'}-\varepsilon^1_{{\bf k}})+i\hbar\eta}+\frac{-F^{'}_0(-\omega)}{-\hbar\omega-(\varepsilon^{3}_{{\bf k}'}-\varepsilon^1_{{\bf k}})+i\hbar\eta}\}\nonumber\\
&&{\hspace {0.7 cm}}-\frac{ie^2\tau E^2\sin^2 \theta_p }{2i\times2i\hbar^2 } \sum_{ {\bf k}'\in II} \frac{U^{13}_{{\bf k}{\bf k}'}U^{41}_{{\bf k}'{\bf k}} {\hat j}\cdot {\cal R}^{34}_{\bf k'} (f^{1}_{{\bf k}}-f^{4}_{{\bf k}'})({\hat j}\cdot{\nabla}_{\bf k}\varepsilon^1_{{\bf k}})}{(\varepsilon^1_{{\bf k}}-\varepsilon^{4}_{{\bf k}'})+i\hbar\eta}\{\frac{-F^{'}_0(\omega)}{[\hbar\omega-(\varepsilon^{3}_{{\bf k}'}-\varepsilon^1_{{\bf k}})+i\hbar\eta]^2}+\frac{-F^{'}_0(-\omega)}{[-\hbar\omega-(\varepsilon^{3}_{{\bf k}'}-\varepsilon^1_{{\bf k}})+i\hbar\eta]^2}\}\nonumber\\
%%m'=3
&&{\hspace {0.7 cm}}+\frac{ie^2\tau E^2\cos^2 \theta_p }{2\times2\hbar^2 }\sum_{ {\bf k}'\in II} {\hat i}\cdot{\nabla}_{\bf k}(\frac{U^{14}_{{\bf k}{\bf k}'}U^{31}_{{\bf k}'{\bf k}} {\hat i}\cdot {\cal R}^{43}_{\bf k'} (f^{1}_{{\bf k}}-f^{3}_{{\bf k}'})}{(\varepsilon^1_{{\bf k}}-\varepsilon^{3}_{{\bf k}'})+i\hbar\eta})\{\frac{F^{'}_0(\omega)}{\hbar\omega-(\varepsilon^{4}_{{\bf k}'}-\varepsilon^1_{{\bf k}})+i\hbar\eta}+\frac{F^{'}_0(-\omega)}{-\hbar\omega-(\varepsilon^{4}_{{\bf k}'}-\varepsilon^1_{{\bf k}})+i\hbar\eta}\}\nonumber\\
&&{\hspace {0.7 cm}}-\frac{ie^2\tau E^2\cos^2 \theta_p }{2\times2\hbar^2 }\sum_{ {\bf k}'\in II} \frac{U^{14}_{{\bf k}{\bf k}'}U^{31}_{{\bf k}'{\bf k}} {\hat i}\cdot {\cal R}^{43}_{\bf k'} (f^{1}_{{\bf k}}-f^{3}_{{\bf k}'}){\hat i}\cdot{\nabla}_{\bf k}\varepsilon^1_{{\bf k}}}{(\varepsilon^1_{{\bf k}}-\varepsilon^{3}_{{\bf k}'})+i\hbar\eta}\{\frac{F^{'}_0(\omega)}{[\hbar\omega-(\varepsilon^{4}_{{\bf k}'}-\varepsilon^1_{{\bf k}})+i\hbar\eta]^2}+\frac{F^{'}_0(-\omega)}{[-\hbar\omega-(\varepsilon^{4}_{{\bf k}'}-\varepsilon^1_{{\bf k}})+i\hbar\eta]^2}\}\nonumber\\
%%%%%
&&{\hspace {0.7 cm}}+\frac{ie^2\tau E^2\sin^2 \theta_p }{2i\times2i\hbar^2 }\sum_{ {\bf k}'\in II} {\hat j}\cdot{\nabla}_{\bf k}(\frac{U^{14}_{{\bf k}{\bf k}'}U^{31}_{{\bf k}'{\bf k}} {\hat j}\cdot {\cal R}^{43}_{\bf k'} (f^{1}_{{\bf k}}-f^{3}_{{\bf k}'})}{(\varepsilon^1_{{\bf k}}-\varepsilon^{3}_{{\bf k}'})+i\hbar\eta})\{\frac{-F^{'}_0(\omega)}{\hbar\omega-(\varepsilon^{4}_{{\bf k}'}-\varepsilon^1_{{\bf k}})+i\hbar\eta}+\frac{-F^{'}_0(-\omega)}{-\hbar\omega-(\varepsilon^{4}_{{\bf k}'}-\varepsilon^1_{{\bf k}})+i\hbar\eta}\}\nonumber\\
&&{\hspace {0.7 cm}}-\frac{ie^2\tau E^2\sin^2 \theta_p }{2i\times2i\hbar^2 }\sum_{ {\bf k}'\in II} \frac{U^{14}_{{\bf k}{\bf k}'}U^{31}_{{\bf k}'{\bf k}} {\hat j}\cdot {\cal R}^{43}_{\bf k'} (f^{1}_{{\bf k}}-f^{3}_{{\bf k}'}){\hat j}\cdot{\nabla}_{\bf k}\varepsilon^1_{{\bf k}}}{(\varepsilon^1_{{\bf k}}-\varepsilon^{3}_{{\bf k}'})+i\hbar\eta}\{\frac{-F^{'}_0(\omega)}{[\hbar\omega-(\varepsilon^{4}_{{\bf k}'}-\varepsilon^1_{{\bf k}})+i\hbar\eta]^2}+\frac{-F^{'}_0(-\omega)}{[-\hbar\omega-(\varepsilon^{4}_{{\bf k}'}-\varepsilon^1_{{\bf k}})+i\hbar\eta]^2}\}\nonumber\\
%%m'=4
&&{\hspace {0.7 cm}}+\frac{ie^2\tau E^2\cos^2 \theta_p }{2\times2\hbar^2 }\sum_{{\bf k}'\in II} {\hat i}\cdot{\nabla}_{\bf k}(\frac{U^{14}_{{\bf k}{\bf k}'}U^{41}_{{\bf k}'{\bf k}}{\hat i}\cdot {\cal R}^{44}_{\bf k'} (f^{1}_{{\bf k}}-f^{4}_{{\bf k}'})}{(\varepsilon^1_{{\bf k}}-\varepsilon^{4}_{{\bf k}'})+i\hbar\eta})\{\frac{F^{'}_0(\omega)}{\hbar\omega-(\varepsilon^{4}_{{\bf k}'}-\varepsilon^1_{{\bf k}})+i\hbar\eta}+\frac{F^{'}_0(-\omega)}{-\hbar\omega-(\varepsilon^{4}_{{\bf k}'}-\varepsilon^1_{{\bf k}})+i\hbar\eta}\}\nonumber\\
&&{\hspace {0.7 cm}}-\frac{ie^2\tau E^2\cos^2 \theta_p }{2\times2\hbar^2 }\sum_{{\bf k}'\in II} \frac{U^{14}_{{\bf k}{\bf k}'}U^{41}_{{\bf k}'{\bf k}}{\hat i}\cdot {\cal R}^{44}_{\bf k'} (f^{1}_{{\bf k}}-f^{4}_{{\bf k}'}){\hat i}\cdot{\nabla}_{\bf k}\varepsilon^1_{{\bf k}}}{(\varepsilon^1_{{\bf k}}-\varepsilon^{4}_{{\bf k}'})+i\hbar\eta})\{\frac{F^{'}_0(\omega)}{[\hbar\omega-(\varepsilon^{4}_{{\bf k}'}-\varepsilon^1_{{\bf k}})+i\hbar\eta]^2}+\frac{F^{'}_0(-\omega)}{[-\hbar\omega-(\varepsilon^{4}_{{\bf k}'}-\varepsilon^1_{{\bf k}})+i\hbar\eta]^2}\}\nonumber\\
&&{\hspace {0.7 cm}}+\frac{ie^2\tau E^2\sin^2 \theta_p }{2i\times2i\hbar^2 }\sum_{{\bf k}'\in II} {\hat j}\cdot{\nabla}_{\bf k}(\frac{U^{14}_{{\bf k}{\bf k}'}U^{41}_{{\bf k}'{\bf k}}{\hat j}\cdot {\cal R}^{44}_{\bf k'} (f^{1}_{{\bf k}}-f^{4}_{{\bf k}'})}{(\varepsilon^1_{{\bf k}}-\varepsilon^{4}_{{\bf k}'})+i\hbar\eta})\{\frac{-F^{'}_0(\omega)}{\hbar\omega-(\varepsilon^{4}_{{\bf k}'}-\varepsilon^1_{{\bf k}})+i\hbar\eta}+\frac{-F^{'}_0(-\omega)}{-\hbar\omega-(\varepsilon^{4}_{{\bf k}'}-\varepsilon^1_{{\bf k}})+i\hbar\eta}\}\nonumber\\
&&{\hspace {0.7 cm}}-\frac{ie^2\tau E^2\sin^2 \theta_p }{2i\times2i\hbar^2 }\sum_{{\bf k}'\in II} \frac{U^{14}_{{\bf k}{\bf k}'}U^{41}_{{\bf k}'{\bf k}}{\hat j}\cdot {\cal R}^{44}_{\bf k'} (f^{1}_{{\bf k}}-f^{4}_{{\bf k}'}){\hat j}\cdot{\nabla}_{\bf k}\varepsilon^1_{{\bf k}}}{(\varepsilon^1_{{\bf k}}-\varepsilon^{4}_{{\bf k}'})+i\hbar\eta})\{\frac{-F^{'}_0(\omega)}{[\hbar\omega-(\varepsilon^{4}_{{\bf k}'}-\varepsilon^1_{{\bf k}})+i\hbar\eta]^2}+\frac{-F^{'}_0(-\omega)}{[-\hbar\omega-(\varepsilon^{4}_{{\bf k}'}-\varepsilon^1_{{\bf k}})+i\hbar\eta]^2}\}\nonumber\\
&&+{\text {other terms}}\nonumber
\end{eqnarray}
Now we would like to mention that there are some terms which are more dominate term when $\nabla_{{\bf k}}$ operates on the expressions. Two terms play important role $\nabla_{{\bf k}} f^{i}_{{\bf k}}$ and $\nabla_{{\bf k}} F^{'}_1(-\omega)$ and expression in denominators, like
\begin{eqnarray}
&&f^{(2)}_{2d,{\bf k}}\sim-\frac{ie^2\tau E^2\cos^2 \theta_p }{2\times2\hbar^2 } \sum_{ {\bf k}'\in II} \frac{U^{13}_{{\bf k}{\bf k}'}U^{41}_{{\bf k}'{\bf k}} {\hat i}\cdot {\cal R}^{34}_{\bf k'} (f^{1}_{{\bf k}}-f^{4}_{{\bf k}'})({\hat i}\cdot{\nabla}_{\bf k}\varepsilon^1_{{\bf k}})}{(\varepsilon^1_{{\bf k}}-\varepsilon^{4}_{{\bf k}'})+i\hbar\eta}\{\frac{F^{'}_0(\omega)}{[\hbar\omega-(\varepsilon^{3}_{{\bf k}'}-\varepsilon^1_{{\bf k}})+i\hbar\eta]^2}+\frac{F^{'}_0(-\omega)}{[-\hbar\omega-(\varepsilon^{3}_{{\bf k}'}-\varepsilon^1_{{\bf k}})+i\hbar\eta]^2}\}\\
%%%
&&{\hspace {0.7 cm}}-\frac{ie^2\tau E^2\sin^2 \theta_p }{2i\times2i\hbar^2 } \sum_{ {\bf k}'\in II} \frac{U^{13}_{{\bf k}{\bf k}'}U^{41}_{{\bf k}'{\bf k}} {\hat j}\cdot {\cal R}^{34}_{\bf k'} (f^{1}_{{\bf k}}-f^{4}_{{\bf k}'})({\hat j}\cdot{\nabla}_{\bf k}\varepsilon^1_{{\bf k}})}{(\varepsilon^1_{{\bf k}}-\varepsilon^{4}_{{\bf k}'})+i\hbar\eta}\{\frac{-F^{'}_0(\omega)}{[\hbar\omega-(\varepsilon^{3}_{{\bf k}'}-\varepsilon^1_{{\bf k}})+i\hbar\eta]^2}+\frac{-F^{'}_0(-\omega)}{[-\hbar\omega-(\varepsilon^{3}_{{\bf k}'}-\varepsilon^1_{{\bf k}})+i\hbar\eta]^2}\}\nonumber\\
%%m'=3
&&{\hspace {0.7 cm}}-\frac{ie^2\tau E^2\cos^2 \theta_p }{2\times2\hbar^2 }\sum_{ {\bf k}'\in II} \frac{U^{14}_{{\bf k}{\bf k}'}U^{31}_{{\bf k}'{\bf k}} {\hat i}\cdot {\cal R}^{43}_{\bf k'} (f^{1}_{{\bf k}}-f^{3}_{{\bf k}'}){\hat i}\cdot{\nabla}_{\bf k}\varepsilon^1_{{\bf k}}}{(\varepsilon^1_{{\bf k}}-\varepsilon^{3}_{{\bf k}'})+i\hbar\eta}\{\frac{F^{'}_0(\omega)}{[\hbar\omega-(\varepsilon^{4}_{{\bf k}'}-\varepsilon^1_{{\bf k}})+i\hbar\eta]^2}+\frac{F^{'}_0(-\omega)}{[-\hbar\omega-(\varepsilon^{4}_{{\bf k}'}-\varepsilon^1_{{\bf k}})+i\hbar\eta]^2}\}\nonumber\\
%%%%%
&&{\hspace {0.7 cm}}-\frac{ie^2\tau E^2\sin^2 \theta_p }{2i\times2i\hbar^2 }\sum_{ {\bf k}'\in II} \frac{U^{14}_{{\bf k}{\bf k}'}U^{31}_{{\bf k}'{\bf k}} {\hat j}\cdot {\cal R}^{43}_{\bf k'} (f^{1}_{{\bf k}}-f^{3}_{{\bf k}'}){\hat j}\cdot{\nabla}_{\bf k}\varepsilon^1_{{\bf k}}}{(\varepsilon^1_{{\bf k}}-\varepsilon^{3}_{{\bf k}'})+i\hbar\eta}\{\frac{-F^{'}_0(\omega)}{[\hbar\omega-(\varepsilon^{4}_{{\bf k}'}-\varepsilon^1_{{\bf k}})+i\hbar\eta]^2}+\frac{-F^{'}_0(-\omega)}{[-\hbar\omega-(\varepsilon^{4}_{{\bf k}'}-\varepsilon^1_{{\bf k}})+i\hbar\eta]^2}\}\nonumber\\
%%m'=4
&&{\hspace {0.7 cm}}-\frac{ie^2\tau E^2\cos^2 \theta_p }{2\times2\hbar^2 }\sum_{{\bf k}'\in II} \frac{U^{14}_{{\bf k}{\bf k}'}U^{41}_{{\bf k}'{\bf k}}{\hat i}\cdot {\cal R}^{44}_{\bf k'} (f^{1}_{{\bf k}}-f^{4}_{{\bf k}'}){\hat i}\cdot{\nabla}_{\bf k}\varepsilon^1_{{\bf k}}}{(\varepsilon^1_{{\bf k}}-\varepsilon^{4}_{{\bf k}'})+i\hbar\eta}\{\frac{F^{'}_0(\omega)}{[\hbar\omega-(\varepsilon^{4}_{{\bf k}'}-\varepsilon^1_{{\bf k}})+i\hbar\eta]^2}+\frac{F^{'}_0(-\omega)}{[-\hbar\omega-(\varepsilon^{4}_{{\bf k}'}-\varepsilon^1_{{\bf k}})+i\hbar\eta]^2}\}\nonumber\\
&&{\hspace {0.7 cm}}-\frac{ie^2\tau E^2\sin^2 \theta_p }{2i\times2i\hbar^2 }\sum_{{\bf k}'\in II} \frac{U^{14}_{{\bf k}{\bf k}'}U^{41}_{{\bf k}'{\bf k}}{\hat j}\cdot {\cal R}^{44}_{\bf k'} (f^{1}_{{\bf k}}-f^{4}_{{\bf k}'}){\hat j}\cdot{\nabla}_{\bf k}\varepsilon^1_{{\bf k}}}{(\varepsilon^1_{{\bf k}}-\varepsilon^{4}_{{\bf k}'})+i\hbar\eta})\{\frac{-F^{'}_0(\omega)}{[\hbar\omega-(\varepsilon^{4}_{{\bf k}'}-\varepsilon^1_{{\bf k}})+i\hbar\eta]^2}+\frac{-F^{'}_0(-\omega)}{[-\hbar\omega-(\varepsilon^{4}_{{\bf k}'}-\varepsilon^1_{{\bf k}})+i\hbar\eta]^2}\}\nonumber\\
&&+{\text {other terms}}\nonumber
\end{eqnarray}

and
\begin{eqnarray}
&&f^{(2)}_{3d,{\bf k}}=\frac{ie^2\tau^2 E^2\cos^2 \theta_p }{2\times2\hbar^2 }{\hat i}\cdot\nabla_{\bf k}\varepsilon^{1}_{{\bf k}}\sum_{{{\bf k}'_2}\in II}[\frac{ {\hat i}\cdot\nabla_{\bf k}f^0}{[\varepsilon^{1}_{{\bf k}}-\varepsilon^{3}_{{\bf k}'_2}+i\eta]^2} U^{13}_{{\bf k}{\bf k}'_2}U^{32}_{{\bf k}'_2{\bf k}}-U^{13}_{{\bf k}{\bf k}'_2}\frac{{\hat i}\cdot\nabla_{{\bf k}'_2}f^0}{[\varepsilon^{1}_{{\bf k}}-\varepsilon^{3}_{{\bf k}'_2}+i\eta]^2} U^{32}_{{\bf k}'_2{\bf k}}][F^{'}_0(\omega)+F^{'}_0(-\omega)]\\
&&{\hspace {0.7 cm}}+\frac{ie^2\tau^2 E^2\sin^2 \theta_p }{2i\times2\hbar^2i }{\hat j}\cdot\nabla_{\bf k}\varepsilon^{1}_{{\bf k}}\sum_{{{\bf k}'_2}\in II}[\frac{ {\hat j}\cdot\nabla_{\bf k}f^0}{[\varepsilon^{1}_{{\bf k}}-\varepsilon^{3}_{{\bf k}'_2}+i\eta]^2} U^{13}_{{\bf k}{\bf k}'_2}U^{32}_{{\bf k}'_2{\bf k}}-U^{13}_{{\bf k}{\bf k}'_2}\frac{{\hat j}\cdot\nabla_{{\bf k}'_2}f^0}{[\varepsilon^{1}_{{\bf k}}-\varepsilon^{3}_{{\bf k}'_2}+i\eta]^2} U^{32}_{{\bf k}'_2{\bf k}}][-F^{'}_0(\omega)-F^{'}_0(-\omega)]\nonumber\\
&&+{\text {other terms}}\nonumber
\end{eqnarray}
On the other hand,
\begin{equation}
 f^{(2)}_{od,{\bf k}}(t)=\int_{-\infty}^t ~dt' e^{-\frac{t-t'}{\tau}} e^{-{i\varepsilon^1_{\bf k}(t-t')/\hbar}}[\frac{e}{\hbar}{\bf E}(t')\cdot \{\nabla_{\bf k}f^{(1)}_{od,{\bf k}}(t')-if^{(1)}_{od,{\bf k}}(t')[{\cal R}^{11}_{\bf k}-{\cal R}^{22}_{\bf k}]\}
-J^{od}_{E}(f^{(1)}_{{\bf k}}(t')) -J^{''}_0(t')]e^{i\varepsilon^{2}_{\bf k}(t-t')/\hbar}   
\end{equation}
\begin{equation}
f^{(2)}_{od,{\bf k}}=f^{(2)}_{1od,{\bf k}}+f^{(2)}_{2od,{\bf k}}+f^{(2)}_{3od,{\bf k}}-\int_{-\infty}^t ~dt' e^{-\frac{t-t'}{\tau}} e^{-{i\varepsilon^1_{\bf k}(t-t')/\hbar}}[
J^{od}_{E}(f^{(1)}_{{\bf k}}(t')) +J^{''}_0(t')]e^{i\varepsilon^{2}_{\bf k}(t-t')/\hbar}
\end{equation}
We define $1/\gamma_{\bf k}=i(\varepsilon^1_{\bf k}-\varepsilon^2_{\bf k})/\hbar+1/\tau$ and calculate the off-diagonal parts. Therefore, we then find
\begin{eqnarray}\label{eq:f21od}
&&-i~ f^{(2)}_{1od,{\bf k}}=\frac{e^2E^2\cos^2 \theta_p }{2\times2\hbar^2}{\hat i}\cdot{\nabla_{\bf k}}(f^0(\varepsilon^1_{\bf k})-f^0(\varepsilon^2_{\bf k}))\gamma_{\bf k}\{F^{'}_1(\omega)+F^{'}_1(-\omega)\}{\hat i}\cdot{\cal R}^{12}_{\bf k}\\
&&{\hspace {0.9 cm}}-\frac{ie^2E^2\cos^2 \theta_p }{2\times2\hbar^2}(f^0(\varepsilon^1_{\bf k})-f^0(\varepsilon^2_{\bf k}))\gamma_{\bf k}{\hat i}\cdot{\nabla_{\bf k}}(\varepsilon^1_{\bf k}-\varepsilon^2_{\bf k})\{F^{' 2}_1(\omega)+F^{' 2}_1(-\omega)\}{\hat i}\cdot{\cal R}^{12}_{\bf k}\nonumber\\
&&{\hspace {0.9 cm}}+\frac{e^2E^2\cos^2 \theta_p }{2\times2\hbar^2}(f^0(\varepsilon^1_{\bf k})-f^0(\varepsilon^2_{\bf k}))\gamma_{\bf k}\{F^{'}_1(\omega)+F^{'}_1(-\omega)\}{\hat i}\cdot{\nabla_{\bf k}}{\hat i}\cdot{\cal R}^{12}_{\bf k}\nonumber\\
&&{\hspace {0.9 cm}}-\frac{ie^2E^2\cos^2 \theta_p }{2\times2\hbar^2}(f^0(\varepsilon^1_{\bf k})-f^0(\varepsilon^2_{\bf k}))\gamma_{\bf k}\{F^{'}_1(\omega)+F^{'}_1(-\omega)\}{\hat i}\cdot({\cal R}^{11}_{\bf k}-{\cal R}^{22}_{\bf k}){\hat i}\cdot{\cal R}^{12}_{\bf k}\nonumber\\
%%%%
&&{\hspace {0.9 cm}}+\frac{e^2E^2\sin^2 \theta_p }{2i\times2i\hbar^2}{\hat j}\cdot{\nabla_{\bf k}}(f^0(\varepsilon^1_{\bf k})-f^0(\varepsilon^2_{\bf k}))\gamma_{\bf k}\{-F^{'}_1(\omega)-F^{'}_1(-\omega)\}{\hat j}\cdot{\cal R}^{12}_{\bf k}\nonumber\\
&&{\hspace {0.9 cm}}+\frac{e^2E^2\sin^2 \theta_p }{2i\times2i\hbar^2}(f^0(\varepsilon^1_{\bf k})-f^0(\varepsilon^2_{\bf k}))\gamma_{\bf k}\{-F^{'}_1(\omega)-F^{'}_1(-\omega)\}{\hat j}\cdot{\nabla_{\bf k}}{\hat j}\cdot{\cal R}^{12}_{\bf k}\nonumber\\
&&{\hspace {0.9 cm}}-\frac{ie^2E^2\sin^2 \theta_p }{2i\times2i\hbar^2}(f^0(\varepsilon^1_{\bf k})-f^0(\varepsilon^2_{\bf k}))\gamma_{\bf k}{\hat j}\cdot{\nabla_{\bf k}}(\varepsilon^1_{\bf k}-\varepsilon^2_{\bf k})\{-F^{' 2}_1(\omega)-F^{' 2}_1(-\omega)\}{\hat j}\cdot{\cal R}^{12}_{\bf k}\nonumber\\
&&{\hspace {0.9 cm}}-\frac{ie^2E^2\sin^2 \theta_p }{2i\times2i\hbar^2}(f^0(\varepsilon^1_{\bf k})-f^0(\varepsilon^2_{\bf k}))\gamma_{\bf k}\{-F^{'}_1(\omega)-F^{'}_1(-\omega)\}{\hat j}\cdot({\cal R}^{11}_{\bf k}-{\cal R}^{22}_{\bf k}){\hat j}\cdot{\cal R}^{12}_{\bf k}\nonumber\\
%%%%
&&{\hspace {0.9 cm}}+\frac{ie^2E^2\cos \theta_p \sin \theta_p}{2i\times2\hbar^2}(f^0(\varepsilon^1_{\bf k})-f^0(\varepsilon^2_{\bf k}))\gamma_{\bf k}\{F^{' 2}_1(\omega)-F^{' 2}_1(-\omega)\}[{\hat i}\cdot{\nabla_{\bf k}}(\varepsilon^1_{\bf k}-\varepsilon^2_{\bf k}){\hat j}\cdot{\cal R}^{12}_{\bf k}]\nonumber\\
&&{\hspace {0.9 cm}}+\frac{ie^2E^2\cos \theta_p \sin \theta_p}{2i\times2\hbar^2}(f^0(\varepsilon^1_{\bf k})-f^0(\varepsilon^2_{\bf k}))\gamma_{\bf k}\{-F^{' 2}_1(\omega)+F^{' 2}_1(-\omega)\}[{\hat j}\cdot{\nabla_{\bf k}}(\varepsilon^1_{\bf k}-\varepsilon^2_{\bf k}){\hat i}\cdot{\cal R}^{12}_{\bf k}]\nonumber
\end{eqnarray}

\begin{eqnarray}
&&-i~f^{(2)}_{2od,{\bf k}}=-\frac{e^2E^2\cos^2 \theta_p }{2\times2\hbar^2} \sum_{{{\bf k}'_1}\in I}\frac{U^{11}_{{\bf k}{\bf k}'}U^{12}_{{\bf k}'{\bf k}}{\hat i}\cdot {\cal R}^{22}_{{\bf k}' }(f^{2}_{{\bf k}}-f^{1}_{{\bf k}'})}{(\varepsilon^2_{{\bf k}}-\varepsilon^{1}_{{\bf k}'})+i\hbar\eta}\gamma_{\bf k}{\hat i}\cdot {\nabla_{{\bf k}}}(-\varepsilon^2_{{\bf k}})\{\frac{F^{'}_1(\omega)}{[\hbar\omega-(\varepsilon^{1}_{{\bf k}'}-\varepsilon^2_{{\bf k}})+i\hbar\eta]^2}+\frac{F^{'}_1(-\omega)}{[-\hbar\omega-(\varepsilon^{2}_{{\bf k}'}-\varepsilon^2_{{\bf k}})+i\hbar\eta]^2}\}\nonumber\\
%%%%%
&&{\hspace {0.7 cm}}-\frac{ie^2E^2\cos^2 \theta_p }{2\times2\hbar^2} \sum_{{{\bf k}'_1}\in I}\frac{U^{11}_{{\bf k}{\bf k}'}U^{12}_{{\bf k}'{\bf k}}{\hat i}\cdot {\cal R}^{22}_{{\bf k}' }(f^{2}_{{\bf k}}-f^{1}_{{\bf k}'})}{(\varepsilon^2_{{\bf k}}-\varepsilon^{1}_{{\bf k}'})+i\hbar\eta}\gamma_{\bf k}{\hat i}\cdot {\nabla_{{\bf k}}}(\varepsilon^1_{{\bf k}}-\varepsilon^2_{{\bf k}})\{\frac{F^{'2}_1(\omega)}{\hbar\omega-(\varepsilon^{1}_{{\bf k}'}-\varepsilon^2_{{\bf k}})+i\hbar\eta]}+\frac{F^{'2}_1(-\omega)}{-\hbar\omega-(\varepsilon^{2}_{{\bf k}'}-\varepsilon^2_{{\bf k}})+i\hbar\eta}\}\nonumber\\
%%%
&&{\hspace {0.7 cm}}-\frac{ie^2E^2\cos^2 \theta_p }{2\times2\hbar^2} \sum_{{{\bf k}'_1}\in I}\frac{U^{33}_{{\bf k}{\bf k}'}U^{32}_{{\bf k}'{\bf k}}{\hat i}\cdot {\cal R}^{11}_{{\bf k}' }(f^{3}_{{\bf k}'}-f^{1}_{{\bf k}})}{(\varepsilon^3_{{\bf k}'}-\varepsilon^{1}_{{\bf k}})+i\hbar\eta}\gamma_{\bf k}{\hat i}\cdot {\nabla_{{\bf k}}}(\varepsilon^1_{{\bf k}}-\varepsilon^2_{{\bf k}})\{\frac{F^{'2}_1(\omega)}{\hbar\omega-(\varepsilon^{1}_{{\bf k}}-\varepsilon^3_{{\bf k}'})+i\hbar\eta}+\frac{F^{'2}_1(-\omega)}{-\hbar\omega-(\varepsilon^{1}_{{\bf k}}-\varepsilon^3_{{\bf k}'})+i\hbar\eta}\}\nonumber\\
%%%
&&{\hspace {0.7 cm}}+\frac{ie^2E^2\cos^2 \theta_p }{2\times2\hbar^2} \sum_{{{\bf k}'_1}\in I} \frac{U^{11}_{{\bf k}{\bf k}'}U^{12}_{{\bf k}'{\bf k}}{\hat i}\cdot {\cal R}^{22}_{{\bf k}' }(f^{2}_{{\bf k}}-f^{1}_{{\bf k}'})}{(\varepsilon^2_{{\bf k}}-\varepsilon^{1}_{{\bf k}'})+i\hbar\eta}\gamma_{\bf k}\{\frac{F^{'}_1(\omega)}{\hbar\omega-(\varepsilon^{1}_{{\bf k}'}-\varepsilon^2_{{\bf k}})+i\hbar\eta}+\frac{F^{'}_1(-\omega)}{-\hbar\omega-(\varepsilon^{2}_{{\bf k}'}-\varepsilon^2_{{\bf k}})+i\hbar\eta}\}{\hat i}\cdot({\cal R}^{11}_{\bf k}-{\cal R}^{22}_{\bf k})\nonumber\\
&&{\hspace {0.7 cm}}+\frac{ie^2E^2\cos^2 \theta_p }{2\times2\hbar^2} \sum_{{{\bf k}'_1}\in I} \frac{U^{33}_{{\bf k}{\bf k}'}U^{32}_{{\bf k}'{\bf k}}{\hat i}\cdot {\cal R}^{11}_{{\bf k}' }(f^{3}_{{\bf k}'}-f^{1}_{{\bf k}})}{(\varepsilon^3_{{\bf k}'}-\varepsilon^{1}_{{\bf k}})+i\hbar\eta}\gamma_{\bf k}\{\frac{F^{;}_1(\omega)}{\hbar\omega-(\varepsilon^{1}_{{\bf k}}-\varepsilon^3_{{\bf k}'})+i\hbar\eta}+\frac{F^{'}_1(-\omega)}{-\hbar\omega-(\varepsilon^{1}_{{\bf k}}-\varepsilon^3_{{\bf k}'})+i\hbar\eta}\}{\hat i}\cdot({\cal R}^{11}_{\bf k}-{\cal R}^{22}_{\bf k})\nonumber
\end{eqnarray}
\begin{eqnarray}
&&-i~f^{(2)}_{3od,{\bf k}}=-\frac{ie^2\tau E^2\cos^2 \theta_p }{2\times2\hbar^2}(-{\hat i}\cdot {\nabla_{\bf k}}\varepsilon^{1}_{{\bf k}})\sum_{{{\bf k}'_2}\in II}[ \frac{{\hat i}\cdot\nabla_{\bf k}f^0}{[\varepsilon^{1}_{{\bf k}}-\varepsilon^{3}_{{\bf k}'_2}+i\eta]^2} U^{13}_{{\bf k}{\bf k}'_2}U^{32}_{{\bf k}'_2{\bf k}}+\frac{{\hat i}\cdot\nabla_{\bf k}f^0}{[\varepsilon^{1}_{{\bf k}}-\varepsilon^{4}_{{\bf k}'_2}+i\eta]^2} U^{14}_{{\bf k}{\bf k}'_2}U^{42}_{{\bf k}'_2{\bf k}}]\gamma_{\bf k}[F^{'}_1(\omega)+F^{1}_1(-\omega)]\\
&&{\hspace {0.9 cm}}-\frac{e^2\tau E^2\cos^2 \theta_p }{2\times2\hbar^2}\sum_{{{\bf k}'_2}\in II}[ \frac{{\hat i}\cdot\nabla_{\bf k}f^0}{\varepsilon^{1}_{{\bf k}}-\varepsilon^{3}_{{\bf k}'_2}+i\eta} U^{13}_{{\bf k}{\bf k}'_2}U^{32}_{{\bf k}'_2{\bf k}}+\frac{{\hat i}\cdot\nabla_{\bf k}f^0}{\varepsilon^{1}_{{\bf k}}-\varepsilon^{4}_{{\bf k}'_2}+i\eta} U^{14}_{{\bf k}{\bf k}'_2}U^{42}_{{\bf k}'_2{\bf k}}]\gamma_{\bf k}[F^{'}_1(\omega)+F^{1}_1(-\omega)]{\hat i}\cdot({\cal R}^{11}_{\bf k}-{\cal R}^{22}_{\bf k})\nonumber\\
%%%%%
&&{\hspace {0.9 cm}}-\frac{ie^2\tau E^2\sin^2 \theta_p }{2i\times2i\hbar^2}(-{\hat j}\cdot {\nabla_{\bf k}}\varepsilon^{1}_{{\bf k}})\sum_{{{\bf k}'_2}\in II}[ \frac{{\hat j}\cdot\nabla_{\bf k}f^0}{[\varepsilon^{1}_{{\bf k}}-\varepsilon^{3}_{{\bf k}'_2}+i\eta]^2} U^{13}_{{\bf k}{\bf k}'_2}U^{32}_{{\bf k}'_2{\bf k}}+\frac{{\hat j}\cdot\nabla_{\bf k}f^0}{[\varepsilon^{1}_{{\bf k}}-\varepsilon^{4}_{{\bf k}'_2}+i\eta]^2} U^{14}_{{\bf k}{\bf k}'_2}U^{42}_{{\bf k}'_2{\bf k}}]\gamma_{\bf k}[-F^{'}_1(\omega)-F^{1}_1(-\omega)]\nonumber\\
&&{\hspace {0.9 cm}}-\frac{e^2\tau E^2\sin^2 \theta_p }{2i\times2i\hbar^2}\sum_{{{\bf k}'_2}\in II}[ \frac{{\hat j}\cdot\nabla_{\bf k}f^0}{\varepsilon^{1}_{{\bf k}}-\varepsilon^{3}_{{\bf k}'_2}+i\eta} U^{13}_{{\bf k}{\bf k}'_2}U^{32}_{{\bf k}'_2{\bf k}}+\frac{{\hat j}\cdot\nabla_{\bf k}f^0}{\varepsilon^{1}_{{\bf k}}-\varepsilon^{4}_{{\bf k}'_2}+i\eta} U^{14}_{{\bf k}{\bf k}'_2}U^{42}_{{\bf k}'_2{\bf k}}]\gamma_{\bf k}[-F^{'}_1(\omega)-F^{1}_1(-\omega)]{\hat j}\cdot({\cal R}^{11}_{\bf k}-{\cal R}^{22}_{\bf k})\nonumber\\
&&+{\text {other terms}}\nonumber
\end{eqnarray}

\subsection{Numerical Scaling}
In order to carry our the numerical calculation of ${\bf J}$ we perform some standard scaling, ${\bar k}=k a_0$,  $\hbar \omega={\varepsilon_0}{\bar \omega}$, ${\varepsilon_0}=\hbar^2/2m_ea^2_0$ and $k_{\rm F}=\sqrt{\pi n}$ where $n$ is the average electron density $\approx 10^{12}$ cm$^{-2}$. Thus:

\begin{eqnarray}
{\bf j} \sim 
&& (\frac{e^3 a_0}{\hbar \varepsilon_0})  {\bar {\bf j}}  EE= (\frac{e^3 a_0} {\hbar \varepsilon_0}){\bar {\bf j}}  \frac{2I_0}  {\epsilon_0 c} \nonumber\\
&&=(\frac{8\pi e a_0\alpha}{\varepsilon_0 }) {\bar {\bf j}} I_0
\end{eqnarray}
when ${\bf J}$ is in units of pA/m and $I_0$ is intensity of light in units of W/m$^2$. We make use of $\epsilon_0 c=\frac{e^2}{4\pi\alpha \hbar}$ with $\alpha=1/137$. Therefore, ${J}^2/I_0$ would be in units of pA m/W. Notice that the vector current is defined by vector ${\cal R}_{kk'}$ therefore by assuming the electric field along the $x$-direction, we do have ${J}^{(2)}_{x}$ and ${J}^{(2)}_{y}$ components. Note that we also use $\tau^{-1}=n_i U^2_0 $, then the results would depend on $\lambda$. Actually the intervalley scattering process is proportional to $\lambda^2$. Figure \ref{fig7} shows the second-order DC photocurrent at the the K-point as a function of energy for circularly polarized light and various strain values. Furthermore, Fig. \ref{fig6} shows the optical response for various trigonal warping.

\begin{figure}
	\includegraphics[width=8.6cm]{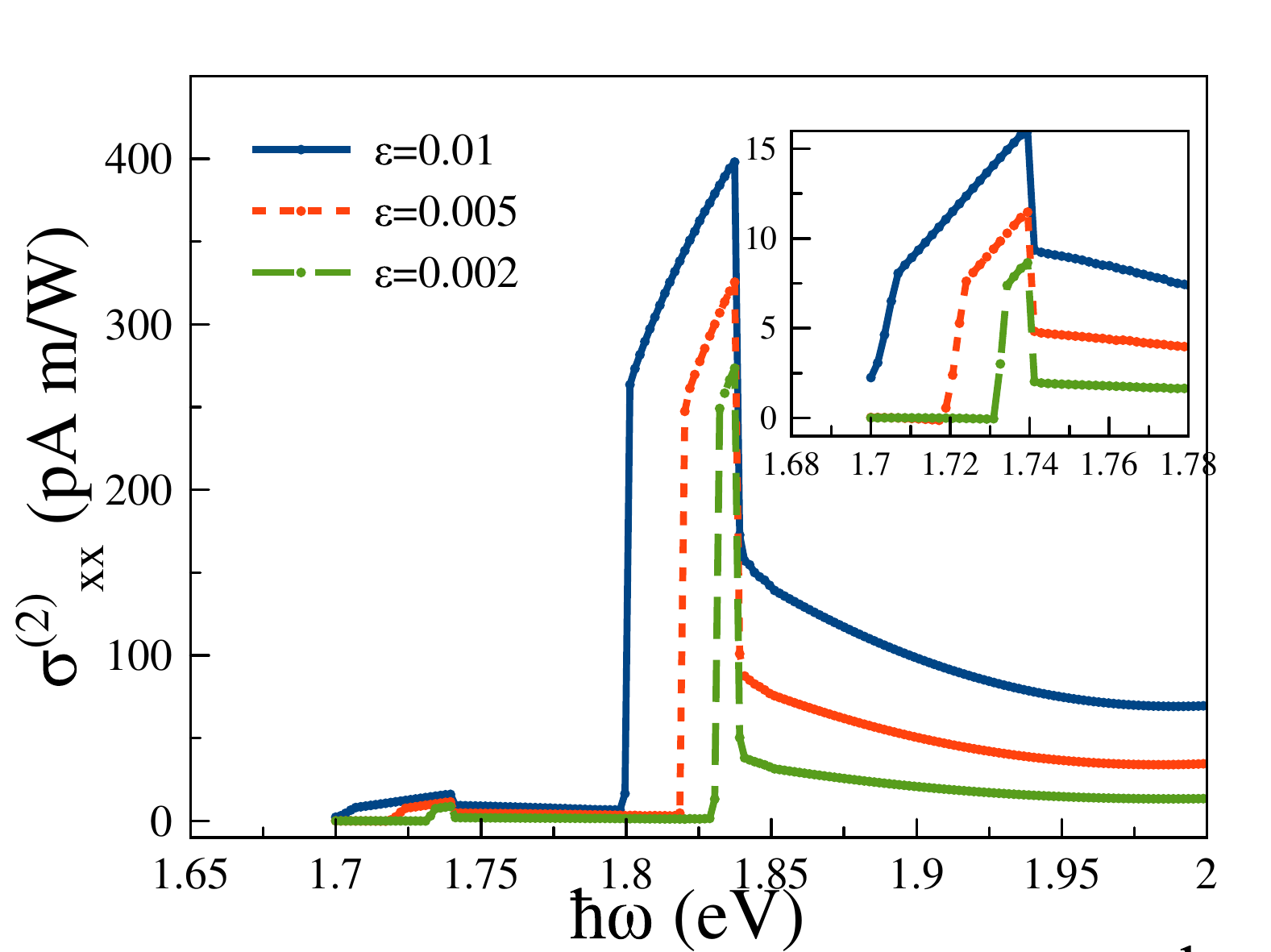}
	\caption{(Color online) $\sigma^{(2)}_{xx}=(J_x(\varepsilon)-J_x(\varepsilon=0))/I_0$ (in units of pA m/W) as a function of $\hbar \omega$ (in units of eV) for a circularly polarized light, and $m=0.02$ eV for different value of strain $\varepsilon$ for given $U_{ele}=-0.02$ eV, $n=5\times 10^{12}$ cm$^{-2}$ and $\tau=2$ ps.  }\label{fig7}
\end{figure}
\subsection{Photovoltaic current}

\begin{figure}
	\includegraphics[width=8.6cm]{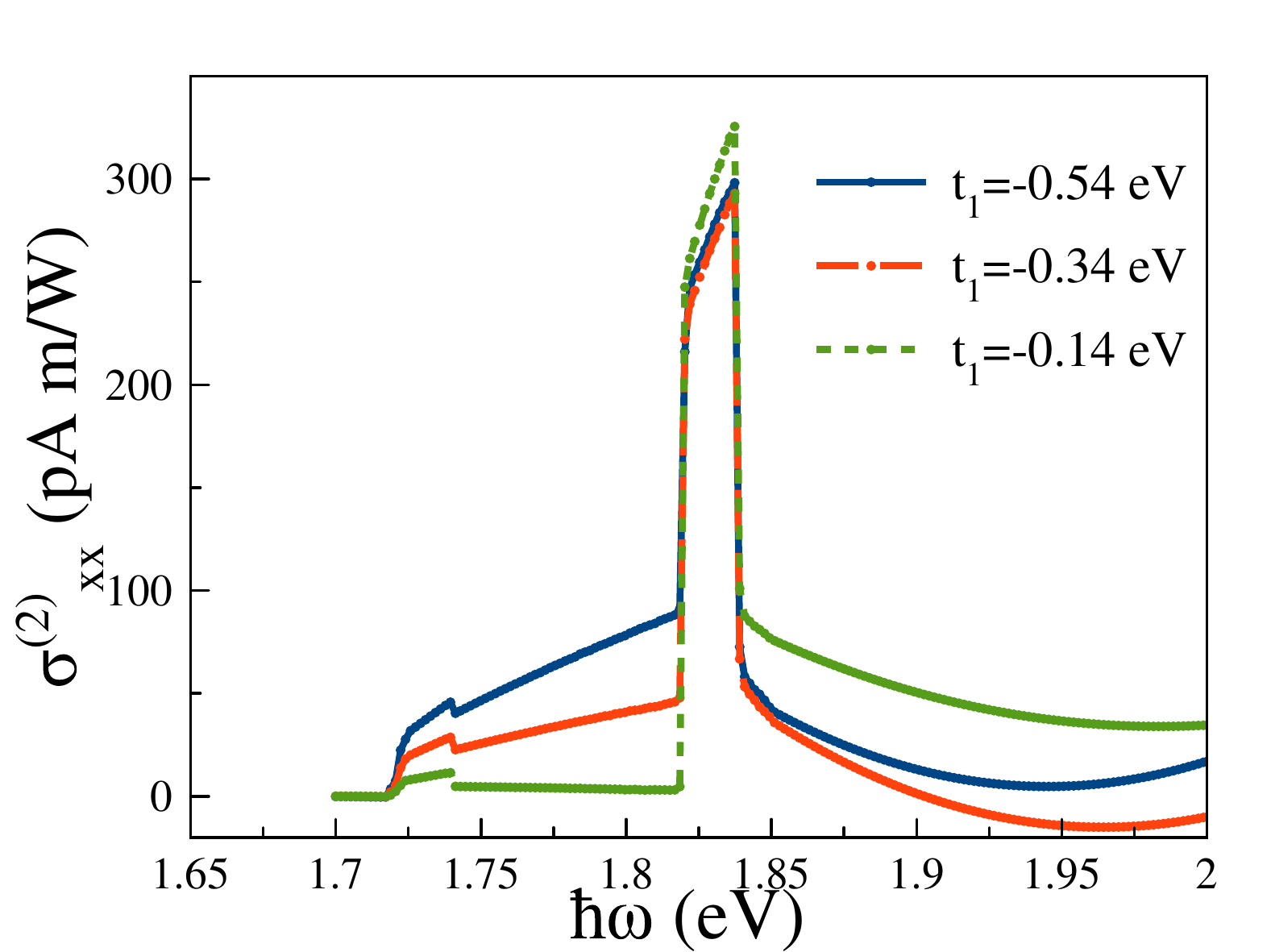}
	\caption{(Color online) The effect of the trigonal warping, through strain, on $\sigma^{(2)}_{xx}$ (in units of pA m/W).  The relaxation time is $\tau=2$ ps, $m=0.02$ eV, $U_{ele}=-0.02$ eV, the strain $\varepsilon=0.005$ and $n=5\times 10^{12}$ cm$^{-2}$. By considering $t_1=t_2=0$, the $J/I_0$ from off diagonal part is totally small.  }\label{fig6}
\end{figure}

\newpage

\section{Bethe-Salpeter equation and Exciton band structure}\label{A:excitons}
Since in an absorption photon a pari of the electron and hole are created thus a new state is emerged owing to Coulomb interaction. This coupled electron-phonon state can be viewed as a non-charged exiton excitation. This new state leads to additional absorption peaks shifted from the fundamental absorption edge by the coupling energies.

It is common believed that the exciton physics of 2D-TMD is controlled by mirror, three-fold rotational and time-reversal symmetries. Finite-momentum excitons are optically inactive. Low-energy exciton states appear both near the Brillouin-zone center and near the Brillouin zone center excitons. Here we are just interested in the BZ corner excitons close to (K, K'). The interacting Hamiltonian is
\begin{equation}
{\cal H}={\cal H}_0+\frac{1}{2}\sum_{{\bf R},{\bf R}'} V(|{\bf R}-{\bf R}'|)a^{\dagger}_{{\bf R}\nu} a^{\dagger}_{{\bf R}'\nu'} a_{{\bf R}'\nu'} a_{{\bf R}\nu}
\end{equation}
where ${\cal H}_0$ is the noninteracting part and the second part is the Coulomb interaction. $a^{\dagger}_{{\bf R}\nu}$ is the electron creation operator for orbital $\nu$ at Mo site ${\bf R}$. The Coulomb interaction is given by the Keldysh form
to account for the finite width of the TMD layer. 

Exciton states with center of mass momentum ${\bf Q}$ can be expanded in terms of electron-hole states as 
\begin{equation}
|\Psi \rangle=\sum_{v, c, {\bf k}} C_{\bf Q}(v, c, {\bf k}) b^{\dagger}_{{\bf k}+{\bf Q},c}b_{{\bf k},v}|\psi^s\rangle
\end{equation}
where $b^{\dagger}_{n{\bf k}}$ is quasiparticle operator for band $n$ at momentum ${\bf k}$. In order to calculate the exciton, we do need to solve Bethe-Salpeter equation where its solution determines the exciton eigenvalue and wave functions. The Hamiltonian matrix is 
\begin{equation}
\langle  {\bf k} |{\cal H} | {\bf k}' \rangle=\delta_{vv'}\delta_{cc'}\delta_{{\bf k}{\bf k}'}(\varepsilon_{{\bf k}+{\bf Q},c}-\varepsilon_{{\bf k},v})-\frac{1}{N}V_{{\bf k}-{\bf k}'} ({\cal U}^{\dagger}_{{\bf k}+{\bf Q}}{\cal U}_{{\bf k}'+{\bf Q}})_{cc'}({\cal U}^{\dagger}_{{\bf k}'}{\cal U}_{{\bf k}})_{v'v}+\frac{1}{N}V_{{\bf Q}} ({\cal U}^{\dagger}_{{\bf k}+{\bf Q}}{\cal U}_{{\bf k}})_{cv}({\cal U}^{\dagger}_{{\bf k}'}{\cal U}_{{\bf k}'+{\bf Q}})_{v'c'}
\end{equation}
where ${\cal U}_{{\bf k}}$ is the unitary matrix which diagonalizes the quasiparticle Hamiltonian ${\cal H}_0$. Notice that in the case ${\bf Q}=0$, the last term of RHS, (exchange term)
vanishes due to the orthogonality property $({\cal U}^{\dagger}_{{\bf k}}{\cal U}_{{\bf k}})_{cv}=0$. To calculate the RHS of above equation, we have
\begin{equation}
|\psi^{s}\rangle=
\begin{pmatrix}
    \psi_1   \\
    \psi_2 
\end{pmatrix}
=\frac{1}{\sqrt{D{^{s}}^2+|h_{12}|^2}}
\begin{pmatrix}
  -h_{12}\\
  D^{s}
\end{pmatrix}
\end{equation}
with $D^{s}=h_{z}-s \sqrt{h_z^2+|h_{12}|^2}$. Therefore, 
\begin{equation}
\langle  {\bf k} |{\cal H} | {\bf k}' \rangle=(\varepsilon^c({\bf k})-\varepsilon^{\nu}({\bf k}))\delta_{{\bf k},{\bf k}'}-\frac{1}{4{\cal A}}V_{{\bf k}-{\bf k}'} 
\{[h^{*}_{12}(k)h_{12}(k') +D^{*}(k)D(k')]_{c} [h^{*}_{12}(k')h_{12}(k) +D^{*}(k')D(k)]_{v}\}
\end{equation}
where ${\cal A}$ is area of 2D system. The Bethe-Salpeter equation now reads 

\begin{equation}\label{A:e-h}
\sum_{{\bf k}'}[(\varepsilon^c({\bf k})-\varepsilon^{\nu}({\bf k}))\delta_{{\bf k}{\bf k}'}-\frac{1}{4{\cal A}}V_{{\bf k}-{\bf k}'} 
\{[h^{*}_{12}(k)h_{12}(k') +D^{*}(k)D(k')]_{s=1} [h^{*}_{12}(k')h_{12}(k) +D^{*}(k')D(k)]_{s=-1}\}]|\Psi_{{\bf k}'}\rangle=
E_{{\bf k}} |\Psi_{{\bf k}}\rangle
\end{equation}
These eigenvalues and eigenvectors might be used to solve Eq. (\ref{A:e-h}) to calculate the impact of the exciton effects on the DC optical current. 
Figure \ref{ren} shows the renormalized band gap exciton in terms of the dielectric constant, calculated from Eq. (\ref{A:e-h}).
\begin{figure}
	\includegraphics[width=8cm]{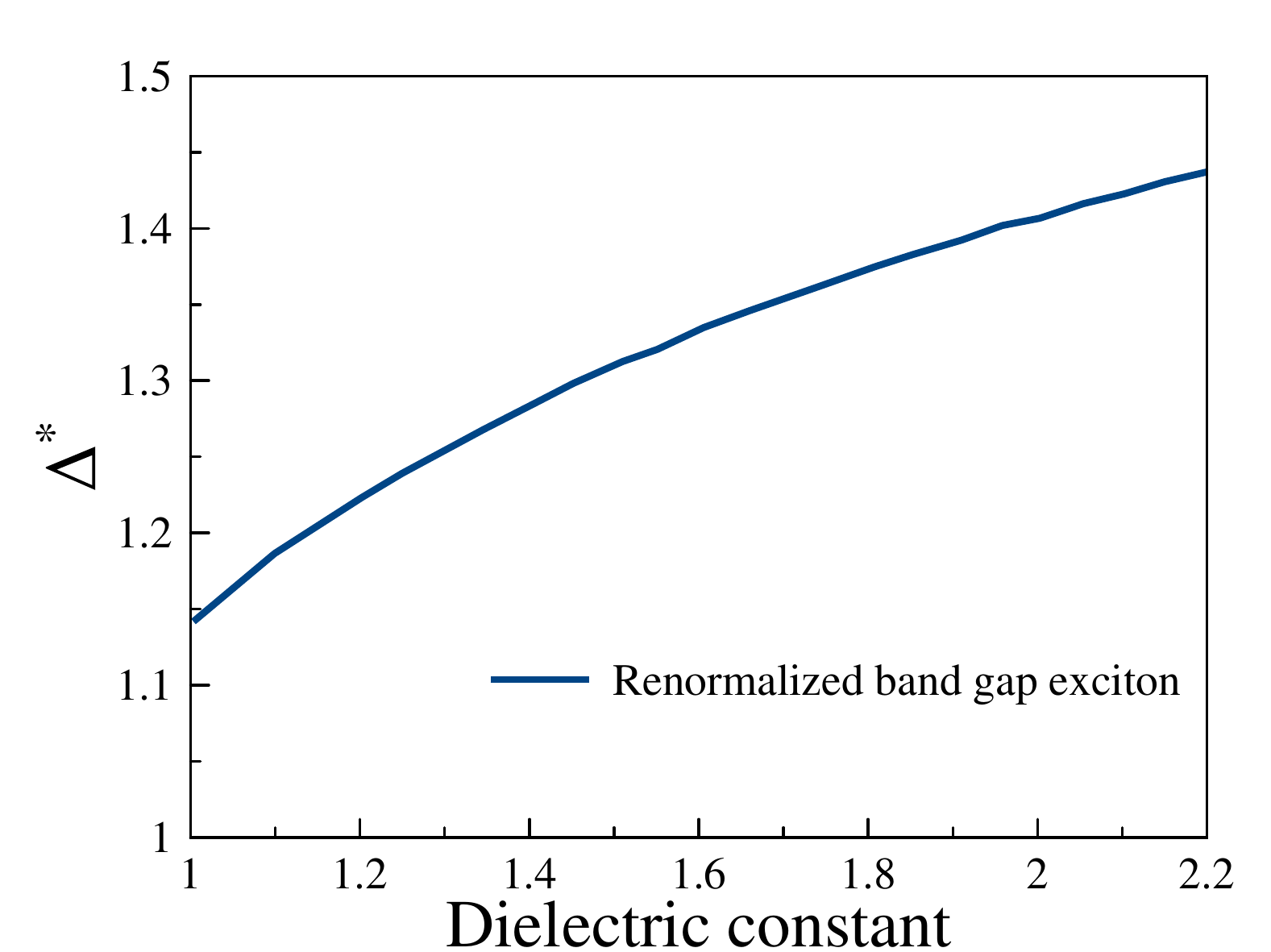}
\caption{The renormalized bandgap exciton, $\Delta^*$, based on the Bethe-Salpeter equation as a function of averaged environment
dielectric constant, $\epsilon$. The bare band gap is 1.82 eV. 
}\label{ren}
\end{figure}

\subsection{Exciton Absorption in density matrix approach}

The interband transitions excited by a homogenous electric field are described by a single-particle operator of perturbation $\delta h \exp(-i\omega t)$ where $\delta h=i(e/\omega)({\bf E}\cdot {\bf v})$ in the basis where eigenvalues and $\langle {\bf v} \rangle=-i{\bf \cal R}^{c \nu }[\varepsilon^{\nu}-\varepsilon^{c}]$ is the interband velocity. From now on, the conduction and valence bands are defined in exciton states. We are interested in the response of the system near the absorption edge when $|\hbar \omega-{\bar \Delta}| \ll {\bar \Delta}$ with the effective bandgap ${\bar \Delta}$. Making use of the eigenstate of noninteracting system, we can obtain the linearized kinetic equation \cite{Vasko2005}
\begin{eqnarray}
-i\omega \delta \rho_{\eta \delta}+&&\frac{i}{\hbar}(\varepsilon_{\eta}-\varepsilon_{\delta})\delta \rho_{\eta \delta}+\frac{i}{\hbar} \delta h (f_{\eta}-f_{\delta})+\nonumber\\
&&+\frac{i}{2\hbar}\sum_{\gamma_1...\gamma_4} \Phi_{\gamma_1...\gamma_4}[\delta_{\gamma_1 \eta}\delta F_{\delta \gamma_2 \gamma_3 \gamma_4}+\delta_{\gamma_2 \eta}\delta F_ {\gamma_1 \delta  \gamma_3 \gamma_4}-\delta_{\gamma_3 \delta}\delta F_{ \gamma_1 \gamma_2 \eta \gamma_4}-\delta_{\gamma_4 \delta}\delta F_{\gamma_1 \gamma_2 \gamma_3 \eta}]=0
\end{eqnarray}
where $\Phi_{\gamma_1\gamma_2\gamma_3\gamma_4}=\int \phi^*_{\gamma_1}({\bf x}')\phi^*_{\gamma_2}({\bf x}) V({\bf x}-{\bf x}')\phi_{\gamma_4}({\bf x})\phi_{\gamma_3}({\bf x}')$
and $\delta F_{\gamma_1 \gamma_2 \gamma_3 \gamma_4}=\delta_{\gamma_2\gamma_3}\delta \rho_{\gamma_4\gamma_1} f_{\gamma_2}+\delta_{\gamma_4\gamma_1}\delta \rho_{\gamma_3\gamma_2} f_{\gamma_4}-\delta_{\gamma_4\gamma_2}\delta \rho_{\gamma_3\gamma_1} f_{\gamma_2}-\delta_{\gamma_3\gamma_1}\delta \rho_{\gamma_4\gamma_2} f_{\gamma_3}$. Notice that $f^{(1)}_{od}({\bf k})=\langle \rho_{12}({\bf k}) \rangle$.

We consider an electron doped system where $f_{\nu}=1$ and $f_c=\theta(\varepsilon_{\rm F}-\varepsilon^c)$ and after some straightforward algebra, the density matrix describing the interband polarization is given by

\begin{eqnarray}
&&(\varepsilon^c-\varepsilon^{\nu}-\hbar \omega-i \hbar \tau )f^{(1)}_{od}({\bf k})+[1-\theta(\varepsilon_{\rm F}-\varepsilon({\bf k})]\delta h({\bf k})+\nonumber\\&&-\frac{1}{4{\cal A}}\sum_{{\bf k}'}V_{{\bf k}-{\bf k}'} 
\{[h^{*}_{12}(k)h_{12}(k') +D^{*}(k)D(k')]_{s=1} [h^{*}_{12}(k')h_{12}(k) +D^{*}(k')D(k)]_{s=-1}\}]f^{(1)}_{od}({\bf k}')=0
\end{eqnarray}
To solve this equation, we do first order integrative approximation and hence
\begin{equation}
f^{(1)}_{od}({\bf k})=\frac{e}{\omega}\frac{[1-\theta(\varepsilon_{\rm F}-\varepsilon({\bf k})] (\varepsilon^c({\bf k})-\varepsilon^{\nu}({\bf k})){\bf E}\cdot {{\bf \cal R}^{c \nu }}({\bf k})}{(\varepsilon^c({\bf k})-\varepsilon^{\nu}({\bf k})-\hbar \omega-i \hbar \tau)}
+\frac{e}{\omega((\varepsilon^c({\bf k})-\varepsilon^{\nu}({\bf k})-\hbar \omega-i \hbar \tau))} {\bf E}\cdot {\bf K}({\bf k})
\end{equation}
where
\begin{eqnarray}
&&{\bf K}({\bf k})=\frac{1}{4{\cal A}}\sum_{{\bf k}'}V_{{\bf k}-{\bf k}'} 
\{[h^{*}_{12}(k)h_{12}(k') +D^{*}(k)D(k')]_{s=1} [h^{*}_{12}(k')h_{12}(k) +D^{*}(k')D(k)]_{s=-1}\}]\times\nonumber\\
&&\times\frac{[1-\theta(\varepsilon_{\rm F}-\varepsilon^c({\bf k}')] (\varepsilon^c({\bf k}')-\varepsilon^{\nu}({\bf k}')) {{\bf \cal R}^{c \nu }}({\bf k}')}{(\varepsilon^c({\bf k}')-\varepsilon^{\nu}({\bf k}')-\hbar \omega-i \hbar \tau)}
\end{eqnarray}
since we are considering the trigonal warping, the last term of the RHS is a complex expression. If we perform another approximation and set $t'=\beta=\beta'=\alpha=\alpha'=0$ only in $K({\bf k})$ and therefore, the trigonal warping carries out by the first term.  Within this simplification, we can have

\begin{eqnarray}
{\bf K}({\bf k})=\int k'dk' &&[(1+\cos \theta_k)(1+\cos \theta_{k'})g(0)+2\sin \theta_k \sin \theta_{k'} g(s)
+(1-\cos \theta_k)(1-\cos \theta_{k'})g(2s)]\nonumber\\
&&\times\frac{[1-\theta(\varepsilon_{\rm F}-\varepsilon^c({k}')] (\varepsilon^c({k}')-\varepsilon^{\nu}({k}')){{\bf \cal R}^{c \nu }}({k}')}{(\varepsilon^c({k}')-\varepsilon^{\nu}({k}')-\hbar \omega-i \hbar \tau)}
\end{eqnarray}
where 
\begin{equation}
g(m)= \frac{1}{2(2\pi)^2}\int_0^{2\pi} d\phi
V(|{\bf k}-{\bf k}'|)\cos (m\phi)
\end{equation}
$\cos \theta_k=\Delta/\varepsilon_k$, and $\cos \phi={\bf k}\cdot{\bf k}'/({|{\bf k}|}{|{\bf k}'|})$.

Again note that $f^{(1)}_{od}({\bf k})$ can be obtained by Green's function in $k-$space. The spectral function is given by the Green's function in real space for which we define the Sommerfeld factor~\cite{Vasko2005}. Sommerfeld factor implies a peak around the optical transition in the spectral function or density of states. Based on that we expect to possess a jump in the current near the optical transition. Put differently, the jump at the optical transition indicates the large density of states or the existence of the Sommerfeld factor.

Let's write down the exciton problem in the relative coordinate with reduced mass $m^*$:
\begin{equation}\label{eq:exciton1}
\{-\frac{\hbar^2}{2m^*}\nabla_{\bf r}-v({\bf r})\}\psi^{n}_{{\bf k}}({\bf r})=\varepsilon^n_{{\bf k}}\psi^{n}_{{\bf k}}({\bf r})
\end{equation}
where the negative of the potential accounts the electron and hole attraction and we use the bare Coulomb potential. In the system we define the effective Bohr radius $a^*_B=\hbar^2 \epsilon/m^* e^2$ where $m^*\sim 0.25 m$. The spectral function of the system is defined as
\begin{equation}
\Psi(\varepsilon({\omega}))=-\Im m \sum_n \frac{|\psi^{n}_{{\bf k}}({\bf r}=0)|^2}{\varepsilon({\omega})-\varepsilon^n+i\eta}=\pi \sum_n|\psi^{n}_{{\bf k}}({\bf r}=0)|^2 \delta(\varepsilon^{n}-\varepsilon({\omega}))
\end{equation}
where $\varepsilon({\omega})=\hbar \omega-(\varepsilon^c-\varepsilon^{\nu})$ and $|\psi^{n}_{{\bf k}}({\bf r}=0)|^2$ called the Sommerfeld factor. 

The solution of Eq. (\ref{eq:exciton1}) for negative energy can be written as a MoS$_2$ wave function (plan wave times to the spinor) and spherical waves. The normalized wave functions are 
\begin{equation}
\psi^{n}_{{\bf k}}({\bf r})=\frac{1}{\sqrt{\cal A}}e^{-{ k}{ r}}\psi_{\text {MoS}_2}\Phi(-n,1; 2kr)
\end{equation}
where ${\cal A}$ is the areal of the system, $k=\sqrt{-2 m^{*} \varepsilon^{n}_{\bf k}/ \hbar}$ and $\Phi$ is the confluent hypergeometric function. $\psi_{\text {MoS}_2}\psi^{\dagger}_{\text {MoS}_2}=1$, hence the Sommerfeld factor is given by
\begin{equation}
|\psi^{n}_{{\bf k}}({\bf r}=0)|^2=\frac{1}{{\cal A}}\frac{8}{\pi a^{*2}_B}
\end{equation}
This expression increases as $m^{*2}$ at a small energy value. Note that the energy is $\hbar^2 k^2/2 m^*$.

In order to calculate the optical current, we de need to evaluate the following expression:
\begin{eqnarray}
\nabla_{\bf k} f^{(1)}_{od}({\bf k})&&=\frac{eE\cos \theta_p}{2\omega}\frac{\delta(\varepsilon({\bf k})-\varepsilon_{\rm F}) \nabla_{\bf k}\varepsilon({\bf k}) (\varepsilon^c({\bf k})-\varepsilon^{\nu}({\bf k}))}{(\varepsilon^c({\bf k})-\varepsilon^{\nu}({\bf k})-\hbar \omega)}(e^{i\omega t}+e^{-i\omega t}){\hat i}\cdot {{\bf \cal R}^{c \nu }}({\bf k})\nonumber\\
&&+\frac{eE\cos \theta_p}{2\omega}\frac{[1-\theta(\varepsilon_{\rm F}-\varepsilon({\bf k})] \nabla_{\bf k} (\varepsilon^c({\bf k})-\varepsilon^{\nu}({\bf k}))}{(\varepsilon^c({\bf k})-\varepsilon^{\nu}({\bf k})-\hbar \omega)}(e^{i\omega t}+e^{-i\omega t}){\hat i}\cdot {{\bf \cal R}^{c \nu }}({\bf k})\nonumber\\
&&-\frac{e \hbar \omega E\cos \theta_p}{2\omega}\frac{[1-\theta(\varepsilon_{\rm F}-\varepsilon({\bf k})] (\varepsilon^c({\bf k})-\varepsilon^{\nu}({\bf k})) \nabla_{\bf k}(\varepsilon^c({\bf k})-\varepsilon^{\nu}({\bf k}))}{[(\varepsilon^c({\bf k})-\varepsilon^{\nu}({\bf k})-\hbar \omega)]^2}(e^{i\omega t}+e^{-i\omega t}){\hat i}\cdot {{\bf \cal R}^{c \nu }}({\bf k})\nonumber\\
&&+\frac{eE\cos \theta_p}{2\omega}\frac{[1-\theta(\varepsilon_{\rm F}-\varepsilon({\bf k})] (\varepsilon^c({\bf k})-\varepsilon^{\nu}({\bf k}))}{(\varepsilon^c({\bf k})-\varepsilon^{\nu}({\bf k})-\hbar \omega)}(e^{i\omega t}+e^{-i\omega t}){\hat i}\cdot \nabla_{\bf k}{{\bf \cal R}^{c \nu }}({\bf k})\nonumber\\
&&+\frac{eE\sin \theta_p}{2i\omega}\frac{\delta(\varepsilon({\bf k})-\varepsilon_{\rm F}) \nabla_{\bf k}\varepsilon({\bf k}) (\varepsilon^c({\bf k})-\varepsilon^{\nu}({\bf k}))}{(\varepsilon^c({\bf k})-\varepsilon^{\nu}({\bf k})-\hbar \omega)}(e^{i\omega t}-e^{-i\omega t}){\hat j}\cdot {{\bf \cal R}^{c \nu }}({\bf k})\nonumber\\
&&+\frac{eE\sin \theta_p}{2i\omega}\frac{[1-\theta(\varepsilon_{\rm F}-\varepsilon({\bf k})] \nabla_{\bf k} (\varepsilon^c({\bf k})-\varepsilon^{\nu}({\bf k}))}{(\varepsilon^c({\bf k})-\varepsilon^{\nu}({\bf k})-\hbar \omega)}(e^{i\omega t}-e^{-i\omega t}){\hat j}\cdot {{\bf \cal R}^{c \nu }}({\bf k})\nonumber\\
&&-\frac{e \hbar \omega E\sin \theta_p}{2i\omega}\frac{[1-\theta(\varepsilon_{\rm F}-\varepsilon({\bf k})] (\varepsilon^c({\bf k})-\varepsilon^{\nu}({\bf k})) \nabla_{\bf k}(\varepsilon^c({\bf k})-\varepsilon^{\nu}({\bf k}))}{[(\varepsilon^c({\bf k})-\varepsilon^{\nu}({\bf k})-\hbar \omega)]^2}(e^{i\omega t}-e^{-i\omega t}){\hat j}\cdot {{\bf \cal R}^{c \nu }}({\bf k})\nonumber\\
&&+\frac{eE\sin \theta_p}{2i\omega}\frac{[1-\theta(\varepsilon_{\rm F}-\varepsilon({\bf k})] (\varepsilon^c({\bf k})-\varepsilon^{\nu}({\bf k}))}{(\varepsilon^c({\bf k})-\varepsilon^{\nu}({\bf k})-\hbar \omega)}(e^{i\omega t}-e^{-i\omega t}){\hat j}\cdot \nabla_{\bf k}{{\bf \cal R}^{c\nu }}({\bf k})\nonumber\\
&&-\frac{eE \cos \theta_p}{2\omega}\frac{\nabla_{\bf k}(\varepsilon^c({\bf k})-\varepsilon^{\nu}({\bf k}))}{[(\varepsilon^c({\bf k})-\varepsilon^{\nu}({\bf k})-\hbar \omega)]^2} (e^{i\omega t}+e^{-i\omega t}){\hat i}\cdot {\bf K}({\bf k})\nonumber\\
&&-\frac{eE \sin \theta_p}{2i\omega}\frac{\nabla_{\bf k}(\varepsilon^c({\bf k})-\varepsilon^{\nu}({\bf k}))}{[(\varepsilon^c({\bf k})-\varepsilon^{\nu}({\bf k})-\hbar \omega)]^2} (e^{i\omega t}-e^{-i\omega t}){\hat j}\cdot {\bf K}({\bf k})\nonumber\\
&&+\frac{eE\cos \theta_p}{2\omega((\varepsilon^c({\bf k})-\varepsilon^{\nu}({\bf k})-\hbar \omega))} (e^{i\omega t}+e^{-i\omega t}){\hat i}\cdot\nabla_{\bf K}({\bf k})\nonumber\\
&&+\frac{eE\sin \theta_p}{2i\omega((\varepsilon^c({\bf k})-\varepsilon^{\nu}({\bf k})-\hbar \omega))}  (e^{i\omega t}-e^{-i\omega t}){\hat j}\cdot\nabla_{\bf K}({\bf k})
\end{eqnarray}
In this stage, we can pluck the above expressions in Eq. (\ref{f}) and calculate second order of the time-independent off-diagonal density matrix component. Before that notice $e^{i\omega t}$ has no physical meaning in the expressions and we should drop that term off.

%\begin{figure}
%	\includegraphics[width=8.6cm]{Jx-exciton}
%	\caption{\textcolor{red}{(Color online) Second-order optical response, $J/I_0$ (in units of pA m/W) as a function of $\hbar \omega$ (in units of eV) for a circularly polarized light, $\theta_p=\pi/4$, the relaxation time is $\tau=2$ ps, the strain $\varepsilon=0.005$ and $m=0.02$ eV for spin-up and down cases in the exciton model for $U_{ele}=-0.02$ eV. We just consider $n=5\times 10^{12}$ cm$^{-2}$.} }\label{fig5}
%\end{figure}

\section{Intrinsic shift-current}
A theory called the shift-current was proposed which attributed the charge separation arising from the asymmetry in the electron and hole wave functions~\cite{morimoto2016topological}. This is an intrinsic effect of the shift current. A two-band model Hamiltonian was considered to explore the interband optical transition. The Floquet Hamiltonian is coupled by time-dependent 
terms, $\langle u_c|\frac{1}{T}\int H({\bf k}-{\bf A}(t))e^{i\omega t}dt |u_{v}\rangle$. The non-linear optical response, $J_j=\sum_{i=x,y}\chi^{ii}_jE_iE_i$, is given by
\begin{equation}\label{eq:na}
{\bf \chi}^{ii}=\frac{\pi e^3}{\hbar^2 \omega^2}\int \frac{d{\bf k}}{(2\pi)^d}\delta(\varepsilon^c_{\bf k} -\varepsilon^v_{\bf k}-\hbar \omega)|v^i_{vc}|^2(\nabla_{\bf k}\phi^i_{cv}+{\cal R}^{cc}-{\cal R}^{vv})
\end{equation}
where $i$ and $j$ are cartesian coordinates and $\phi^i_{cv}=\Im m (\log v_{cv})$ maintains the gauge invariant.
The interband velocity is also defined as $v_{cv}=i(\varepsilon^c_{\bf k} -\varepsilon^v_{\bf k}){\cal R}^{vc}$.
Figure \ref{fig8} shows the shift current response as a function of energy for a circularly polarized 
light calculated from Eq. (\ref{eq:na}).
\begin{figure}
	\includegraphics[width=8.6cm]{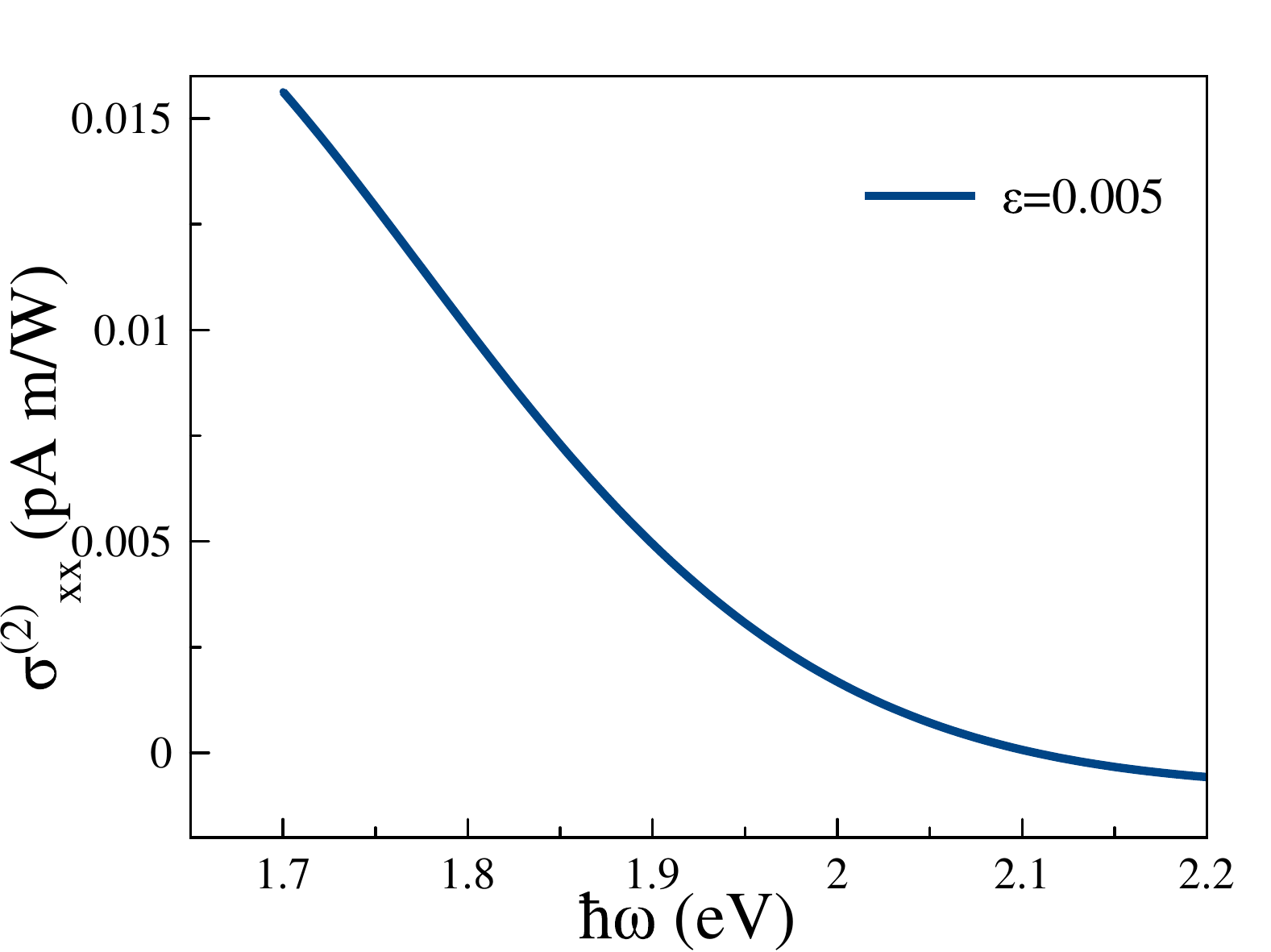}
	\caption{(Color online) Non-linear optical response (in units of pA m/W) as a function of $\hbar \omega$ (in units of eV) for a circularly polarized light, $n=5\times 10^{12}$ cm$^{-2}$ and $m=0.02$ eV, $U_{ele}=-0.02$ eV using Nagaosa's formalism.  }\label{fig8}
\end{figure}

\end{widetext}
\end{document}